\documentclass[aps,rmp,twocolumn]{revtex4-1}
\usepackage{tabularx}
\usepackage{bm}
\usepackage{subfigure}
\usepackage{euscript}
\usepackage{epsfig,psfrag,subfigure}
\usepackage{graphicx,psfrag,subfigure}
\usepackage{color}
\usepackage{amsfonts}
\usepackage{exscale}
\usepackage{amsbsy}
\usepackage{subfigure}

\def\be{\begin{equation}}
\def\ee{\end{equation}}
\def\ba{\begin{eqnarray}}
\def\ea{\end{eqnarray}}
\begin{document}

\bibliographystyle{apsrmp4-1}

\title{Anomalous metals -- failed superconductors}

\author{Aharon Kapitulnik}

\address{Department of Physics and Department of Applied Physics, Stanford University, Stanford CA 94305}

\author{Steven A. Kivelson}

\address{Department of Physics, Stanford University, Stanford CA 94305}

\author{Boris Spivak}

\address{Department of Physics, University of Washington, Seattle, WA 98195, USA}

\begin{abstract}

The observation of metallic ground states in a variety of two-dimensional electronic systems poses a fundamental challenge for the theory of  electron fluids.  Here, we analyze evidence for the existence of a  regime, which we call the ``anomalous metal regime,'' in diverse 2D superconducting systems driven through a quantum superconductor to metal transition (QSMT) by tuning  physical parameters such as the magnetic field, the gate voltage in the case of systems with a MOSFET geometry, or 
  the degree of disorder. The principal phenomenological observation is that in the anomalous metal, as a function of decreasing temperature, the resistivity first drops as if the system were approaching a superconducting ground state, but then saturates at low temperatures to a value that can be  orders of magnitude smaller than the Drude value.  The anomalous metal 
  also  shows a giant positive magneto-resistance. Thus, it behaves as if it were
 a ``failed superconductor.'' This behavior is observed in a 
   broad range of parameters.  We moreover exhibit, by theoretical solution of a model of superconducting grains embedded in a metallic matrix,  that as a matter of principle such anomalous metallic behavior can occur in 
   the neighborhood of a QSMT.  However, we also 
   argue that the robustness and ubiquitous nature of the observed phenomena are difficult to reconcile with any existing theoretical treatment, and speculate about the character of a more fundamental theoretical framework.

 \end{abstract}

\maketitle
\tableofcontents

\section{Introduction}
\label{sec:1}

A metallic state  is defined  as a state in which the conductivity $\sigma(T)$  remains finite as $T \to 0$.  There is an extraordinarily successful 
 Fermi liquid theory of  clean 3D metals with $k_{F}\ell\gg 1$ and relatively weak interactions (i.e. with a small  ratio, $r_s$, of the potential to the kinetic energy).  Here $k_{F}$ and $\ell $ are the Fermi momentum and the electron mean-free-path respectively. In the Fermi liquid theory there are two types of excitations, fermionic and bosonic: Fermionic excitations (quasiparticles) have a finite density of sates at the Fermi level. Bosonic excitations (e.g. zero sound)  roughly can be divided in two groups:  Those associated with the charge excitions have a plasmon spectrum.  Those associated with spin fluctuations have a sound wave spectrum.\footnote{Phonons are also a class of ubiquitous bosonic modes.  They are in a sense neutral, although they can make a contribution to charge transport through the mechanism of ``phonon drag.''}  In principle electric current can be carried by both fermionic and   bosonic excitations.  (See for example Refs.~\cite{BrazovskiiNozieres,nayakorgad}) However,  at low temperatures the contribution of the bosonic excitations to the current is negligible 
 due to  their vanishing density of states. Thus the electronic transport properties are controlled by  the  Fermionic excitations (quasiparticles). 

The low-temperature conductivity of relatively pure 3D metals  is determined by  impurity scattering, and is given by the Drude formula $\sigma_{D}=e^{2}D\nu$.  Here $\nu$ is the electron density of states at the Fermi energy,
$v_F$ is the Fermi velocity,
and $D=v_{F}\ell /3$ is the  diffusion coefficient. 

Another well established paradigm 
  is the BCS theory of superconductivity.   It is based on the idea that under some circumstances the electron attraction can dominate the electron repulsion so that at low temperatures electrons form bosonic Cooper pairs which 
  can condense.  It is this condensate 
that carries the supercurrent.   
 As parameters controlling the electronic environment (e.g. band structure, interactions, or external magnetic field)  change, the system may exhibit  a superconductor to metal transition,  
which at $T=0$ is a quantum transition (QSMT).
As we will discuss in detail, it follows from the conventional theory of metals that in zero magnetic field, the QSMT that occurs as the effective interactions between electrons changes from attractive to repulsive {\em does not have a  quantum critical regime}. In other words, as the system approaches the BCS superconducting state from the metallic side, its properties in no way reflect the proximity of another phase. In particular, the conductivity of the system is controlled by the femionic excitations (quasiparticles) 
everywhere in the metallic phase.

{This picture is supported by a large number of experiments on a variety of systems.
However,  
there exists a variety of experimental systems  
which exhibit a zero temperature transition
from a superconducting state to an  ``anomalous metallic regime'' with  $T\to 0$ electronic properties that cannot be understood on the basis of Fermi liquid/Drude theory.  
  Specifically, the $T\to 0$ conductivity in the anomalous metallic regime can be orders of magnitude larger than the Drude conductivity, there is a giant positive magneto-resistance, 
  and (as has been  observed in at least one case) the Hall response is   anomalous. 
Such behavior has been observed for transitions 
tuned by changing a variety of parameters including the magnetic field, gate voltage,
and  degree of disorder. 

The properties of such anomalous metals is the focus of this article.  
We will argue below that the dramatic signatures in the anomalous metal  are due to the fact that it behaves as a ``failed superconductor,'' a  state in which there are significant  superconducting correlations but nonetheless the system fails to condense even as $T\to 0$.
In other words  in the anomalous metal regime current is carried by bosonic quantum fluctuations of the superconducting order parameter.  Contrary to popular belief, this anomalous metal appears robust even in two spatial dimensions, $d=2$.  It represents a new paradigm for the electronic properties of a metal that is very different from a Fermi liquid.

\subsection{Background}

\subsubsection{Experimentally observed properties of the anomalous metal} 
  
A typical early
observation of an anomalous metal was 
in a 
study of the onset of superconductivity in ultra-thin granular metal films by Jaeger {\it et al.} \cite{Goldman2}
in which the  resistance was observed to level off as $T \to 0$ 
to a value much below the Drude (normal state) value. 

That this represents an anomalous metallic phase emerging from a QSMT was first 
  identified in experiments on the magnetic-field driven transition in relatively low-resistance ($k_F\ell \gg 1$) 
  amorphous Mo$_{1-x}$Ge (a-MoGe)  films \cite{Ephron,MasonKapitulnik1,MasonKapitulnik2}.  There, the anomalous metal was observed
   over a broad range of magnetic fields, exhibiting a low $T$ resistivity that is as much as 3 orders of magnitude smaller than the Drude value. 
  Since then, such a metallic phase proximate to a QSMT 
  has been  found in  many different systems with different tuning knobs.  
  Below we 
  discuss the main experimental observations in the anomalous metal regime, their robustness,  and their significance.  Representative results are shown in Sec. \ref{experiments}. 
 Generic features seen in a wide variety of material-systems and experimental platforms can be summarized as follows:

\begin{itemize}

\item [i)] Most of the evidence for an anomalous metal proximate to a QSMT
 comes from studies of two dimensional systems.    Non-thermal parameters that have been used to  tune  from the superconducting to a non-superconducting state 
 include microscopic and/or macroscopic disorder, carrier density (typically varied by tuning a gate voltage), screening properties of a nearby ground-plane, and a magnetic field (see Sec. \ref{experiments}). 
(Note,  in Sec. \ref{theory} we show that the theoretical rational for the existence of such a state applies as well in 3D.  There have also been numerous experiments on superconducting wires, but since the physics in 1D is quite different than in higher dimensions, we will not survey these results in the present review.) 

\item [ii)] The anomalous metallic state is ubiquitously found in metallic films with  normal state conductance $\sigma_{D}^{(2d)}$ 
that  is significantly higher than the quantum of conductance $
e^{2}/h$. 
  Here   the dimensionless conductance per square of the 2d system $\sigma_{D}^{(2d)}$ is determined either by applying a sufficiently high magnetic field to suppress superconductivity and then extrapolating the measured conductivity to $T\to 0$, or from the  value of $\sigma$ somewhat above the mean-field $T_c$ .

\item [iii)] 
 The anomalous metal  appears  
 as an intermediate regime; 
 when the systems are tuned further from the QSMT, they 
 either exhibit a crossover to a  ``normal metallic phase,'' or a further metal-insulator transition (MIT).
 The range of parameters in which the anomalous metal is observed is often broad (order 1).

\item [iv)] While disorder may be present, or even used as a tuning parameter, there is no obvious dependence of the observed phenomena on the detailed morphology of the disorder. 
 The anomalous metallic state has been observed in strongly non-uniform systems, including  naturally granular systems and  artificially prepared arrays of superconducting ``dots'' on 2D semiconductors or metals.  It is also observed in what otherwise seem to be homogeneous films, both crystalline and amorphous.

\item [v)] In some cases, typically characterized by strong disorder, a direct superconductor to insulator transition (SIT) is observed.  However, we note that much of the published literature that exhibit an anomalous metallic phase has been interpreted  in terms of a putative SIT.  This (sometimes incorrect) interpretation was, in turn, motivated by the theoretical belief that metallic phases are forbidden in 2D. The observation that some films exhibit a SIT while others undergo a QSMT can be understood if one posits the existence of a critical disorder strength\cite{Steiner} (corresponding to $k_F\ell \sim 1$) such that there is a SIT in more disordered films and a QSMT (possibly followed by a 
 MIT) in less.  

\item [vi)] 
Measurements of the Hall effect and finite frequency
conductivity also can reveal
distinguishing characteristics of the anomalous metallic phase. 

\end{itemize}

 \subsubsection{Summary of the theoretical situation.}
 
 We will argue that there is currently no  satisfactory theory of  anomalous metals that accounts for the full set of key  experimental facts,  in particular the  robustness of the anomalous metallic state. We view this as one of the major open problems in condensed matter theory. However, there are  circumstances in which controlled theory is possible and where the existence of an extended $T=0$ quantum critical regime beyond a QSMT has been established.\cite{OretoKivSp,Hruska,LarkinFeigelman}

 Such theoretical considerations are discussed in Sec. \ref{theory}, where we analyze  a ``model system''  of superconducting puddles  embedded in a ``good'' metal. 
  
 The finite temperature  superconductor-metal transition is driven by classical fluctuations, and it takes place when the inter-grain  Josephson exchange energy is comparable to the temperature, $J_{ij}\sim T$. 
 Neglecting quantum fluctuations of the order parameter, one would infer that such a system is always a superconductor  at sufficiently low $T$.  Quantum fluctuations of the phase of an isolated superconducting grain are associated with the charging energy.  However, 
 there is no charging energy for a grain embedded in a metal;  nonetheless, 
 provided the effects of electron-repulsion in the metal are taken into account,
it can be shown that there exists a critical concentration of puddles below which long-range phase coherence is destroyed by quantum fluctuations.
\footnote{A corollary of this analysis is that with attractive interactions only, an electron fluid can only undergo a SIT and will never exhibit an intermediate metallic phase.}
 
In the 
 neighborhood of the 
 resulting QSMT, a substantial fraction of the current is carried by bosonic fluctuations of the  superconducting order parameter.  Thus, as a point of principle,  such a granular system can have an anomalous metallic ground-state ({\it i.e.} without superconducting long-range phase coherence) with a $T=0$ conductivity that diverges upon approach to the quantum critical point (QCP).  Such a system   also exhibits a large positive magnetoresistance. 
 
However, while these considerations address the point of principle, they do not  account for the broad range of temperatures and tuning parameters over which anomalous metallic behavior is observed.  At issue is the fact that  the Ginzburg-Levanyuk parameter \cite{Levanyuk1959,Ginzburg1961}, which typically characterizes the width of the fluctuational regime near a critical point, is very small in most 
 of the relevant experimental systems.  This seemingly implies a narrow range of parameters where significant quantum fluctuations exist.
 In 
 Sec. \ref{conclusion} we consider other possible 
  mechanisms for the QSMT that 
 could  pertain even to uniform 
 systems.

  We also  discuss  the role of ``rare events'' on  the quantum critical transition in Subsection \ref{griffiths}. 
Systems that exhibit only slight non-uniformities in their electronic structure when far from criticality show amplified effects of small inhomogeneities when tuned close to a QCP.  In the case of quantum phase transitions involving a discrete symmetry breaking, such considerations\cite{DSFisher1,DSFisher2} can lead to a  ``quantum Griffiths phase,'' {\it i.e.} a range of parameters of finite measure in the vicinity of the QCP in which the appropriate thermodynamic susceptibility diverges.  
It was shown in Ref. \cite{OretoKivSp} that 
in the case of the QSMT, while there
exist 
 circumstances in which the effect of rare regions 
are highly amplified, 
  they can never be strong enough to produce a true quantum Griffith phase.

\subsubsection{ Is 2D  localization relevant in the anomalous metal regime?}

Experimentally,  most reports of anomalous metals involve two-dimensional (2D) samples. Thus, a natural question arises concerning the relevance of 2D localization effects, which have been a key feature 
of the theory of transport phenomena in the presence of disorder.  2D localization theory is based on the observation that in the absence of interactions, 
 the first correction to Drude theory in powers of $1/k_{F}\ell$ diverges logarithmically 
as $T\to 0$.   Renormalization group analysis, assuming one parameter scaling, 
 leads to the inference that $\sigma(T) \to 0$ as $T\to 0$.
\cite{GangofFour,LarkinKhmelnitskiiGorkov} (For a review see \cite{LeeRamakrishnan}.)  This hypotheses has been confirmed by numerical solution of  the Schr\"odinger equation for a single particle in a disordered  medium.  (See, for example Refs.~[\cite{Schreiber,Markos06}]). In other words, in the absence of interactions and spin-orbit scattering, 2D metals do not exist.  The question of 2D localization in disordered metals with electron-electron interactions is more complicated, and in spite of  extensive theoretical effort there is still no full understanding of the problem. 

 In our opinion, 
 2D localization in the presence of interactions has never been unambiguously proven experimentally. 
 In order to see the predicted crossover to insulating behavior  for systems with  $k_{F} \ell \gg 1$, one would have to measure the conductance at exponentially low temperatures,
 \be
T<T^\star\sim E_F \exp[-\pi k_F \ell].  
\label{Tstar}
\ee

In any case, for the purposes of the present paper, we   
can ignore 
``localization'' effects, including  interactional ones \cite{AronovAltshuler,LeeRamakrishnan,Finkelstein2}, for several reasons:   i) In 
most cases the experiments we are interested in are carried out in the range $T \gg T^\star$. 
ii) The fact that the low $T$ conductivity 
is typically orders of magnitude {\it larger} than $\sigma_{D}^{(2d)}$ implies that 
 the starting point of the perturbative RG consideration in Refs.~\cite{GangofFour,LarkinKhmelnitskiiGorkov} is  inapplicable in the present circumstances.
iii)  Finally, bosonic excitations are not subject to weak localization.

On the other hand, weak localization effects are cut off by the superconducting gap, $\Delta_0$.   Therefore the  superconducting state is robust for $k_F \ell \gg 1$ and $T=0$
 so long as $\Delta_0 > T^\star$.

 \section{
 Experiment}
 \label{experiments}

In this section, we discuss multiple examples of 
experimental systems in which an anomalous metallic phase is found to exist
 as $T\to 0$ proximate to a superconducting phase.   Various ``knobs'' are used to tune these systems from a superconducting to a non-superconducting ground state including gate voltage, film thickness, or an applied magnetic field.  
 It is important to stress that the nature of the anomalous metal is roughly similar in all cases.

The systems we discuss in this section are all in some essential way two dimensional (2D).   
One reason for this is that it is relatively easier to tune 2D systems through a QSMT.  However, on the basis of the theoretical considerations of Sec. \ref{theory}, there does not appear to be any reason that similar phenomena are excluded in 3D. 
In relatively pure 3D samples,  magnetic field tuned superconductor-metal transitions 
have been studied for decades. For most 3D superconductors, the transition to the ``normal state" can be satisfactorily described by the usual mean-filed description 
 of the upper critical field $H_{c2}$. However, 
 in some circumstances classical melting of the vortex lattice is observed before the mean-field $H_{c2}$, and as $T \to 0$, this may cross over to quantum melting. While the thermal melting transitions are now reasonably well understood \cite{Blatter1994}, the transition to the quantum-dominated regime is  not fully understood.  We will return to the issue of the QSMT in 3D in Sec. \ref{conclusion}.

\subsection{Distinguishing insulators, metals, and superconductors} 
The defining feature that distinguish metals from insulators is  the value of the conductivity in the limit $T\to 0$;  it vanishes in an insulator and approaches a finite limit in a metal.  The resistivity vanishes in a superconductor -- in some cases below a non-zero critical temperature, but in other cases (as in a 2D superconductor in the presence of a magnetic field) only in the limit $T\to 0$.  

Alas, experiments are always confined to non-zero temperatures.  We are thus always faced with the task of inferring the character of  ground-state phases based on low temperature measurements.  In doing this, it is important to pay attention both to the magnitude and the temperature dependence of the resistivity.  Even in conventional, 3D metals, there are circumstances in which the resistivity is an increasing or a decreasing function of $T$ at  low temperatures, so the sign of $d\rho/dT$ cannot be taken (as it sometime is in the literature) as the defining feature of a metal.  Rather, the relevant analysis involves fitting the measured $T$ dependence of $\rho$ to an appropriate functional form, and then using this fit to extrapolate the results to $T\to 0$.  

In most cases we will be reviewing, $\rho(T)$ in the anomalous metallic phase is essentially $T$ independent for a range of accessible low temperatures, so the extrapolation to $T=0$ is obvious.  In other cases, where the $T$ dependence remains strong down to the lowest observable temperatures, this extrapolation is more dangerous.  In all cases, it is also important to pay attention to the magnitude of the resistivity:  when $\rho(T) \gg h/e^2$ at the lowest temperatures, it is a priori reasonable to expect that it will diverge as $T\to 0$, while conversely it would be rather unexpected to encounter a low temperature regime in which $\rho(T) \ll h/e^2$ in a system that is tending toward an insulating ground-state.

\subsection{SIT vs QSMT} 
\label{SIT}

\begin{figure}
\begin{center}
\includegraphics[scale=0.15]{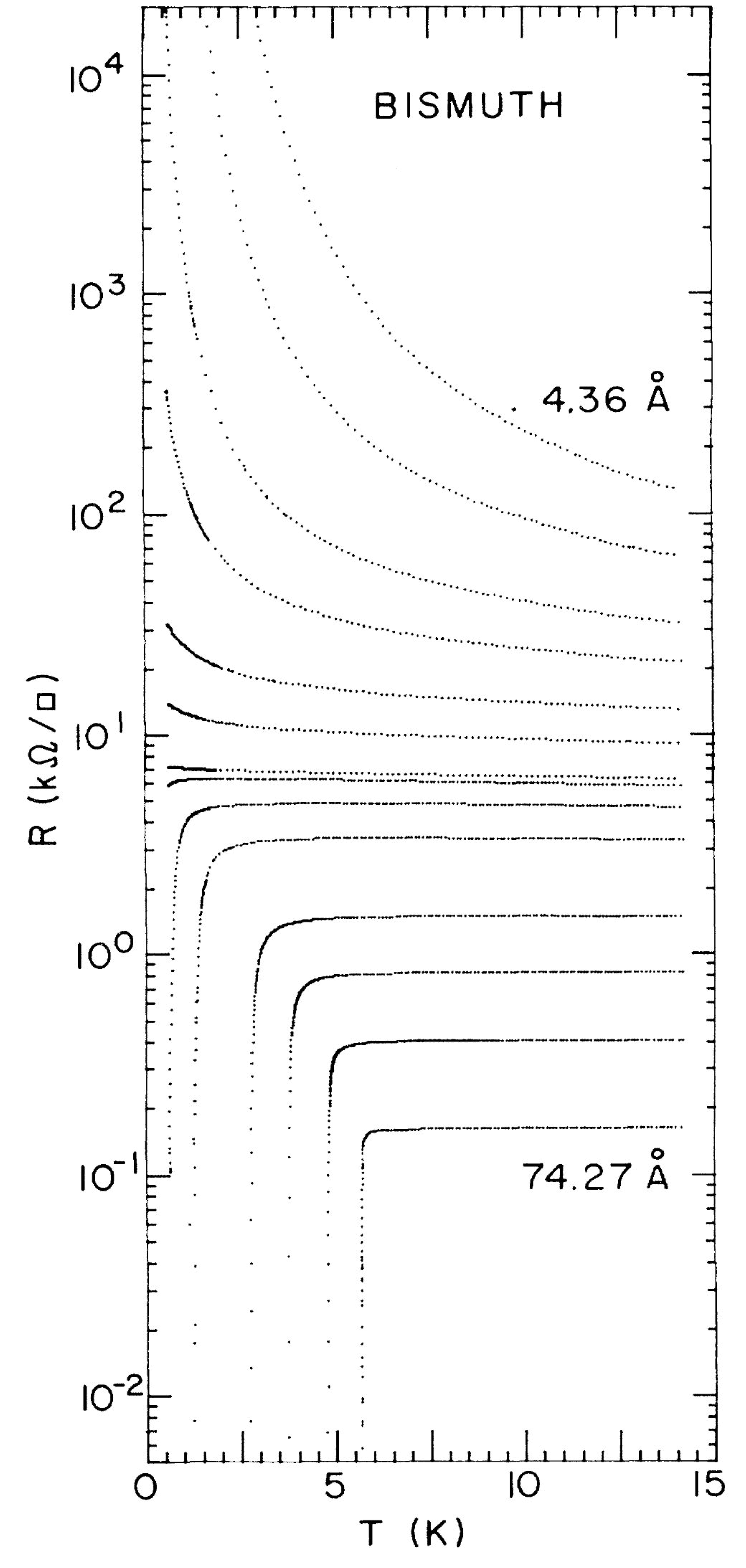}
\caption{The iconic figure of a superconductor to insulator transition in amorphous Bi films of varying film-thickness.  (Thicker films have lower resistance.)  From Ref.~\cite{GoldmanIconic}.}
\label{iconic}
\end{center}
\end{figure} 

As mentioned in the introduction,  early studies of 2D systems were interpreted in the context of a  scaling theory\cite{MPAFisher1990} of the  SIT.  Some of this data appears in the present review, but now interpreted as showing evidence of a QSMT.  
To avoid confusion, we begin with a discussion of this ``historical'' point.  

Typically, in  studies of a putative SIT, a state was identified as ``superconducting'' if the resistivity at low $T$ was an increasing function of $T$ and ``insulating'' if a decreasing function.   In some cases  the experimental data  
  may be consistent with the assumption that there is a direct transition  
  with no intermediate metallic phase.  
  For example, the early study of 
  Ref. \cite{GoldmanIconic} (see Fig.~\ref{iconic}) shows an evolution of the temperature dependence of the sheet resistance R(T) with increasing  thickness of an amorphous Bi film  deposited onto Ge. 
At the separatrix, the resistance is $T$ independent and has a value $\rho\approx (1/4) h/e^2$ corresponding $k_F \ell \approx 4.$
\footnote{The value of the resistance on the separatrix was identified in Ref. \cite{GoldmanIconic} as the Cooper pair quantum of resistance -- $h/(2e)^2$ -- in agreement with the prediction of Ref. \cite{MPAFisher1990}
 which was based on an idea that of localization of Cooper pairs. We would like to note however that  the resistivity at the separtrix  is $T$ independent up to  10 times or more than the maximal $T_c$ where the Cooper pairs do not exist. Therefore it should be associated with a more conventional Drude theory rather than with quantum critical diffusion of charge $2e$ bosons.  Note, in other cases, where a clear separation between the normal state and the regime of superconducting fluctuations is observed, the notion that the critical conductivity is associated with a self-dual point of charge $2e$ bosons - and hence has a value $h/(2e)^2$ - has some experimental support. See, for example, \cite{KapKivDualityPaper} }

 \begin{figure}
\begin{center}
\includegraphics[scale=0.15]{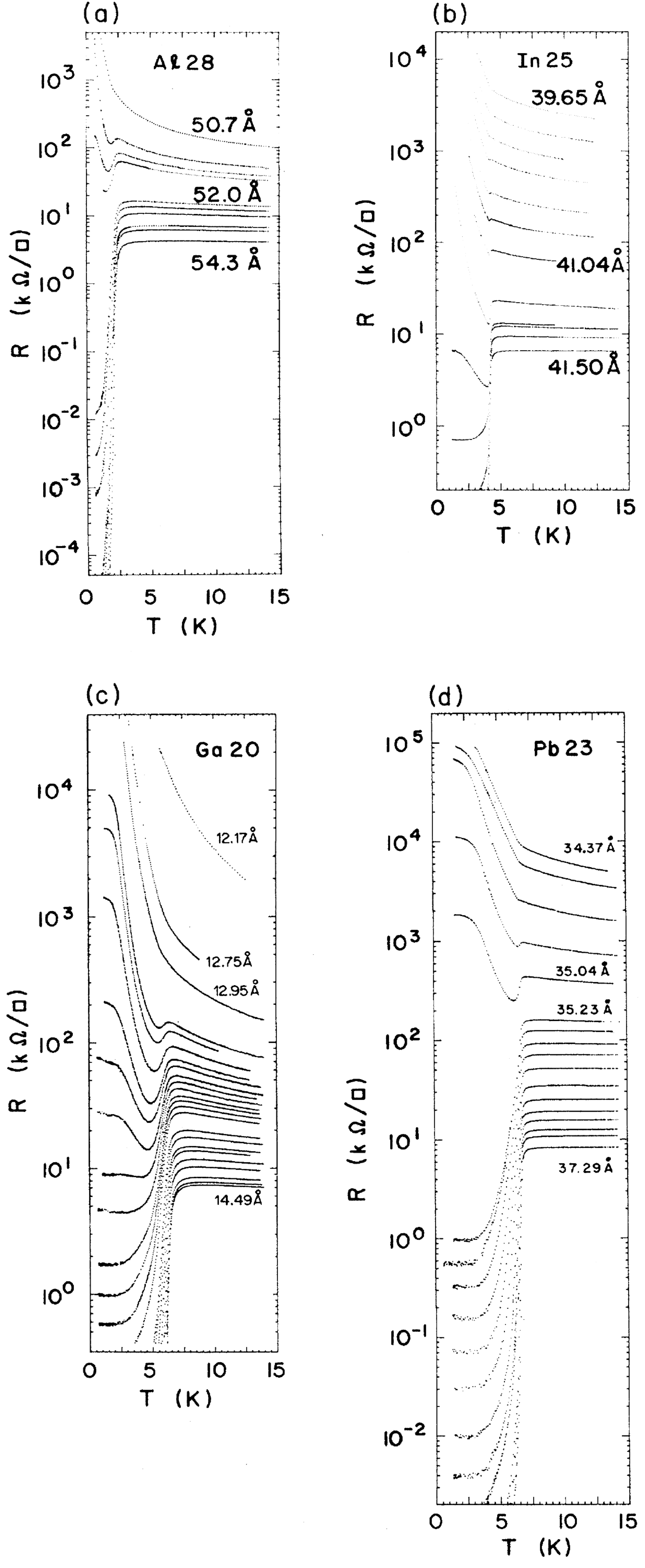}
\caption{Resistance (on a logarithmic scale) vs. $T$ for a sequence of ``granular'' films of a) Al, b) In, c) Ga, and d) Pb where for each subsequent film, a small amount of metal is added to the previous film increasing the nominal thickness of the film.  From \cite{Goldman2}.}
\label{Goldman2}
\end{center}
\end{figure}
Another example of a set of data that was so interpreted is shown in Fig. \ref{Goldman2}.  Here,
using a similar technique to gradually increase a film's thickness by depositing at very low temperatures, 
the evolution of the 
sheet resistance R(T) with thickness was studied  for 
various metallic elements.  Qualitative differences in the $T$ dependences are   apparent between the thicker films (with lower normal state sheet resistance) and thinner films --  as $T$ decreases below a characteristic scale (presumably associated with the onset of local superconducting pairing),  the resistivity of the thicker films drops percipitously while in the thinner films it increases.  
The  existence of an approximately thickness independent pairing scale was (reasonably) taken as indication of a granular morphology of the films.  However, importantly from the current perspective, at still lower temperature, the resistance of the near critical films does not vanish at a well-defined finite temperature transition, but rather levels off to a value well below the normal state value.

Similar results are apparent in data of Ref. \cite{WhiteDynes} for thin layers of Sn and Pb on helium-cooled glass substrate.
(See Fig.~\ref{White}.)
\begin{figure}
\begin{center}
\includegraphics[scale=0.4]{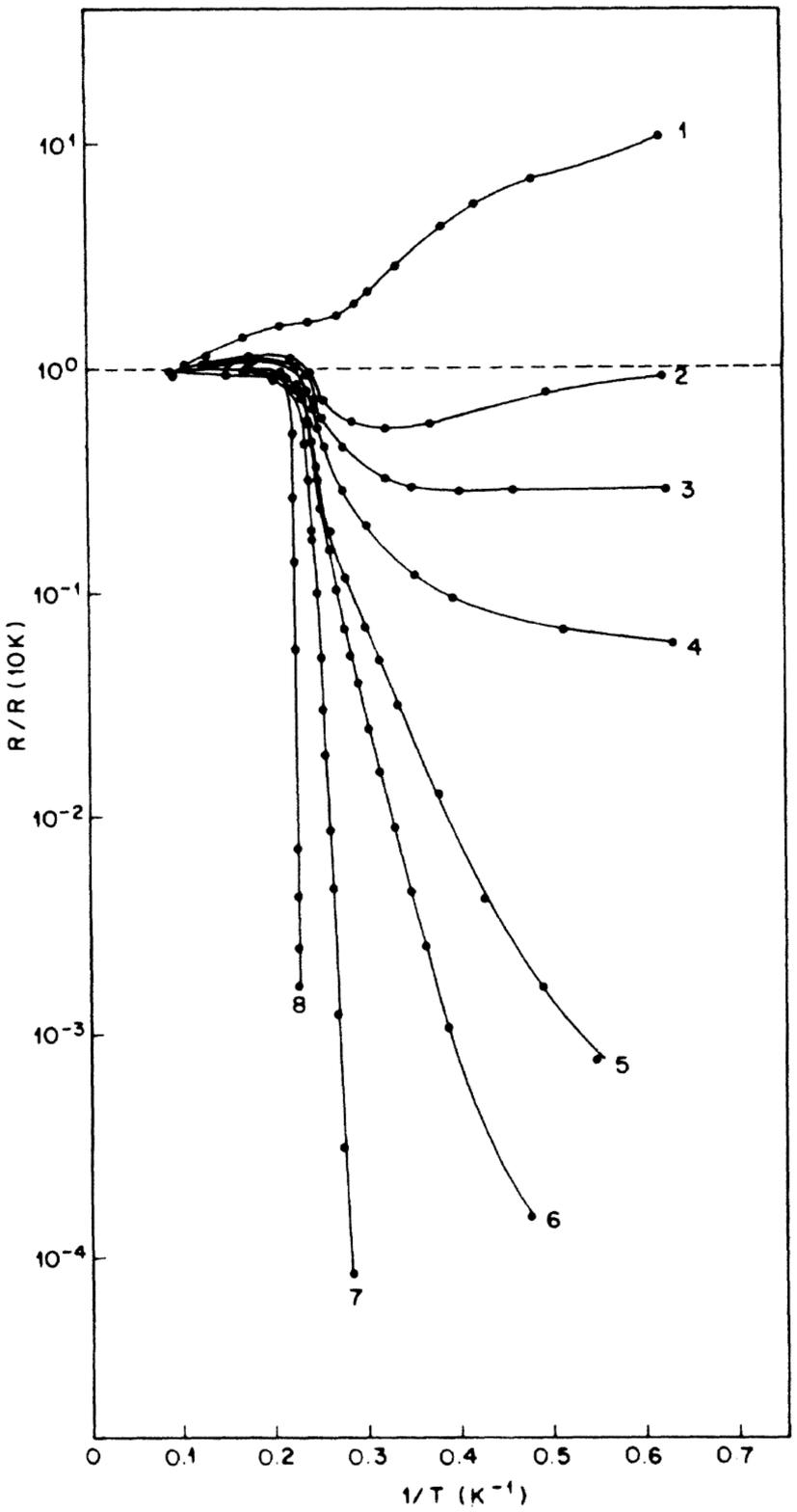}
\caption{Resistance (on a logarithmic scale) vs. $1/T$ for a sequence of  Sn films of varying thicknesses.  From \cite{WhiteDynes}.}
\label{White}
\end{center}
\end{figure}

Data that approximately satisfies scaling relations\cite{MPAFisher1990} expected in the critical regime of a magnetic field driven quantum SIT were obtained in Ref. \cite{HebardPaalanen} (not shown).  Here, a transverse 
 magnetic field was used to tune a thin film of disordered (mostly amorphous) indium-oxide (InO$_x$) from a state in which the resistance decreases as a function of decreasing temperature    to a state where the resistance increases in an activated fashion. 
The SIT was associated with the existence of a thermal ``crossing point,'' corresponding to a  critical field at which the sign of the temperature derivative of $R$ goes from positive to negative.  (An example of such a crossing point is shown in the inset of Fig. \ref{YazdaniEphron}a.)
\begin{figure}
\begin{center}
\includegraphics[scale=0.3]{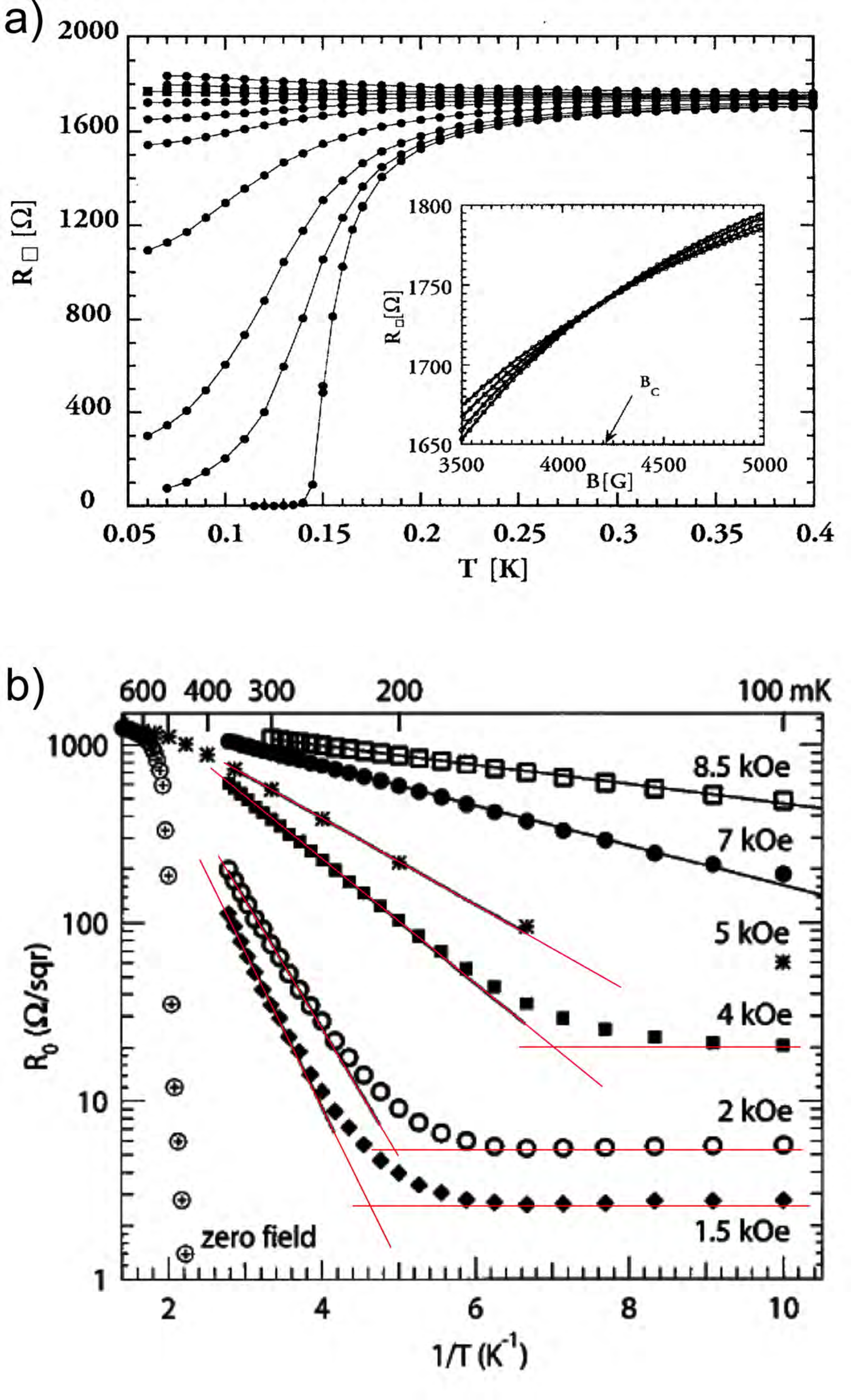}
\caption{a) Resistance  vs. temperature for an a-MoGe film at a sequence of fixed magnetic fields. (Inset shows the putative crossing point of the isotherms from \cite{Yazdani}.) 
b) Resistivity on a logarithmic scale vs $1/T$ for an a-MoGe film at various values of magnetic field, $H$;
the low $T$ saturation is evidence of the existence of an 
anomalous metallic phase.}
\label{YazdaniEphron}
\end{center}
\end{figure}

It is an important open question that should be revisited under what circumstances a direct SIT can occur without a (possibly narrow) intervening metallic phase.  In the remainder of this Section, we focus  exclusively on experiments in which the existence of an anomalous metallic phase  is clear.  In some cases, 
 these studies involve films with $k_F\ell \gg 1$. 
As far as we know, whenever $k_F \ell \gg 1$ at the point of the quantum phase transition from the superconducting state, the proximate phase is always  a metal.  
No similarly catagorical statement can be made concerning systems with $k_F \ell \sim 1$;  
 however, as we shall see, many such systems also exhibit clear anomalous metallic phases.
 \footnote{A way to reconcile the differences between 
 systems that exhibit a QSMT vs 
 a SIT was proposed in Ref.~\cite{Steiner}.  Rather than focussing on the value of $k_F\ell$, they proposed that there are two distinct behaviors depending on the value of the critical conductivity, $\sigma_c$, defined as the $T\to 0$ limit of the conductivity at the point at which superconductivity is destroyed.  Where $\sigma_c  < 4e^2/h$, there is generally a SIT.  On the other hand, where $\sigma_c \gg h/4e^2$   there is a QSMT.}

 \subsection{
 Magnetic Field Driven QSMT}

The fact that magnetic fields can be tuned continuously, and that in almost all cases superconductivity can be quenched in  accessible field ranges, makes the magnetic field driven transition particularly suitable for experimental study.   However, there are possibly special aspects that are associated with field-induced vortices, and with the breaking of time-reversal symmetry that could, in principle, distinguish the field-induced QSMT from other cases.  
Nonetheless, in Sec. \ref{gate} we will show that many aspects of the problem appear to be the same whether or not a magnetic field is present.

Figs.~\ref{YazdaniEphron} and \ref{exp1MassonKapitulnik} show data 
from a field driven transition in highly metallic a-MoGe  
from Refs.~\cite{Ephron} and \cite{MasonKapitulnik1,MasonKapitulnik2}. 
The ``normal state'' resistivity of these films, $\rho_N=1/\sigma_D^{(2d)}$, measured at temperatures somewhat above the zero field $T_c$, or at $T=0$ and large $H$, is small compared to the quantum of resistance, implying that $k_F\ell \gg 1$ ( Typical values of the Drude conductivity  in this case are in the range $\sigma_{D}^{(2d)}\sim 20- 40 \times e^{2}/h$.)  Moreover, the high field  resistance is only weakly $H$ and $T$   dependent, as is expected in a range of fields in which  $\omega_c \tau \ll 1$. (Here $\omega_{c}$ is the cyclotron frequency, $\tau = \ell/v_F$ is the transport lifetime.)    At  smaller $H$, there exists a broad range of intermediate fields in which the resistivity first drops dramatically with decreasing temperature, and then saturates at a low $T$ ``plateau'' value  that can be as much as 3-4 orders of magnitude smaller than $\rho_{D}$. 
Assuming that a $T$ independent $\sigma(H,T)$ can be extrapolated to $T=0$,
this data implies the existence of a well defined  metallic quantum phase of matter. 
Moreover, the extent of this phase can be explicitly delimited:  On the high field side, it is bounded by the above mentioned crossing point that was previously associated with a SIT, but which is now to be associated with either a metal-insulator transition (MIT), or possibly a crossover from an anomalous to a more conventional metal.  On the low field side, a later study in Ref. \cite{MasonKapitulnik1,MasonKapitulnik2} identified a critical field that marks the phase transition between a fully superconducting phase (in which, within experimental uncertainty, $\rho \to 0$ as $T\to 0$) at low field and the anomalous metal at higher fields.  This situation is sketched in the qualitative phase diagram in Fig. \ref{Hphase}.

\begin{figure}[h]
\begin{center}
\includegraphics[scale=0.27]{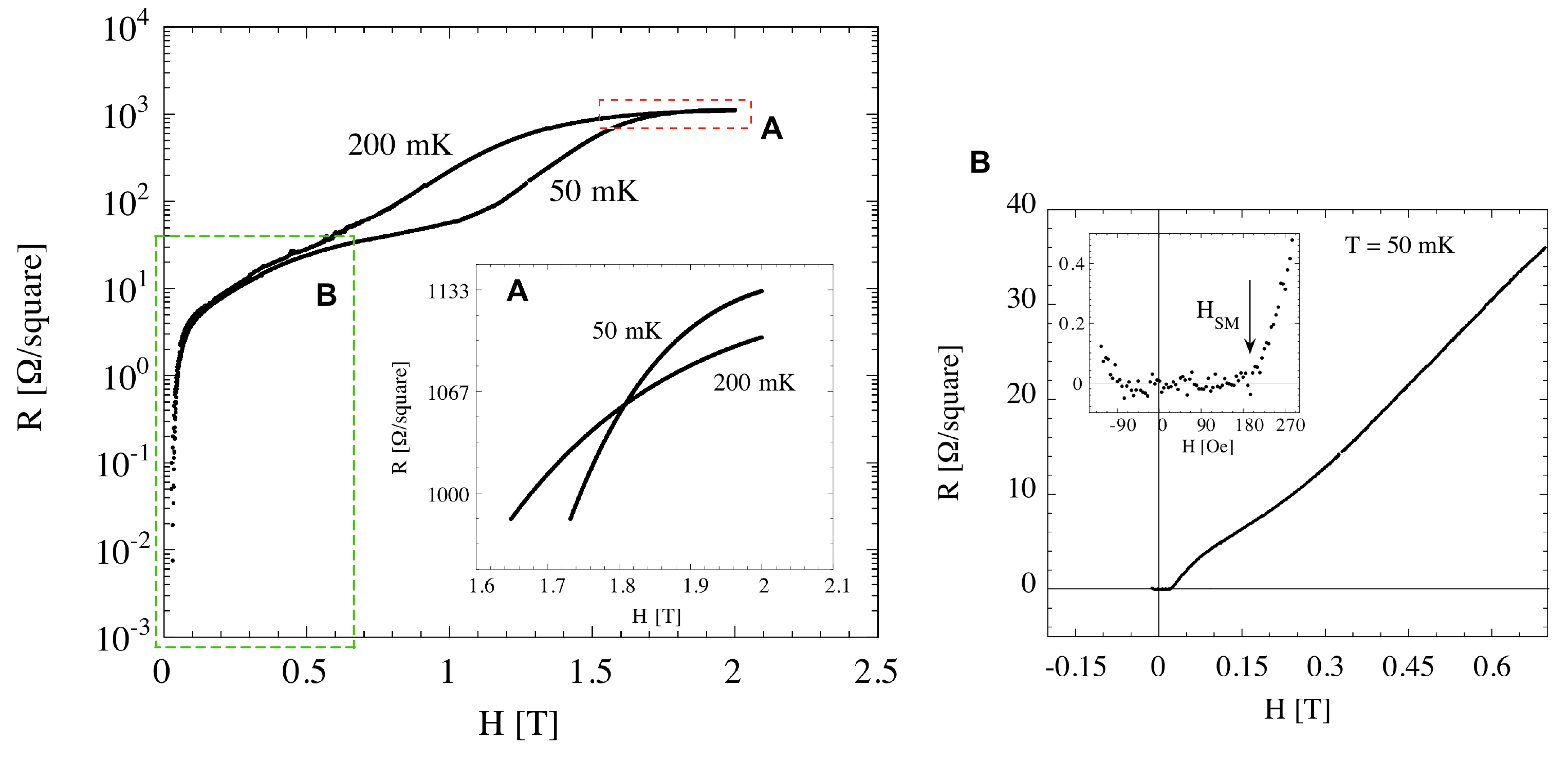} 
\caption{
The magnetic field dependence of the low temperature resistivity of  a highly metallic an amorphous MoGe film, shown on on a logarithmic.  (Over most of this field range, $R$ is essentially temperature independent below 100mK.)  The inset shows a ``crossing point'' at an apparent critical field of approximately 1.8T.  B)  A expanded version of the lowest temperature curves shown as the dashed rectangle in A.  The inset shows that, within experimental error,  a zero resistance state is found below a QSMT at $H \approx 0.18$T.   From Ref.~\cite{MasonKapitulnik1,MasonKapitulnik2}.}
\label{exp1MassonKapitulnik}
\end{center}
\end{figure}

\begin{figure}[h]
\begin{center}
\includegraphics[scale=0.4]{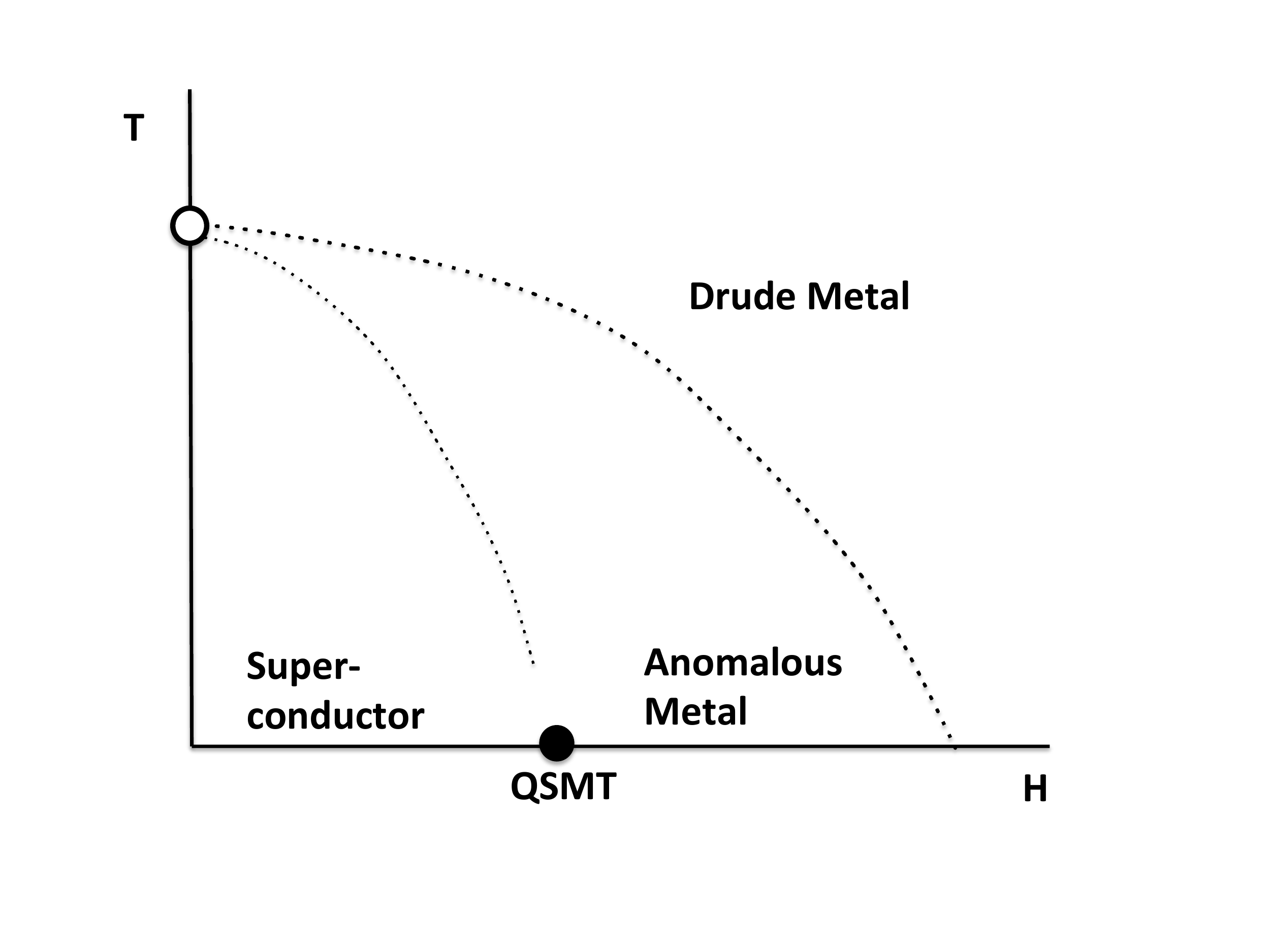} 
\caption{
Schematic phase diagram for the magnetic field driven QSMT.  The open circle represents the thermal transition at $H=0$ and the solid circle the QCP associated with the QSMT.  The dashed curves represent possible crossovers.  The anomalous metal may be bounded at high $H$ by an insulating phase, in which case there would be a QCP associated with a MIT at the end of the upper crossover line.  Alternatively, there could be a quantum crossover to a metallic phase dominated by fermionic excitations.}
\label{Hphase}
\end{center}
\end{figure}

Similar field driven QSMTs with an anomalous metal regime have been  observed in a diverse range of material systems with different morphologies.  Below we show data on field-tuned anomalous metal phases for homogeneously disordered superconducting tantalum thin films (Fig.~\ref{Yoon}, from  \cite{Qin2006}), 
 amorphous tantalum-nitride (TaN$_x$) and  indium-oxide (InO$_x$) films (Fig.~\ref{nick}, from \cite{Breznay2017}).
\begin{figure}[h]
\begin{center}
\includegraphics[scale=0.15]{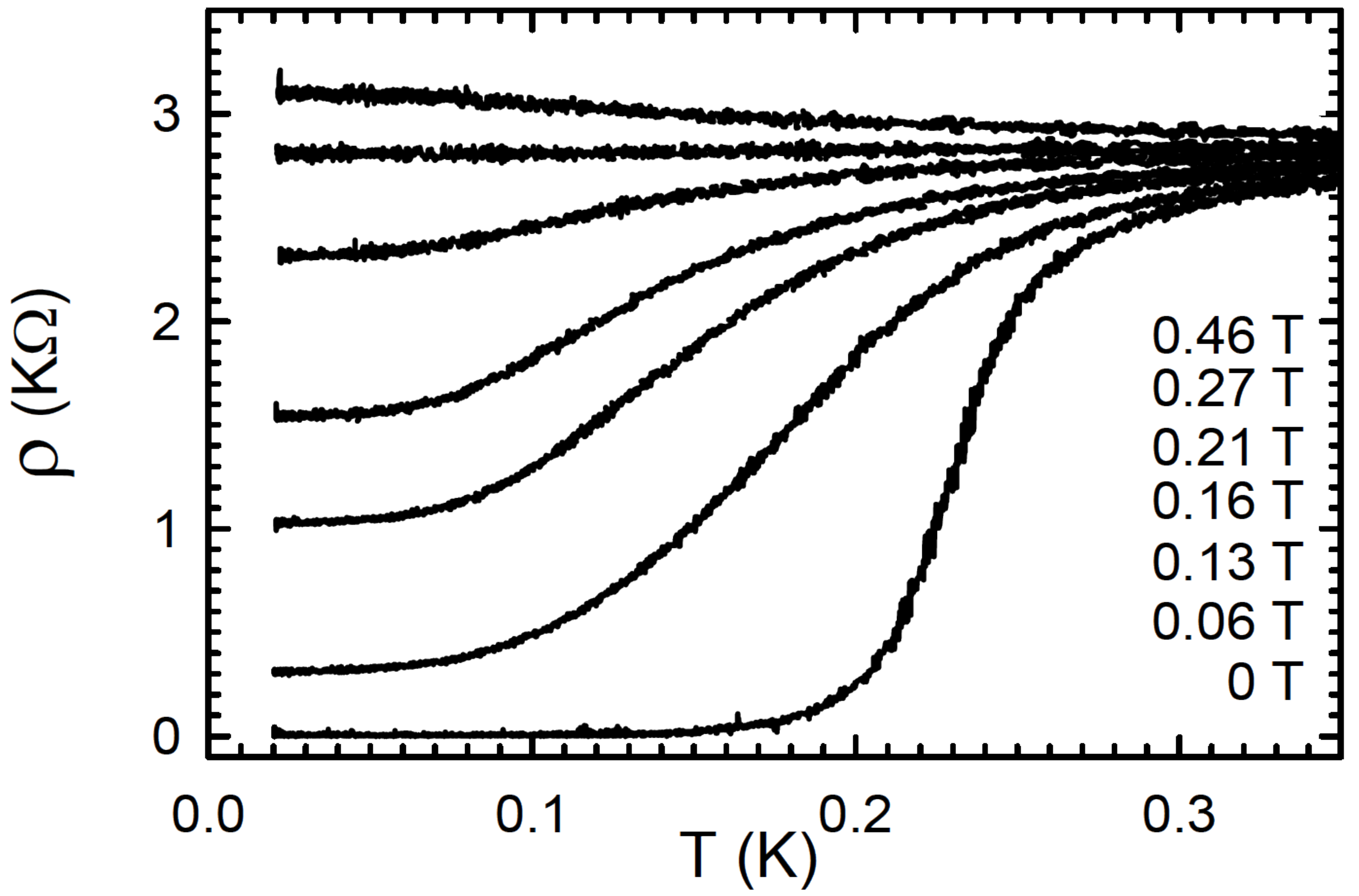} 
\caption{Resistivity as a function of $T$ for a Ta film for distinct magnetic fields equally spaced between 0T and 5T (from \cite{Qin2006}).}
\label{Yoon}
\end{center}
\end{figure}
\begin{figure}[h]
\begin{center}
\includegraphics[scale=0.17]{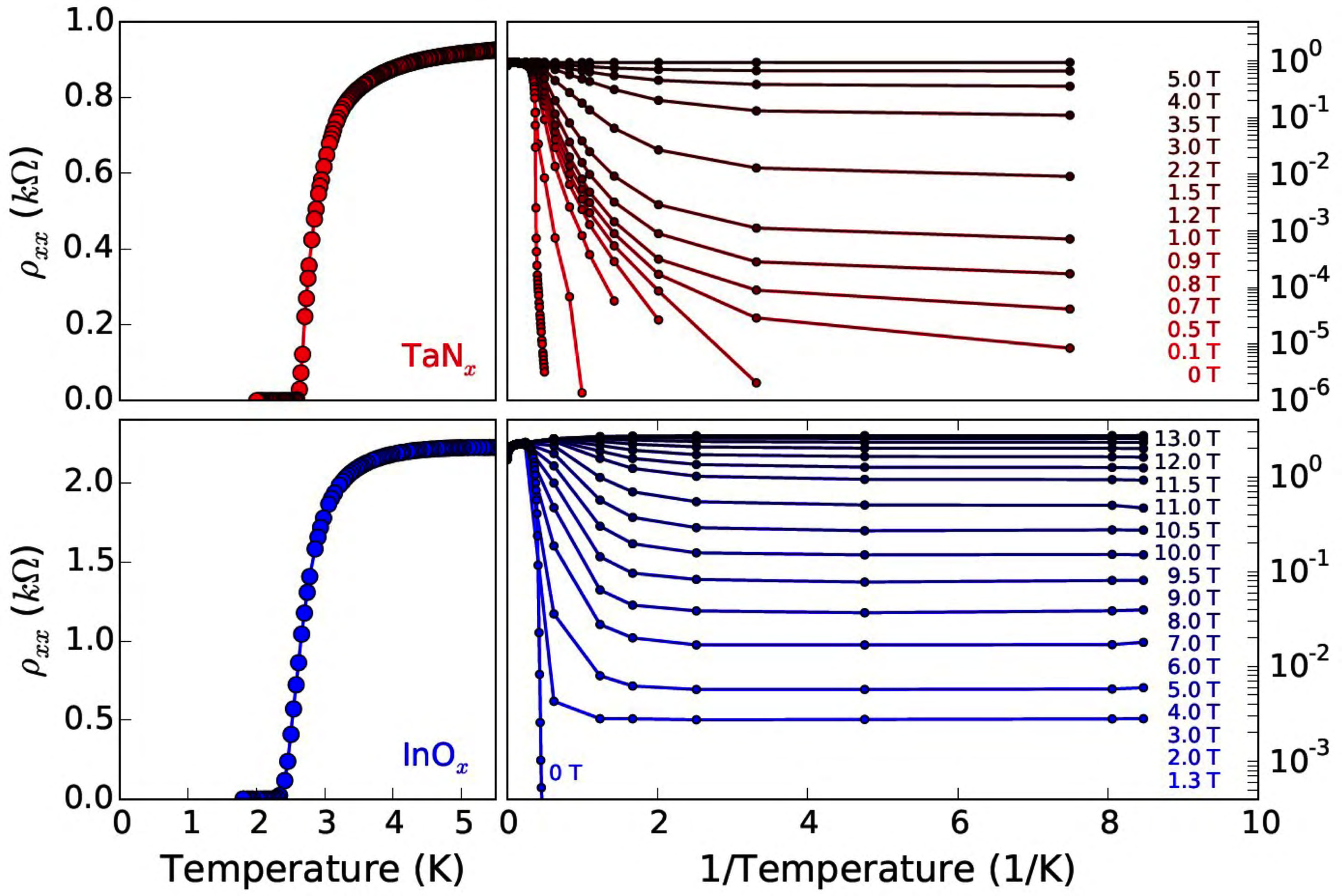} 
\caption{Resistivity as a function of $T$ for a TaN$_x$ and a InO$_x$ films.  The left-hand panels show the superconducting transition in resistance vs $T$  for $H=0$.  The right-hand panels show the resistance on a logarithmic scale as a function of $1/T$ for various values of the applied magnetic field (from \cite{Breznay2017}).}
\label{nick}
\end{center}
\end{figure}

While early field-tuned measurements demonstrated the emergence of a metallic phase in ``homogeneously disordered'' films, recent results on highly crystalline materials reinforce the idea that the important parameter is the initial high conductance of the films (that is, $k_F\ell \gg 1$), rather than the disorder per se. For example, 
Ref. \cite{Saito2015} reported transport studies on a single-crystalline flake of ZrNCl, which is ion-gated, hence allow for the tuning of the interface carrier density.  In particular, they found that the zero resistance state is destroyed by the application of finite out-of-plane magnetic fields, and  a metallic state is stabilized in a wide range of magnetic fields (Fig.~\ref{Saito}). It is interesting to note how remarkably similar is this data to the 
  measurements on a-MoGe \cite{Ephron}.
\begin{figure}
\begin{center}
\includegraphics[scale=0.15] {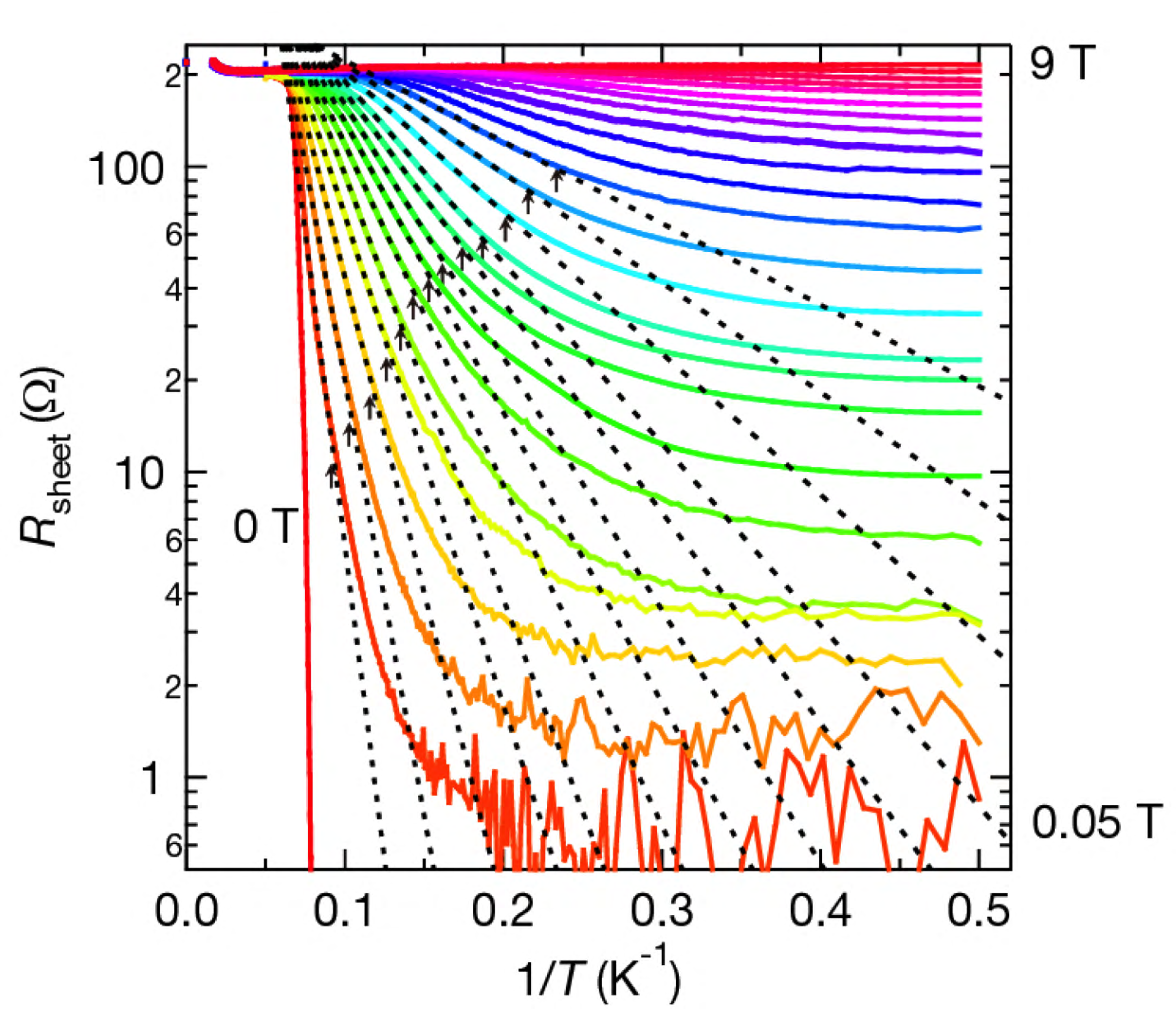} 
\caption{Arrhenius plot of the sheet resistance of an electric double layer transistor (EDLT) of ZrNCl at gate voltage $V_G = 6.5$ V for different magnetic fields perpendicular to the surface. The black dashed lines demonstrate the activated behavior with activation energy $U(H)\propto {\rm ln}(H_0/H)$, similar to Ephron {\it et al.} \cite{Ephron}. The arrows separate the thermally activated state in the high-temperature limit and the saturated state at lower temperatures. From  \cite{Saito2015}}
\label{Saito}
\end{center}
\end{figure}

Also recently, layered transition-metal dichalcogenides (TMD), which often show superconductivity, either naturally, or upon intercalation, have been thinned down (exfoliated) to study the occurrence of superconductivity in the 2D limit. In Fig.~\ref{exp2Pasupathy2015} we show results on NbSe$_2$ by Tsen {\it et al.} \cite{Tsen2016} where the resistance saturates at least at two-orders of magnitudes lower value than the ``Drude resistance.'' 

\begin{figure}
\begin{center}
\includegraphics[scale=0.15]{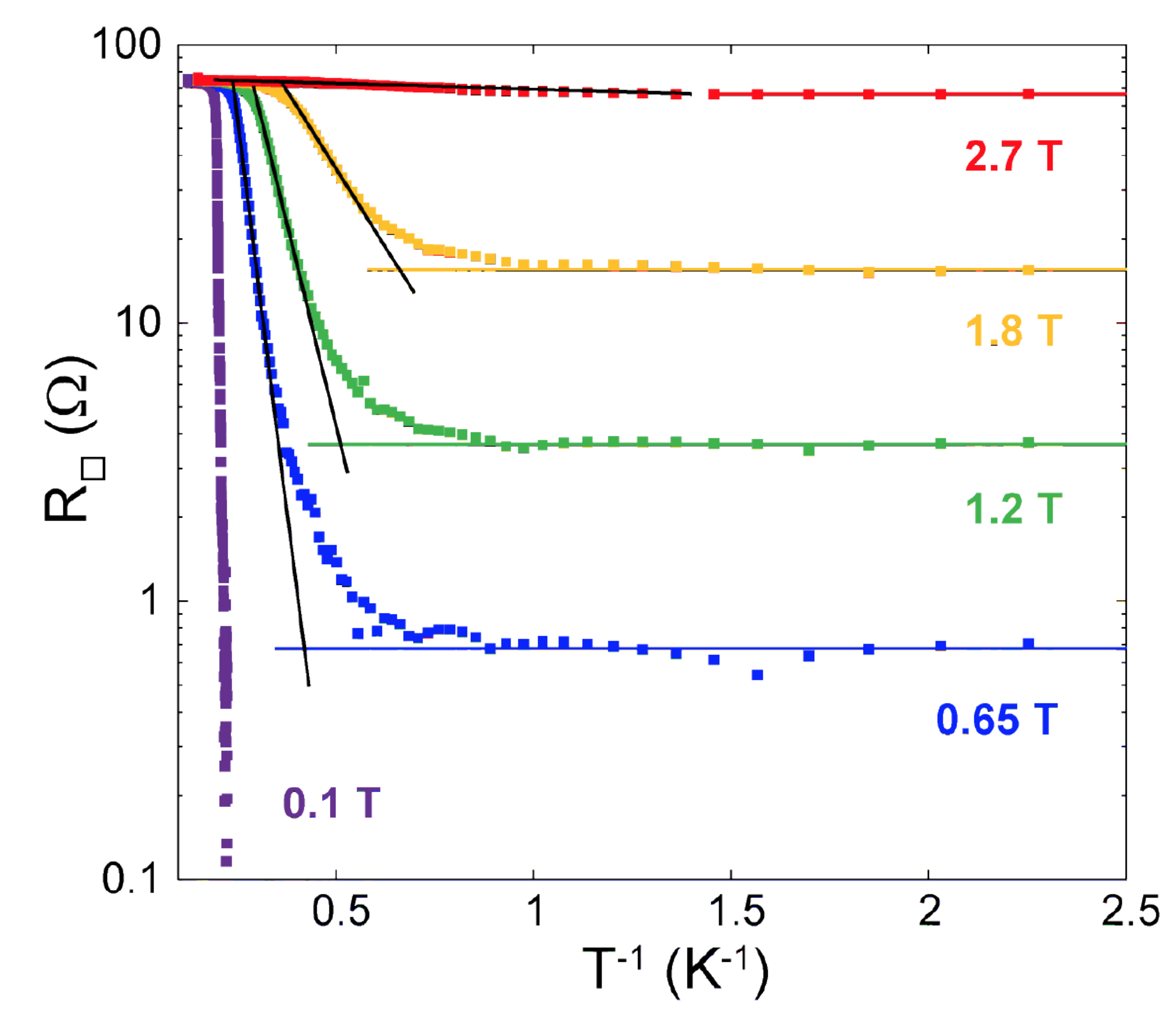} 
\caption{Resistivity (on a logarithmic scale) vs $1/T$ at various magnetic fields spanning the QSMT in highly crystalline films of NbSe$_2$.  From \cite{Tsen2016}}
\label{exp2Pasupathy2015}
\end{center}
\end{figure}

 \subsection{
 QSMT 
  at zero magnetic field}
 \label{gate}

Nominally, a disordered superconductor in the presence of a magnetic field forms a glassy state, which implies slow dynamics and even history dependent properties.
 Indeed,   some experiments on  field-induced anomalous metals exhibit hysteretic behavior.\cite{MasonKapitulnik1,MasonKapitulnik2}  Moreover, one can wonder whether the fact that $H$ breaks time-reversal symmetry is essential for the existence of the anomalous metal.
Therefore it is important to study the same phenomena in  cases in which a zero field transition can be driven by other means.  Certainly, one difference with the field driven case is that, so long as the groundstate is superconducting, one expects there to be a finite temperature phase transition, as shown in the schematic phase diagram in Fig. \ref{gatePD}.

\begin{figure}[h]
\begin{center}
\includegraphics[scale=0.4]{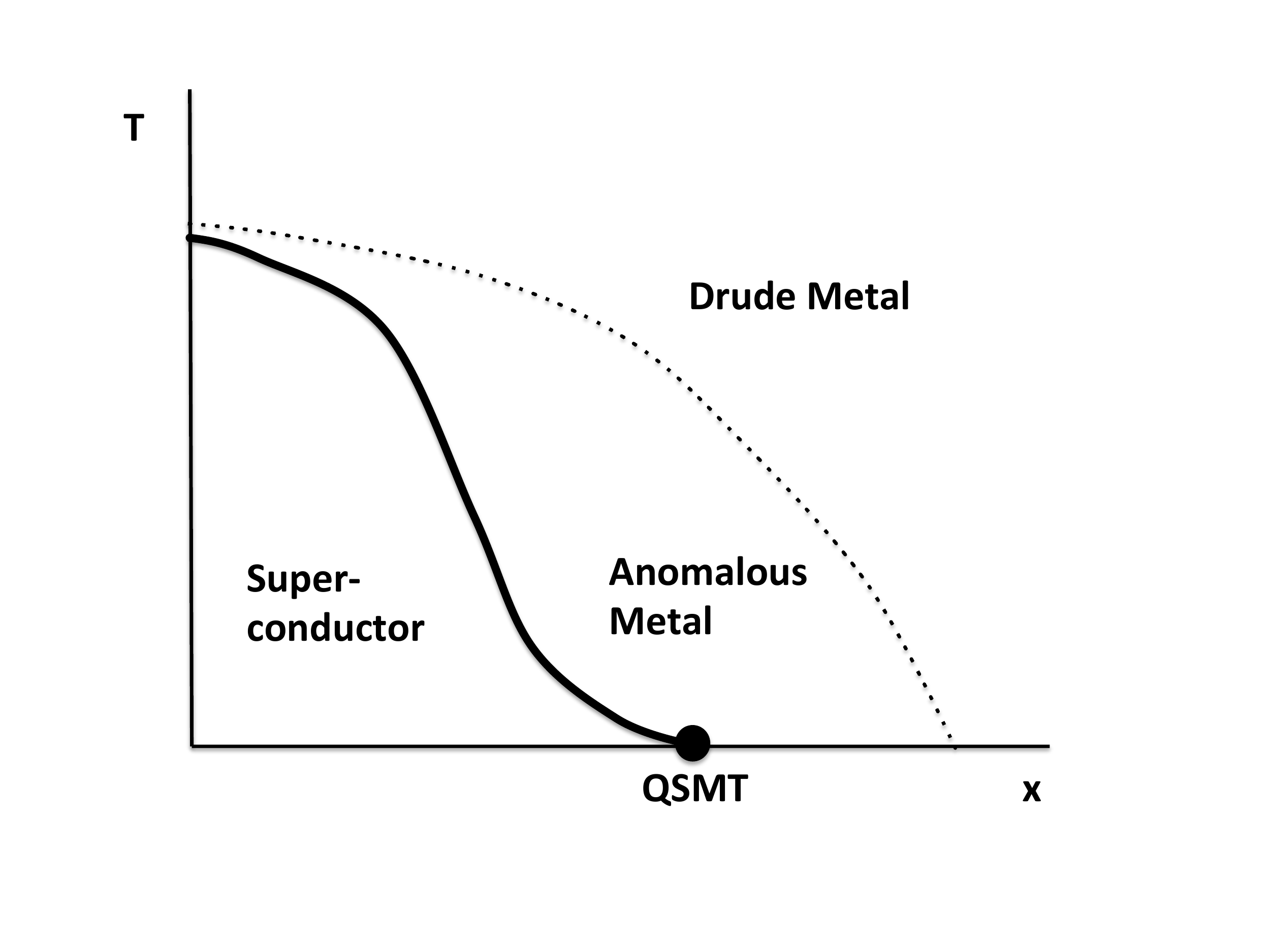} 
\caption{
Schematic phase diagram for the gate-tuned QSMT.  (More generally, $x$ represents a quantum tuning parameter that does not break time-reversal symmetry.)  The solid line represents the superconducting phase boundary and the solid circle the QCP associated with the QSMT.  The dashed curve represents a crossover.  As in the field tuned case,  the anomalous metal may be bounded at large $x$ by an insulating phase, in which case there would be a QCP associated with a MIT at the end of the  crossover line, or there could be a quantum crossover to a metallic phase dominated by fermionic excitations.}
\label{gatePD}
\end{center}
\end{figure}

Electrostatic gating is an effective method to introduce doping at the interface of a conducting material. As gating involves introducing a nearby metallic electrode, 
it also affects the screening of Coulomb interactions, and so introduces an additional dissipation channel.

 Probably the first study of a gate-controlled QSMT was performed on 
 an array of Al-Al$_2$O$_x$-Al Josephson-junctions fabricated on a GaAs/Al$_{0.3}$Ga$_{0.7}$As heterostructure in which a 2D electron gas (2DEG) was located approximately 100 nm from the surface \cite{RimbergClarke}. In this study, the 2DEG was presumably only coupled capacitively to the Josephson junction array; however so long as the conductivity of the 2DEG was sufficiently large,   screening provided by the 2DEG caused the array to show superconducting behavior despite a large junction resistance. Gating was then used to change the resistance of the 2DEG, and hence the dissipation in the electrodynamic environment of the array.   
As shown in Fig. \ref{Rimberg}, the temperature dependence of the array is different depending on the resistance of the 2DEG.  In all cases, the resistance of the array decreases with decreasing temperature for $T\sim 0.2$K, presumably reflecting the local superconducting order in the array.  However, at lower temperatures, the resistance of the array continues to drop and then  to saturate at the lowest temperatures when the resistance of the 2DEG is small.  Conversely, the resistance of the array increases  strongly with decreasing temperature when the resistance of the 2DEG is large.  This behavior is  suggestive of the existence of an anomalous metallic state. 

\begin{figure}[h]
\begin{center}
\includegraphics[scale=0.35]{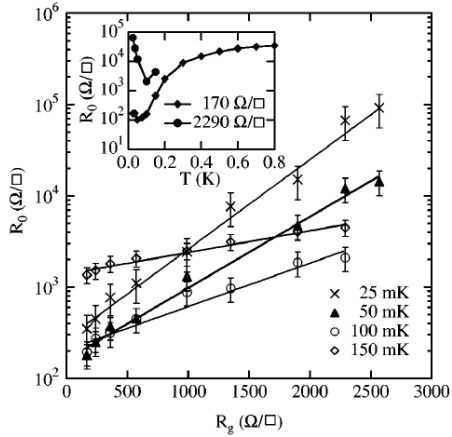} 
\caption{Resistance of the Josephson array $R_0$ (on a logarithmic scale) vs resistance $R_g$ of a ground-plane  (also on a logarithmic scale) to which the array is capacitively coupled.  The main figure shows results for a set of increasingly low temperatures.  The inset shows the temperature dependence of $R_0$ for $R_g = 170\Omega/\square$ (which exhibits anomalous metallic behavior) and 2290 $\Omega/\square$ which exhibits insulating tendencies, presumably  due to quantum fluctuations of the order parameter phase in the Josephson junction array. From Ref. \cite{RimbergClarke} }
\label{Rimberg}
\end{center}
\end{figure}

There have been a number of other studies of gate-tuned QSMTs. In contrast to the early experiments of \cite{RimbergClarke}, in  these other studies the gate primarily serves  to tune an intergrain Josephson coupling: 
In Fig.~\ref{Bouchiat} we show data from experiments \cite{BouchiatFeigelman} on artificially prepared samples where a regular array of superconducting Sn disks were placed in a regular lattice on a graphene substrate.
The density of electrons in the graphene can be varied by varying the voltage applied by a back gate.  There is a proximity effect coupling between the superconducting droplets and the graphene, so the gate voltage (among other things) tunes the effective Josephson coupling between neighboring disks. The distinct colors in the figure represent the resistance as a function of $T$ for  various values of the gate voltage.  The initial drop in the resistance  is associated with the onset of superconductivity within the droplets.  For large values of the gate-voltage (large electron densities in the graphene), the resistance drops sharply at a somewhat lower temperature, extrapolating to zero at a superconducting transition temperature  that varies depending on voltage.  However, in the   anomalous metal regime which appears for somewhat smaller gate voltages, as $T$ decreases, the resistance drops by as much as 3-4 orders of magnitude, but then  saturates at a finite plateau value that can be 3 to 4 orders of magnitude smaller than the Drude value. This behavior is very similar to that seen in the vicinity of the field driven QSMT.   In this range of gate-voltages, a  magnetic field applied at low $T$ produces a giant increase in  the resistance which saturates  at high fields -- a gigantic positive magneto-resistance that recovers  the Drude value of the resistance, presumably by suppressing any remnant superconducting coherence.
 \begin{figure}
\begin{center}
\includegraphics[scale=0.18]{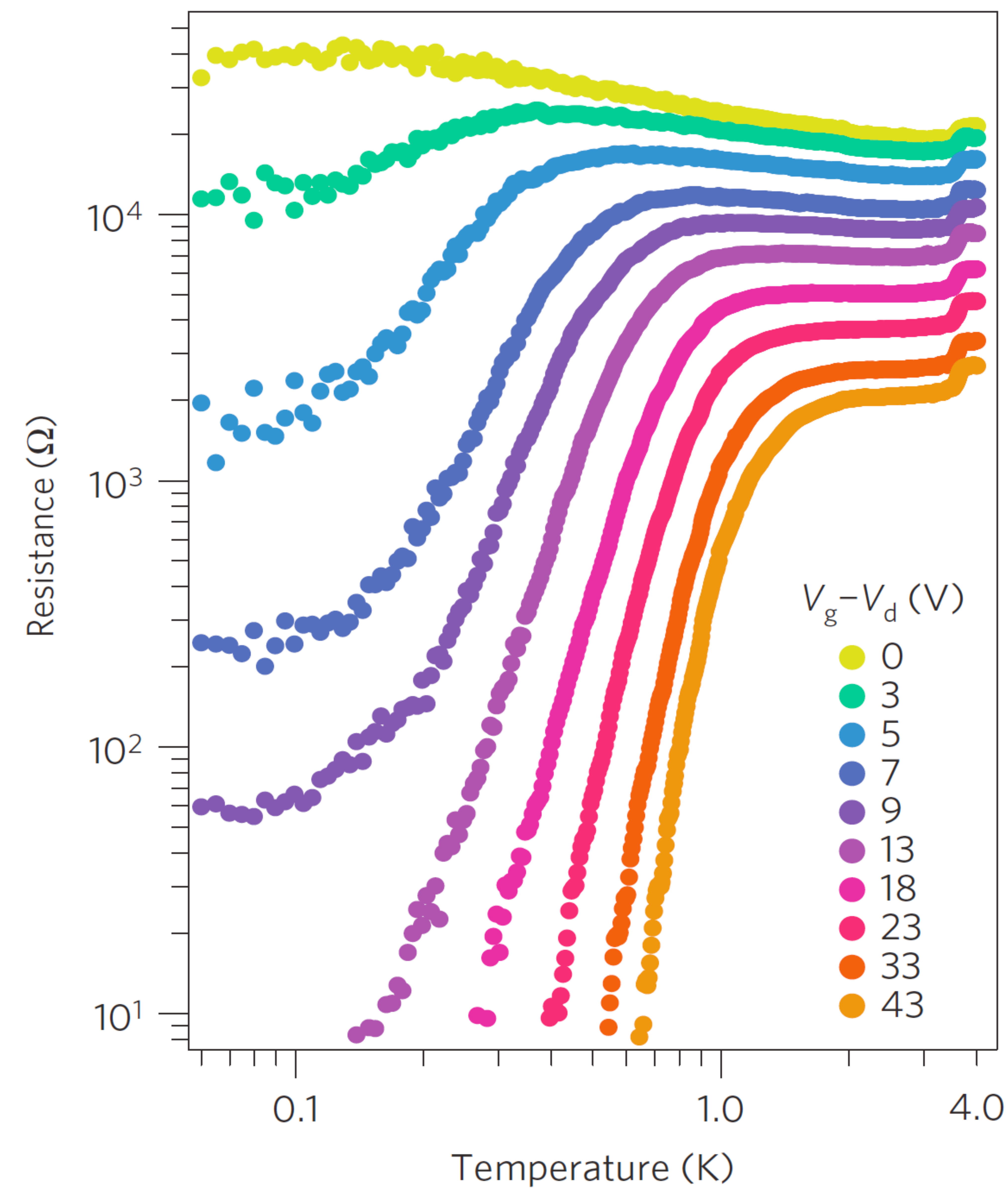}
\caption{Ther esistance vs $T$ on a log-log scale of an ordered array of Sn discs on a graphene substrate;  the density of electrons in the graphene is controlled by adjusting the voltage with a back gate.  For the largest gate voltages (highest electron densities) there is a clear finite temperature transition to a superconducting state.  However, for a broad range of lower gate voltages, we see the familiar several orders of magnitude drop in the resistance that terminates in a temperature independent plateau.  From \cite{BouchiatFeigelman}.}
\label{Bouchiat}
\end{center}
\end{figure}

Fig. \ref{Marcus} shows  results for the  system  studied in  
\cite{Bottcher2017}.  Here,  
a gated semiconductor heterostructure with epitaxial Al was patterned to form a regular array of superconducting islands connected via a InAs quantum well. Gating the quantum well allowed for variation by 
many orders of magnitude in resistance, thus unveiling a 
 range of anomalous metal behavior. 
 \begin{figure}
\begin{center}
\includegraphics[scale=0.43]{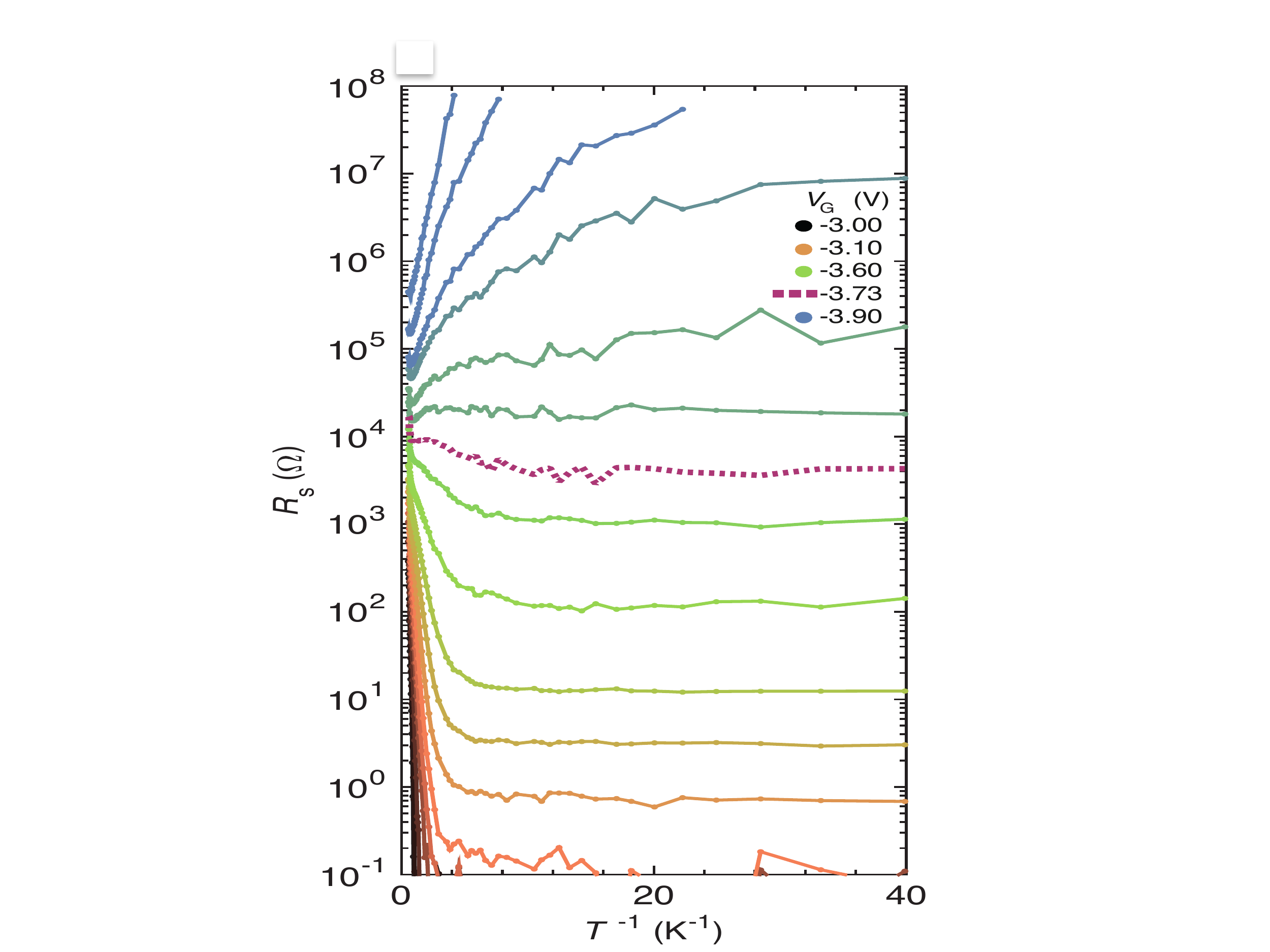}
\caption{Log of the sheet resistance, $R_s$, as a function of the inverse temperature,  $T^{-1}$, in a gated InAs heterostructure with epitaxial Al  patterned to form a regular array of superconducting islands.  Data is shown for  a range of gate voltages, $V_G$, from $-3.0$ V to $-3.9$ V. The dashed curve corresponds to $V_G=-3.73$ V; 
the tendency of the curves with $V_G \geq -3.73$ V to saturate at low $T$ is indicative  of the occurrence of a metallic phase. From \cite{Bottcher2017}.}
\label{Marcus}
\end{center}
\end{figure}

All the examples presented so far of a gate-tuned QSMT involved artificially fabricated granular systems, where the gate affects the properties of the intergranular (substrate) electronic structure.  

However, anomalous metallic states have also been observed in 2D films and interfaces that are considered homogeneous. For example, the same ZrNCl system for which the field-tuned transition is shown in Fig. \ref{Saito}, 
can also be tuned by tuning an ionic-gate voltage \cite{Saito2015}. 

Devices made of exfoliated single crystalline  
transition metal dichalcogonides (TMD), such as MoS$_2$ \cite{Iwasa} and  WTe$_2$ \cite{JoshFolk} have shown a transition from a superconducting state to an anomalous metallic state upon varying the gate voltage.  A recent example on WTe$_2$ is shown in Fig.~\ref{JoshFolk} (see ref.~\cite{JoshFolk}).

\begin{figure}
\begin{center}
\includegraphics[scale=0.8]{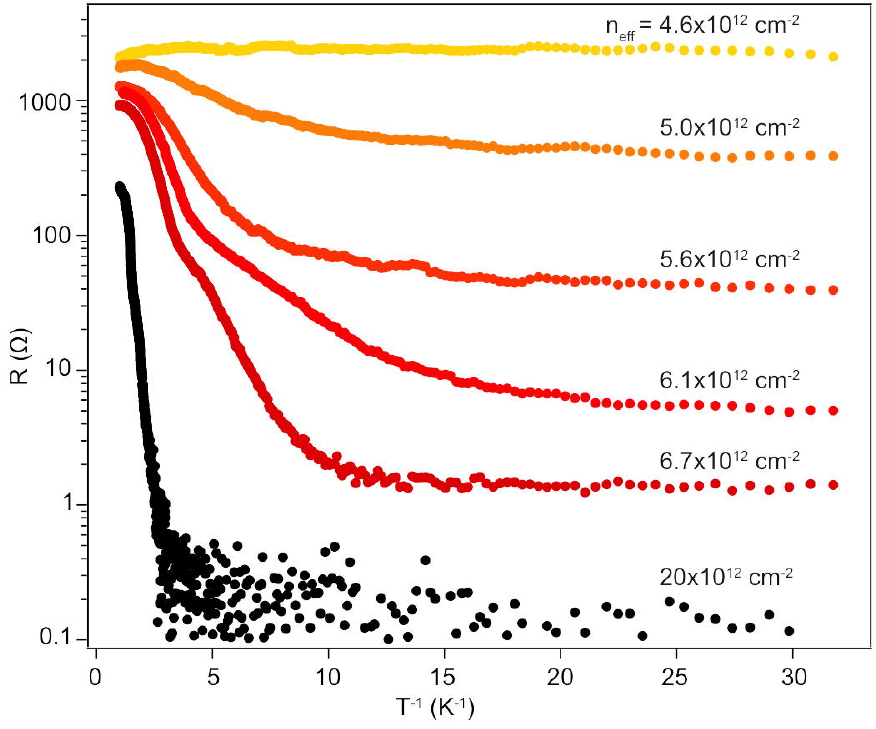}
\caption{ Four-probe resistance (on a logarithmic scale) vs inverse temperature  of a monolayer WTe$_2$ flake, 
which is tuned by application of a gate voltage from  
a superconducting state (seen for   the highest electron density $n_{eff} =20 \times 10^{12}$ cm$^{-2}$), to an anomalous metallic state (when  $n_{eff}=  12, 8.5, 6.7, 6.1, 5.6$, and $5 \times 10^{12}$ cm$^{-2}$), and finally to what appears to be a normal metal when $n_{eff}=4.6 \times 10^{12}$ cm$^{-2}$.  From \cite{JoshFolk}.}
\label{JoshFolk}
\end{center}
\end{figure}

An  advance in gate controlled coupling has been achieved 
in Ref. \cite{Chen2017} utilizing dual electrostatic gates, which, as shown in Fig. \ref{Hwang}, were used to manipulate 
both the mean depth and the asymmetry of the quantum well in a  SrTiO$_3$-LaAlO$_3$ heterostructure.  
Notably, the large (exceeding 20,000) and nonlinear dielectric constant of the SrTiO$_3$ 
 greatly enhances the tunability of this system as compared to conventional gating experiments. On one side, the 2DEG is bounded by the wide-gap LaAlO$_3$, where a  top-gate ($V_{TG}$) predominantly control the density of carriers confined close to the SrTiO$_3$/LaAlO$_3$ interface. A back-gate ($V_{BG}$) is then used to control the thickness of the conduction layer at the interface, hence the interfacial scattering rate and mobility of the 2DEG.  
\begin{figure}[h]
\begin{center}
\includegraphics[scale=0.6]{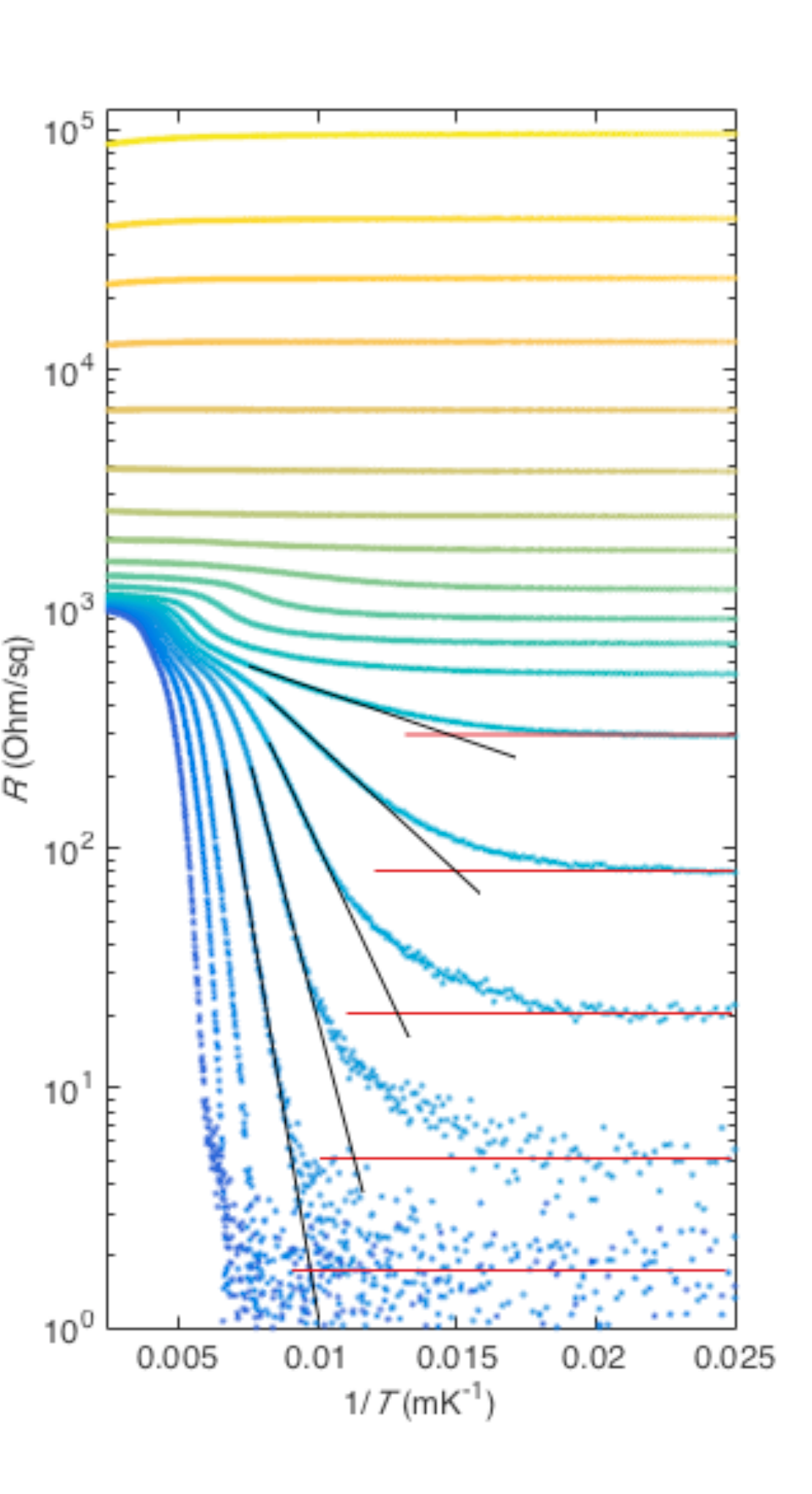}
\caption{Resistivity (on a logarithmic scale) versus inverse temperature  as a function of top-gate voltage, $V_{TG}$ from -0.36 V to 1.80 V with fixed bottom gate voltage $V_{BG} = 0$ V in a  SrTiO$_3$-LaAlO$_3$ heterostructure.  The solid lines have been added as guides to the eye.  The asymptotic approach of the measured curves to the temperature independent red lines as $T\to 0$ shows the existence of an
 anomalous metallic phase.  From \cite{Chen2017}.}
\label{Hwang}
\end{center}
\end{figure}

Turning to   ``unconventional''  superconducting states, Fig. \ref{GoldmanLSCO} shows the resistivity as a function of $T$  for a liquid ion gated film of the cuprate superconductor, La$_2$CuO$_{4+\delta}$.    
At small gate voltage, the film exhibits  clear insulating behavior, while at large gate voltage it is superconducting below a  
non-zero superconducting transition temperature.  However, as shown in the inset, at intermediate values of the gate voltage, while the resistance drops below a well-defined crossover temperature, it appears to saturate at low $T$ to a small value (sometimes four orders of magnitude smaller than $\rho_D$). Note that with the higher $T_c$ of this system, also the temperature where saturation is apparent increases. For example, at the highest gate voltage in Fig.~\ref{GoldmanLSCO}, saturation occurs below $\sim10$ K. This issue will become important in arguing against a simple heating as explanation of the resistance saturation.
\begin{figure}[h]
\begin{center}
\includegraphics[scale=0.2] {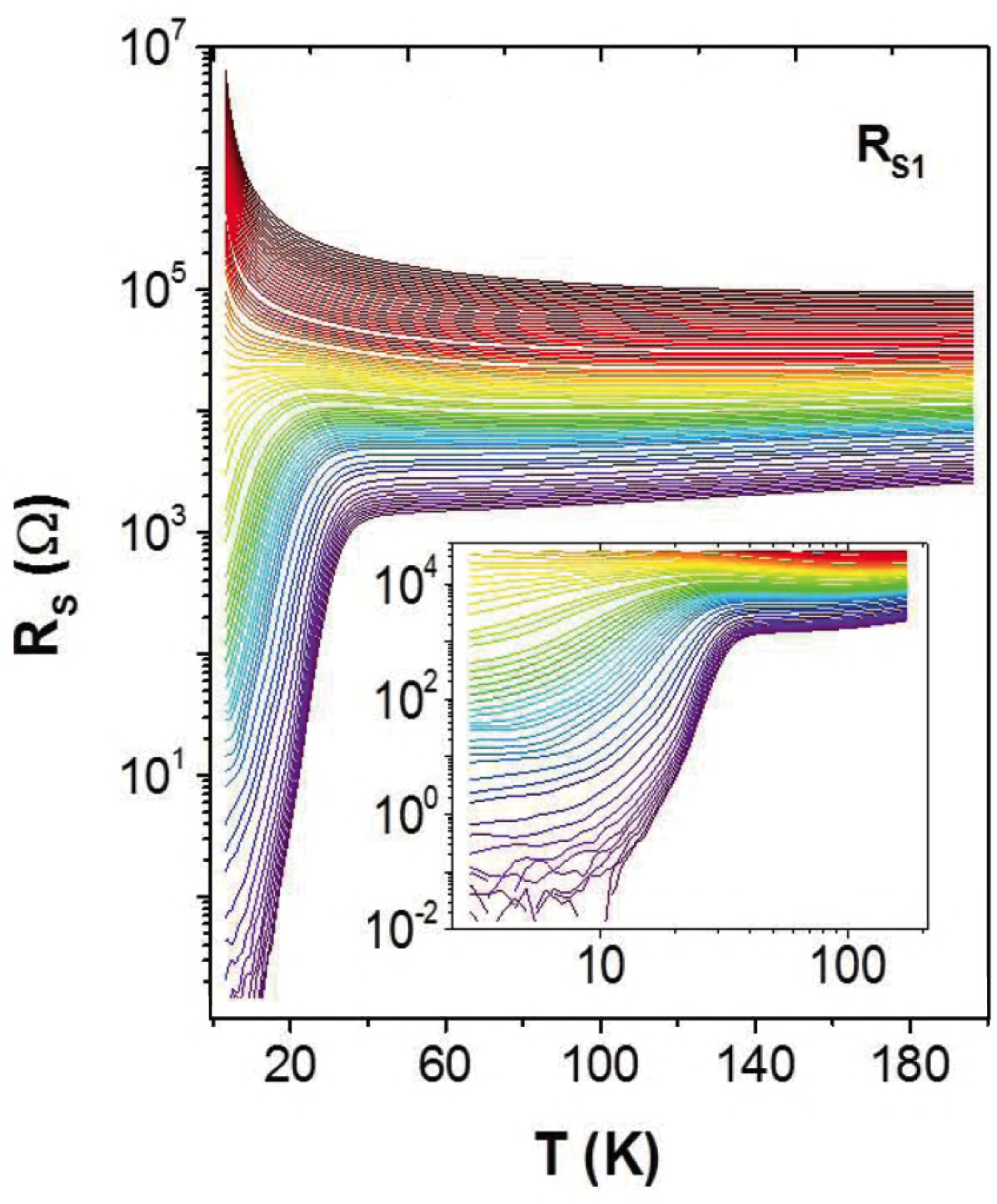} 
\caption{Sheet resistance (on a logarithmic scale) as a function of temperature for different values of gate voltage $V_G$ measured on a $\sim5$ nm thick La$_2$CuO$_{4+\delta}$ film. Gate voltages ranged from   1.2 V for the most insulating sample to 3V for the most superconducting one. Inset shows 
the less resistive samples on  a log-log scale, which expands the low temperature portion of the curves thereby making clear the  saturation of the resistance at low temperatures. From \cite{GoldmanLSCO}.}
\label{GoldmanLSCO}
\end{center}
\end{figure}

While gate or magnetic field tuning 
have the advantage that they can be varied continuously, 
other approaches to the QSMT have been successfully explored as well.
 
 Early studies in which a sequence of (presumably granular) films are studied for various  film thickness have already been presented in Fig. \ref{Goldman2}.  In Fig. \ref{Crauste} we show representative data\cite{Crauste2009} from a  study on (presumably homogeneous) flims of Ni$_x$Si$_{1-x}$ of various thicknesses, in which, as a function of decreasing $T$, the thinnest films show a strong divergence of the resistivity indicative of approach to an insulating groundstate, the thickest films show a  finite temperature transition to a zero resistance state, but films of intermediate thickness show the  familiar signatures of an anomalous metal.
\begin{figure}[h]
\begin{center}
\includegraphics[scale=0.12]{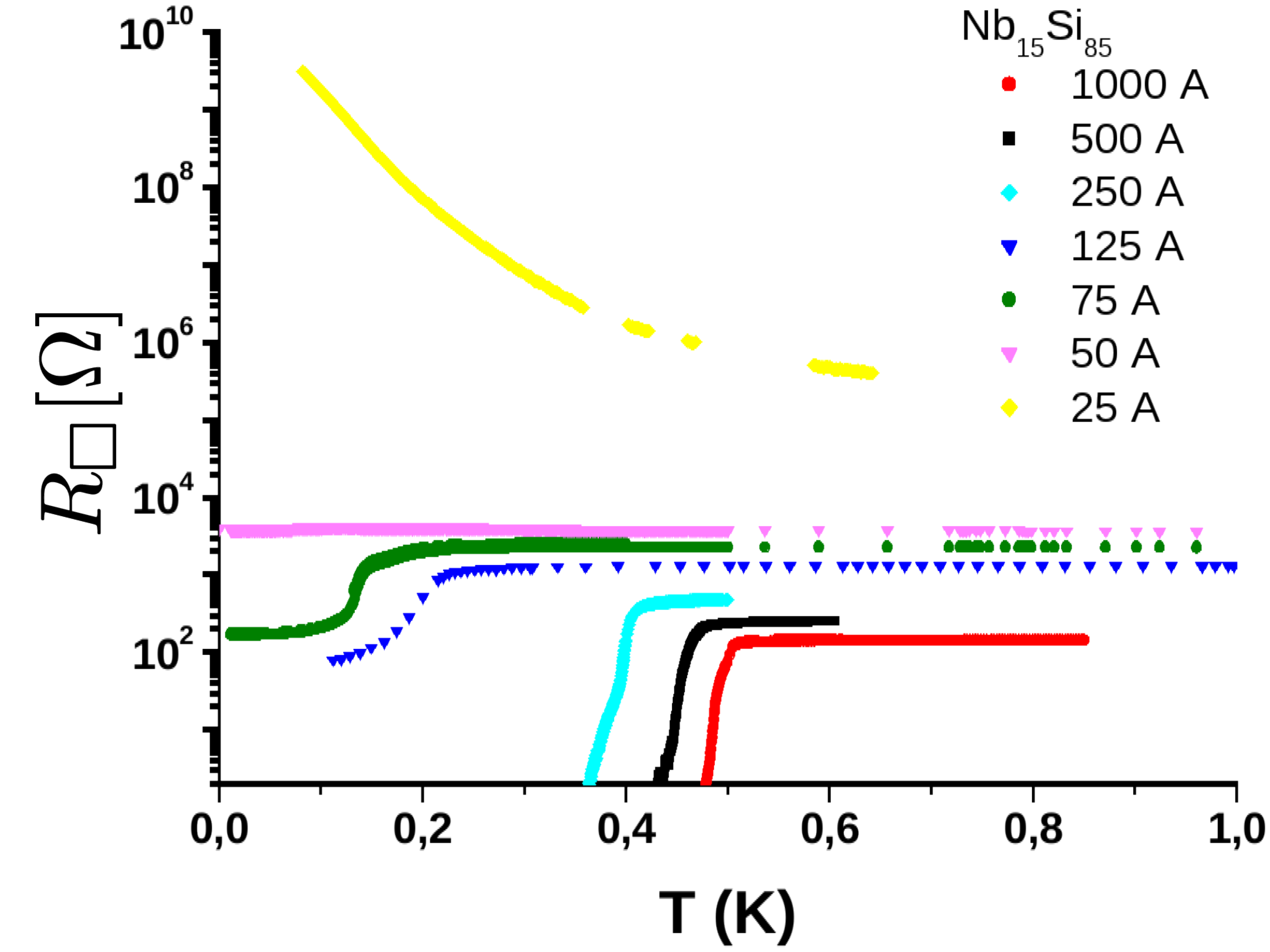}
\caption{Resistance of Ni$_ x$Si$_{1-x}$ films with $x = 15$\% and varying film thickness (given in $\AA$). Samples were assumed to be homogeneously disordered; the blue and green curves suggest the existence of an intermediate anomalous metallic phase between a superconducting phase (seen in thicker films) and an insulating phase (in thinner).  From Ref. \cite{Crauste2009}
}
\label{Crauste}
\end{center}
\end{figure}

 Since  both the Josephson coupling and charging energies depend on the size and distance between grains, similar tunability can be achieved by preparing samples with different grain size and periodicity. Indeed, this approach was taken 
 in Ref. \cite{Eley2013}, where  an array of Nb dots was deposited on a gold substrate. As seen in Fig.~\ref{Mason3}, a low-temperature metallic state is clearly revealed at a wide range of distance between grains.
\begin{figure}[h]
\begin{center}
\includegraphics[scale=0.15]{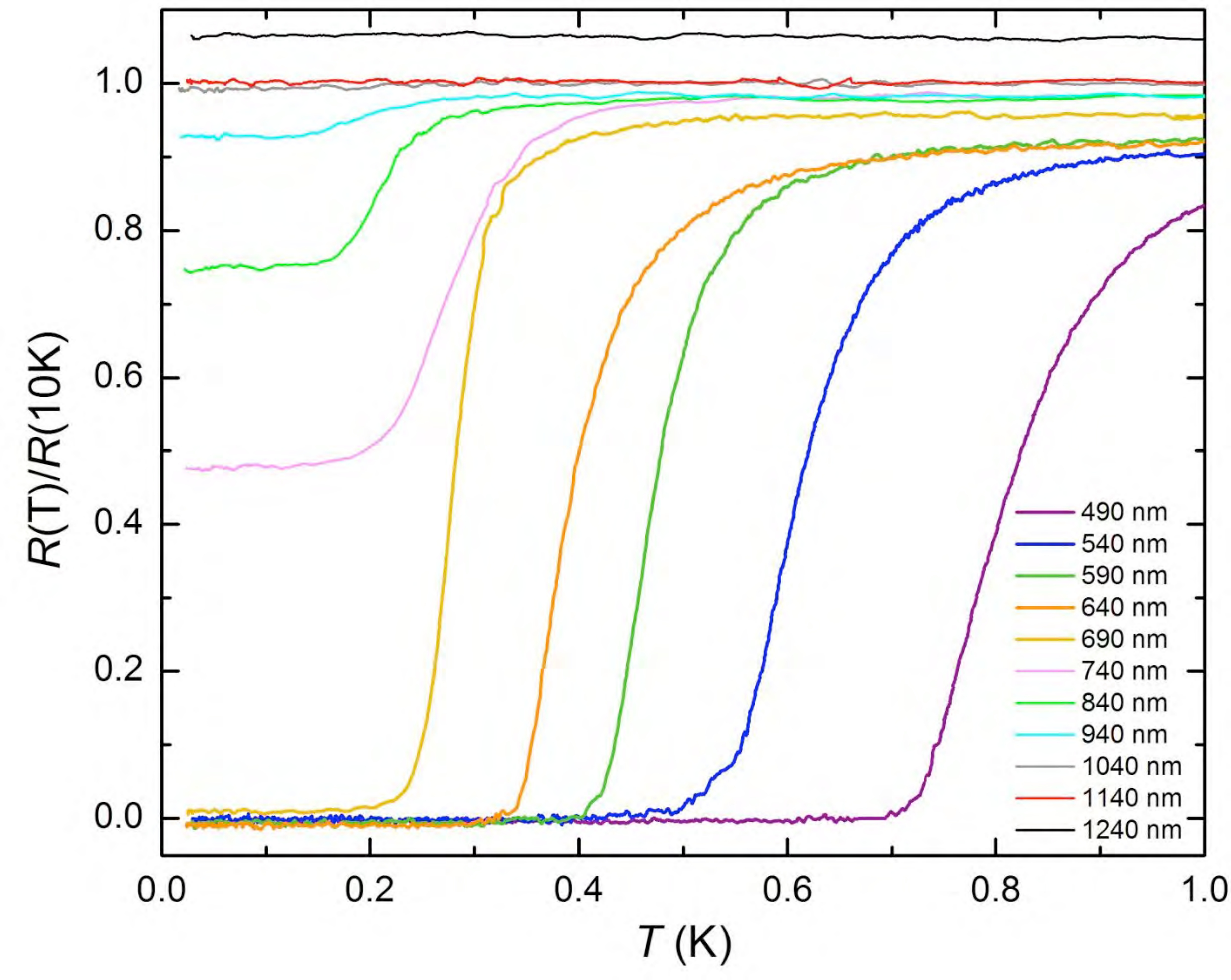}
\caption{  Normalized resistance as a function of temperature for 
 arrays of widely spaced Nb islands on a Au substrate. For spacings exceeding 700 nm, the BKT transition is interrupted by a low-temperature metallic state. The data for $d \leq 690$ nm and for $d\geq 740$ nm  come from systems with Nb island heights of 125 nm and 145 nm, respectively. From \cite{Eley2013}.}
\label{Mason3}
\end{center}
\end{figure}

\subsection{
Not just the resistivity}

Other features of the anomalous metal that illustrate its character as a failed superconductor have been measured in a limited number of cases.

 1. The emergent particle-hole symmetry of the  superconducting state suggests  
  that it is natural to expect a reduction of the Hall and thermoelectric responses in the anomalous metal and a tendency for them to vanish upon approach to the QSMT.  
 To explore this issue, simultaneous measurements of $\rho_{xx}$ and $\rho_{xy}$ were performed on  InO$_x$ and TaN$_x$ films and these were used to calculate $\sigma_{xy}$.  Sample data at 120 mK, which is
far below the zero field $T_c$, is shown in Fig.~\ref{HallBreznay} (adopted from 
 \cite{Breznay2017}),
   with various characteristic fields indicated by vertical dotted lines. 
  Panels A and B show plots of the  longitudinal and 
Hall resistivities vs. magnetic field on a linear scale,  panel C shows the same data on a semi-log scale, and panel D shows $\sigma_{xy}$ obtained by inverting the measured resistivity tensor.  
The point of the QSMT at $H_{M1}$ is identified from the $T$ dependence of $\rho_{xx}$ in much the same way as in the data already shown in
 Fig.~\ref{exp1MassonKapitulnik} -- for the fixed low $T$ shown in the present figure, $H_{M1}$ is apparent as the field below  which $\rho_{xx}$ drops sharply.
 However, both $\rho_{xy}$ and $\sigma_{xy}$ remain immeasurably small for a broader range of fields, including  both the superconducting range, $0<H <H_{M1}$ and the anomalous metallic range $H_{M1} < H < H_{M2}$.  
The clearest  signature of $H_{M2}$ appears in the $H$ dependence of $\sigma_{xy}$ which appears to grow (approximately as $H^{-1}$) with decreasing $H$ for a range of $H$ below $H_{c2}$, before turning around and dropping sharply toward zero on approach to $H_{M2}$.
(The solid lines in Panels B-D represent a linear in $H$ extrapolation of the high field Hall response - which serves as a putative ``normal state'' value for purposes of comparison.)  
\begin{figure}[h]
\begin{center}
\includegraphics[scale=0.1]{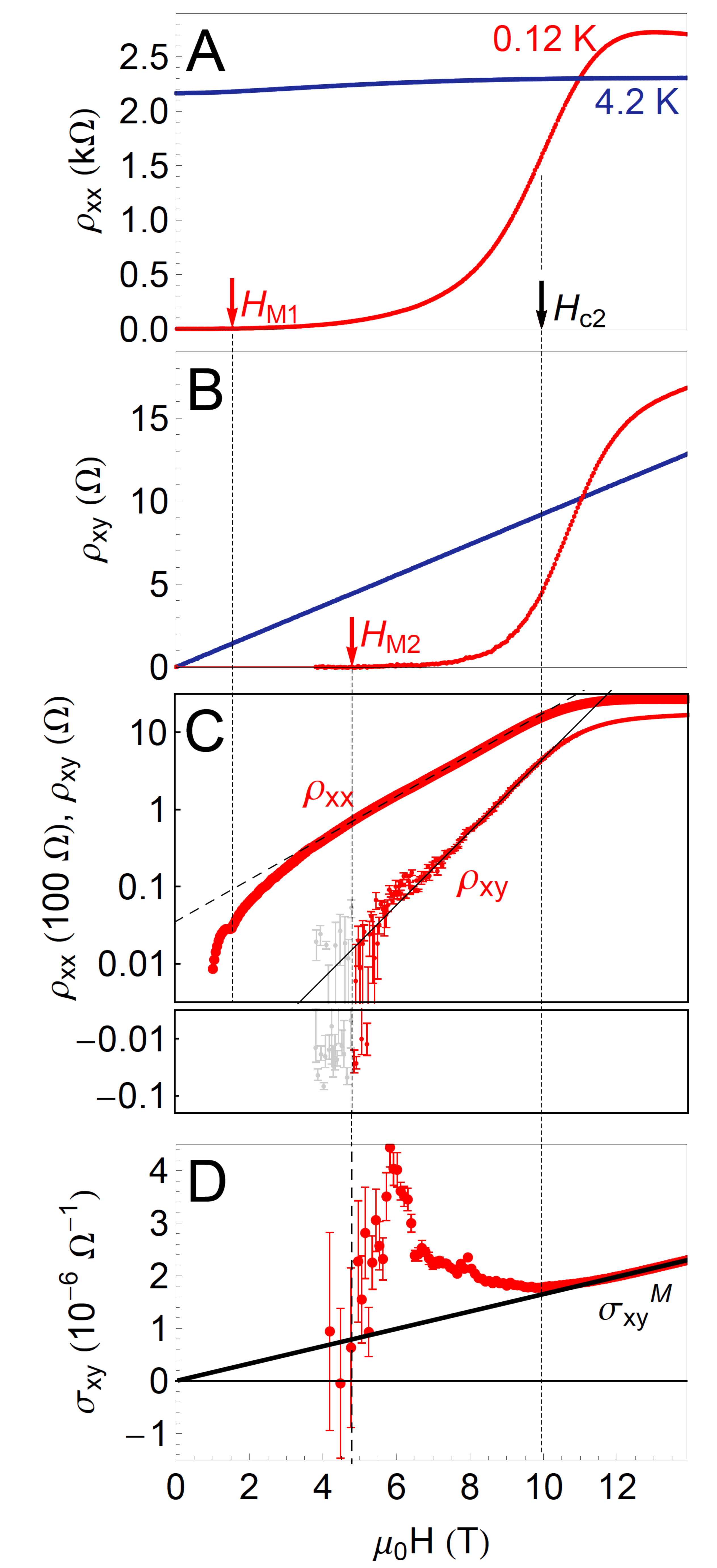} 
\caption{Resistivity and Hall resistance vs. magnetic field at $T=0.12$ K, of the same InO$_x$ film as in Fig.~\ref{nick}.
(A) and (B) show 
 $\rho_{xx}$ and $\rho_{xy}$ respectively, on a linear scale together with the ``normal state'' values as recorded at $T=4.2$ K. $H_{M1}$ marks the superconductor to anomalous metal transition, $H_{M2}$ is the field above which the Hall resistivity becomes measurable, and  $H_{c2}$ 
is roughly the point of the 
mean-field transition. The three distinct magnetic fields 
 can be identified with various features in the field dependences shown in panel (C) where the logarithm of the low $T$ resistivities are plotted. 
 Panel D shows $\sigma_{xy}$ computed from the measured resistivity tensor; 
 as $H$ decreases below $H_{c2}$, $\sigma_{xy}$ at first increases approximately in proportion to $H^{-1}$, but then drops sharply upon approaching $H_{M2}$. 
For $H<H_{M2}$, both $\sigma_{xy}$ and $\rho_{xy}$ vanish within experimental uncertainty. From \cite{Breznay2017}.}
\label{HallBreznay}
\end{center}
\end{figure}

2. The presence of significant superconducting  fluctuations in a system in which long-range superconducting phase coherence has been lost can often be apparent in the finite frequency response, $\sigma(\omega)$.  Notionally,   assuming that some form of dynamical scaling applies, the finite $\omega$ response probes correlations at finite length scales. This strategy has been successfully employed to establish the existence of substantial finite-range superconducting correlations in 
   a-MoGe films  \cite{YazdaniThesis,Yazdani1993}  in a magnetic field, where the high-frequency dynamics was shown to behave as expected upon approach to an ideal classical vortex lattice melting transition. Using a similar approach, finite range superconducting correlations were also established for the cuprate high temperature superconductors in a range of temperatures above $T_c$ \cite{Corson1999,Bilbro11} and in the insulating state proximate to an SIT in highly disordered InO$_x$ films \cite{Crane2007}.  

Recently, the finite $\omega$ response of  a magnetic field induced anomalous metallic state in weakly-disordered InO$_x$ films was  measured by \cite{Liu2013,Wang2017}. Here, broadband microwave measurements were performed in the frequency  range from 50 MHz to 8 GHz and  the temperature and magnetic field dependences of the complex microwave conductance determined. Strongly non-Drude features are clearly visible in the anomalous metal regime, including the presence of a sharp low frequency ``superfluid-like'' peak in the longitudinal response at small $\omega$ and the absence of any cyclotron resonance at the expected field-dependent frequency see Fig.~\ref{Armitage}.) 
\begin{figure}[h]
\begin{center}
\includegraphics[scale=0.3]{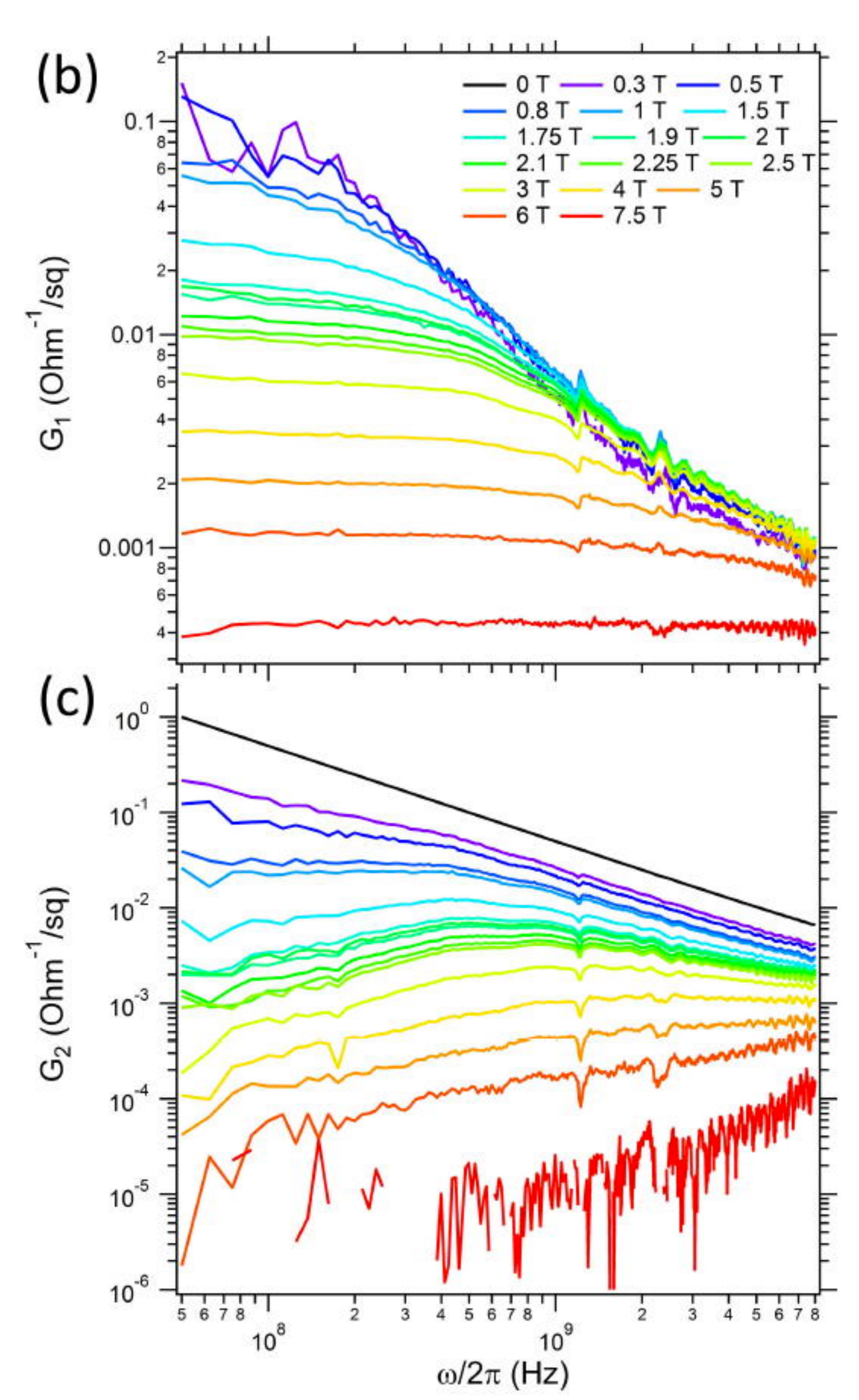} 
\caption{ The frequency dependence of the real (b) and imaginary (c) parts of the complex optical conductivity (G$_1$ and G$_2$ in the figure) measured on a weakly disordered InO$_x$ film at multiple values of the magnetic field.  
From DC transport it was estimated that the QSMT in this film occurs at $H\approx 2T$.  
The Drude response is essentially frequency independent over the range of (low) frequencies probed here,  so the entire frequency dependent signal shown can presumably be associated with  bosonic (superconducting) fluctuations. Special attention was drawn in this study to the absence of a finite frequency peak in G$_1$ of the sort that would indicate the presence of a  cyclotron resonance.
From \cite{Wang2017}.}
\label{Armitage}
\end{center}
\end{figure}
}
While intuitively these features support the identification of this regime as a failed superconductor, as far as we know no explicit theoretical account of these observations currently exists. Further study, both theoretical and experimental, of the finite frequency response is clearly warranted.

\subsection{The strange case of ``granular'' films}

For the most part, the properties of the anomalous metal seen in all the studies so far discussed are similar, independent of system morphology, degree of order, and whether or not a magnetic field is applied.  There is, however, another class of systems -- which are granular films in some not entirely well defined sense -- which also show evidence of an anomalous metallic phase, but of a very different character.  
Granular films can be synthesized in various ways.~\cite{Abeles1975,Kapitulnik1982,Deutscher1983}   While local superconductivity 
can occur within a single grain, global superconducting phase coherence necessarily involves Josephson (pair) tunneling between grains, 
  and thus is sensitive to  various details of the grain morphology and the nature of the material between grains.  (See e.g. refs.~\cite{Entin1981,Ioffe1981,ImryStrongin}). 
 Here, for completeness, we briefly discuss some such experimental observations.

\begin{figure}
\begin{center}
\includegraphics[scale=0.2]{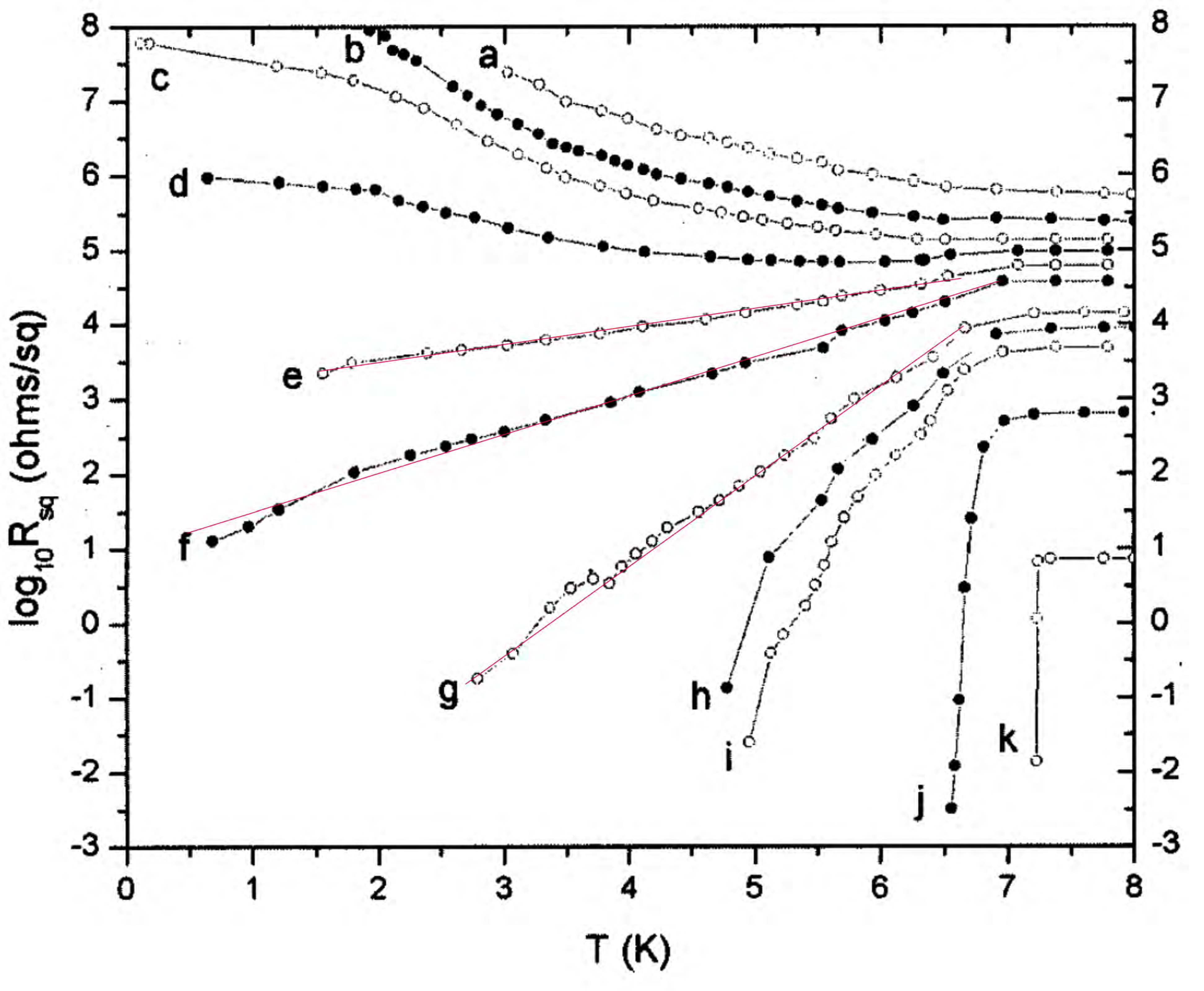} 
\caption{Resistivity (on a logarithmic scale) as a function of $T$ for a sequence of granular Pb films.  Each curve corresponds to data at fixed Pb coverage, and the tuning from one curve to the one immediately below it is accomplished by depositing a small additional quantity of Pb on the previous film.  Even in the ``insulating'' films with the least Pb coverage (i.e. films a - d), tunneling spectra reveal a well-developed superconducting gap, and this gap is more or less the same even as additional Pb is added, including the ``metallic films'' (i.e films e, f, and g) and the clearly superconducting films (i.e. films j and k).    From \cite{Merchant}. }
\label{Merchant}
\end{center}
\end{figure}

Pb films 
are a particular well studied model system.\cite{ImryStrongin,Goldman2,Merchant}. 
Figure~\ref{Merchant}  
shows data from \cite{Merchant} on Pb films.  As in the data on other ``granular'' materials shown in Fig.~\ref{Goldman2}, above, there is a clear signature of the onset of local superconductivity within a ``grain'' at a relatively high $T$, but then depending on the distance between grains (or more particularly the Pb coverage), the system evolves from a globally insulating  to globally superconducting state.    Elegantly, in the present case, tunneling studies (not shown here) on the same films show a clean BCS-like superconducting gap opening up at around the same temperature, largely independent of the Pb coverage.   Here, the resistivity of films with low Pb coverage (a, b, c, and d in the figure) show a clear tendency to diverge in the $T\to 0$ limit, and thus can be characterized as insulating. However, there is an intermediate regime of concentrations (films e, f, and possibly g) in which the resistivity decreases strongly with decreasing $T$, but it does so in a manner such that  
the $T$ dependence of 
$\rho$ approximately follows the  phenomenological relation
\be
\rho(T) \approx \rho_0 \exp[ T/T_0]
\label{linearT}
\ee
where $\rho_0$ and $T_0$ are $T$ independent functions of the concentration of Pb grains. In the case of film g, for example, $\rho_0$ is roughly 4 orders of magnitude smaller than the normal state resistivity.  (See lines fit to the data in Fig. \ref{Merchant}.)

Indeed, such behavior has been widely observed in thin superconducting films.  For instance, in Ref. \cite{Merchant}, very similar behavior was seen when the coupling between grains of an insulating granular Pb film was gradually increased by addition of a thin layer of Ag.  Behavior of this sort was seen long ago in thin Al films 
in Ref. \cite{Masker1969}).
 While the fit to Eq. \ref{linearT} implies the existence of an intermediate 
 metallic state (that is, the resistance extrapolates to a finite value as $T \to 0$), it is difficult to rule out a power-law temperature dependence of the prefactor that could lead to a vanishing resistance at some much lower temperature.  
 More importantly for present purposes, our present understanding of the anomalous metallic phase does not include a satisfactory explanation of the ${\rm exp}(T/T_0)$ phenomenon.    We will therefore (reluctantly)  not discuss it further in this article.

\subsection{On the Issues of Heating and Nonequilibrium Effects}
Since the anomalous metal is typically observed  at low temperatures, it is important to make certain that the measurements have been performed within linear response, and that the nominal temperature corresponds to the actual electron temperature. While  these issues have been discussed by the authors of each of the experiments we have reported, we feel it is still important to briefly review possible experimental issues. 
 
Heating and other non-equilibrium effects can  arise as a consequence of uncontrolled external ``noise.'' Time-varying electric fields couple via mutual capacitances (i.e. electrostatic coupling), and thus inject noise into the system as a current source, while magnetic or inductive interference, which arises from time-varying magnetic fluxes passing through sections of the measurement circuit, can induce fluctuating voltage sources. 
At elevated temperatures, where electron-lattice relaxation rates are high, the effects of such noise is negligible. Even at low temperatures, the corresponding electronic relaxation times are typically  short enough that non-equilibrium effects can  be thought of as producing a local effective electron temperature, which can be somewhat different from the lattice temperature. 
 
Some of the systems reported above exhibit saturation at sufficiently high  temperatures that contamination by non-equilibrium effects is a priori unlikely.
 Examples in this category include the magnetic-field tuned QSMT in ZrNCl where the saturation temperatures are $\sim 7$ K (see Fig.~\ref{Saito}), and the gate-tuned QSMT in La$_2$CuO$_{4+\delta}$ where saturation starts at around 10K  (see Fig. \ref{GoldmanLSCO}). Similar results have also been obtained on magnetic-field tuned thin MgB$_2$ films, also in this range of  temperatures (unpublished data related to~\cite{Siemons2008}.) 

When the resistivity saturation occurs at lower temperatures, it is essential to establish that it is intrinsic, and does not simply represent a point below which the electron temperature ceases to decrease.  Thus, precautions are taken in each experiment to address this issue, typically employing proper shielding and filtering of the input and output signals to avoid the effects of electromagnetic interference. For example,  the experiments described in  \cite{Ephron}, \cite{MasonKapitulnik1}, and \cite{MasonKapitulnik2} used all analog equipment in a  shielded room, with efficient low-pass filtering at both, top and bottom of the dilution refrigerator, while signal lines to outside the shielded room were all optical with high degree of isolation (see e.g. \cite{MasonThesis}.)

Besides physical precautions, sometimes the most convincing evidence that heating or extrinsic noise do not affect the measurements is in the systematics of the data itself. As an example of such analysis, we  follow a line of reasoning introduced in Ref. \cite{Ephron}.  It is often observed (see, e.g.  Fig. \ref{YazdaniEphron})  that the onset of superconducting (pairing) correlations leads to an exponential drop of the resistivity as a function of $1/T$.  Thus, if one were to suppose that the low $T$ saturation were due to a failure of the electron fluid to equilibrate at the lattice temperature, one could infer a minimal ``effective electron temperature'' from the saturation value of the resistivity from the implicit relation
\begin{equation}
R_0(T,H)=\tilde{R}_0(H){\rm exp}\{ -U(H)/k_BT_{eff}
\}
\end{equation}
where 
$\tilde{R}_0$ and $U(H)$ can be determined 
accurately from a fit to   the initial drop in the resistivity to this expression.    The limiting value of $T_{eff}$ as a function of  $H$ would then determined graphically as the intersection between the Arrhenius fit and a horizontal (constant) curve corresponding to the low $T$ value of the resistivity.  
The value of
 $T_{eff}$
 inferred  in Fig. \ref{YazdaniEphron}  in this way is  a decreasing function of $H$, {\it i.e.} $\rho(T)$ for the more resistive samples (which would be more prone to heating due to any external source of current noise) continues to evolve with temperature to lower temperatures than for the less resistive samples.
The above arguments can be applied more generally.  
In Figs. \ref{Saito}, \ref{exp2Pasupathy2015},  and \ref{Hwang},  which show resistance data for current-bias measurements in the form of ${\rm ln}R$ vs. $1/T$, we include lines showing the activation and saturation lines which clearly demonstrate, in all the exhibited data, that the deviation from activated behavior occurs at higher temperatures in the presence of  tuning that results in lower saturated resistance.  
This is inconsistent with heating by current noise.

The  issue of voltage bias heating is more delicate. In this case the average dissipated power would be proportional to $\langle V ^2\rangle/R$, so it could  account for the just discussed leveling of the resistance at higher temperatures for lower values of the resistance as the external parameter varies.  However, the fact that within a small range of magnetic fields or gate voltages, without any other change in the sample environment, the sample can either exhibit  anomalous metallic or  superconducting behavior  appears to be inconsistent with  the notion that the effective electron-temperature is bounded by voltage noise.  Specifically, were there significant voltage noise, the delicate superconducting state observed proximate to the QSMT would  not be expected to survive; rather, the saturation temperature should simply shift slowly (logarithmically) down  as the control parameter is varied.
 
  Additional evidence can sometimes be obtained by using the same experimental protocol to study the  thermal evolution of different systems, or the same system tuned from the superconducting state in a different manner.  For instance,  in the system of Josephson coupled superconducting dots studied  in  \cite{Bottcher2017} (the same system shown in Fig. \ref{Marcus}), where low $T$ saturation of $\rho(T)$ is observed for a range of gate voltages and perpendicular magnetic fields, in the same temperature range, as a function of an increasing parallel field  (not reproduced here), the system did not show any range in which the resistivity saturates at low $T$. The authors  concluded that the absence of an anomalous metallic regime when the transition is driven by an in-plane magnetic field indicates that it is unlikely that uncontrolled heating is responsible for the observed saturation in a perpendicular field or in the gate-tuned transition. Similarly, in other cases, measurements in the same cryostat system, on non-superconducting samples with similar resistance in the same measurement circuit, and on the same type of substrate yield consistent results with no signature of heating (see e.g. \cite{EphronThesis,MasonThesis}.) 

Finally, it is important to stress that these issues are significant and complex. 
The electron-lattice relaxation rates vanish as $T$ tends to 0, and that this could lead in some circumstances to an apparent saturation of the $T$ dependence of $\rho$ below a crossover temperature at which the effective electron temperature ceases to decrease.  In light of the importance of  the issue, it is important that continuing efforts be made to directly measure the electron temperature in the anomalous metallic regime, and to mitigate the effects of any external noise in {\it each} system in which such behavior is observed.

\section{Theory}
\label{theory}
\subsection{The inadequacies of various ``obvious'' approaches}
To begin with, we discuss a variety of theoretical approaches, to examine why they are {\it not} consistent with the observed phenomena. 

\subsubsection{The inadequacy of classical percolation}

One might think to account for the anomalous metallic phase from considerations of classical percolation.  Imagine a system that consists of a macroscopic mixture of superconducting regions (with typical radius  large compared to the superconducting coherence length, $\xi$) and metallic regions with conductivity $\sigma_{D}$.  
 The conductivity is then given by $\sigma=\sigma_D\  F(x)$, where $x$ is the volume fraction of superconductor, and $F(x)$ is a  dimensionless function. 
 Obviously, above percolation, $x>x_c$,  the conductivity is infinite for any $x$. However, 
 for $x<x_c$, the conductivity is finite.  
 While some aspects of $F$ depend on the details of the ensemble being studied, in general\cite{percolation}  $F(x) \to 1$ as $x\to 0$ and $F(x)$ diverges as $F(x) \sim (x_c -x)^{-s}$ as $x$ approaches 
 $ x_c$ from below, with $s=4/3$ in 2D and $s\approx 0.73$ in 3D.

The conductivity of an almost percolating superconductor, 
while finite, can be arbitrarily large.

There are several reasons such an explanation cannot successfully be invoked to account for the observed anomalous metallic phases:

a) In the framework of the classical percolation to satisfy the condition that $\sigma/\sigma_{D}\gg 1$, it is necessary that the system be fine-tuned to the very close vicinity of the percolation threshold, $x_c$.  
In the experiments reviewed above, 
$\sigma/\sigma_{D}$ can be as large as $10^{4}$ which would require $(x-x_{c})\sim 10^{-3}$. This is difficult to reconcile with the relatively broad range of parameters and circumstances over which the anomalous metal is observed.

b)  For a classical percolation picture to hold, the distance between superconducting puddles must be larger than $min[L_{T}, L_{B}]$, where $L_{T}=\sqrt{\hbar D/k_BT}$ is the normal metal coherence length, and $L_{B}=\sqrt{\Phi_0/2\pi B}$ is the magnetic length. ( $\Phi_0$ is the flux quantum.)  An inevitable corollary of this picture is that at low enough temperatures, such that $L_T$ grows to be larger than the typical spacing between superconducting regions, global superconducting coherence will be established,
leading to a further growth of $\sigma$ and  
a superconducting ground-state. 
 Manifestly, to describe the quantum superconductor-metal transition at $T=0$ one has to take into consideration quantum fluctuations of the order parameter. 

c) 
It can be shown\cite{Stroud1984} that the effective Hall conductivity in a 2D metal-superconductor mixture is the same as that of the metallic component, independent of $x$ for $x < x_c$. 
Where this expectation has been tested in InO$_x$ and TaN$_x$ films \cite{Breznay2017}, 
  it has been found that $\sigma_{xy}$ of the anomalous metal is much smaller than its Drude value.

d) Finally, there is good reason to 
doubt that such macroscopic inhomogeneities occur in 
many of the systems discussed above.  Some of these systems consist of ordered arrays of superconducting dots on metallic substrates, and others consist of metallic films whose structural and chemical homogeneity has been scrutinized using various probes.  It 
seems unlikely  that there is a hidden inhomogeneity in the structures of 
 these systems on the requisite length-scales to justify a percolation analysis.

\subsubsection{The inadequacy of ``conventional'' fluctuation superconductivity}

The theory of classical superconducting fluctuations upon approach to a transition with a finite $T_c$ is well developed.  (See for a review  Ref. ~\cite{LarkinVarlamov}).
In some sense this would seem to provide a prototype for the properties of an anomalous meal. 
Indeed, the fact that the growing superconducting correlations allow an increasing portion of the current to be carried by collective Cooper pair fluctuations leads to a 
 contribution to the conductivity that diverges as $T\to T_{c}$.
  Moreover, since bosonic fluctuational Cooper pairs have a size which diverges as $T\to T_{c}$,  they are not subject to the single particle interference effects that lead to the weak localization. 
  
 There are problems with using this approach to explain properties of the anomalous metal regime:  
 
 The width of the regime in which fluctuational effects are significant $\delta T\sim T_{c}\ {\cal G} 
 \ll T_{c}$ is controlled by the Gilzburg-Levanyuk parameter \cite{Levanyuk1959,Ginzburg1961}
 ${\cal G}\equiv 1/N_{\xi}\ll 1$. 
 Here
  \be
 N_\xi = \nu \Delta \xi^D
 \ee
can be interpreted to be   the number of electrons per coherence volume that are paired upon entering the superconducting state,  $\nu$ is the metallic density of states at the Fermi energy and $\Delta$ and $\xi$ are, respectively,  the typical gap magnitude and the superconducting coherence length in the superconducting ground-state. 
In many conventional superconductors ${\cal G}$ is  small. 
For example, in quasi-2D samples with statistically uniform disorder,
${\cal G}\sim e^{2}/\hbar \sigma_{D}^{(2d)}$.
(See, for example, Ref.~\cite{LarkinVarlamov}) Note that the celebrated Aslamazov-Larkin \cite{AslamazovLarkin}, and Maki-Thompson \cite{Maki,Thompson} corrections to the Drude conductivity  are  calculated in the temperature interval $(T-T_{c})/T_{c}\gg {\cal G}$, where they are small. 
 Moreover, these fluctuations corrections exhibit strong temperature dependence as $T\to T_{c}$,  while the measured conductivity in the anomalous metal regime is temperature independent at the lowest temperatures.

 \subsubsection{The inadequacy of local Bosonic  theories}\label{bosonictheories}

A theoretical treatment of the transition to a superconducting state can always be treated in terms of an effective action, $S^{eff}[\Delta]$, that is a functional of a charge $2e$ complex scalar field, 
$\Delta$.  Formally, $S^{eff}$ can be obtained from a microscopic electronic Hamiltonian by introducing $\Delta$ as a Hubbard-Stratonovich field and then integrating out the Fermionic electronic degrees of freedom. 
However, physically there is an important distinction between cases in which $S^{eff}$ is a local functional, when it can be expressed in terms of 
an integral over $\Delta({\bf r},t)$ and its derivatives, or a non-local functional. 
In the former case, the low energy long-wave-length degrees of  freedom can be thought of as 
 ``purely bosonic.''  In the latter case, the non-locality reflects the existence of gapless, delocalized fermionic degrees of freedom that need to be taken into account in one way or another;  under these circumstances, the procedure of integrating out the fermionic modes is a formal trick that can be 
misleading. 

The conventional Landau-Ginzburg-Wilson treatment of classical finite temperature phase transitions is an example of a purely bosonic theory.\footnote{While strictly speaking the notion of statistics does not enter the discussion of classical critical phenomena,  order parameters always correspond to an even number of electron creation operators, and so are ``bosonic.'' }
$S^{eff}$ is  ``local'' in an interval of temperatures near $T_{c}$, and on spatial scales larger than  
the coherence length of the normal metal $L_{T}=\sqrt{\hbar D/k_BT}$  evaluated at   $T=T_c$. In other words $S^{eff}$  can be expanded in terms of
 $\Delta({\bf r},t)$ and its time and space derivatives.
This follows from the fact that the various fermionic response functions that enter $S^{eff}$ decay exponentially on scales larger than $L_{T}$.
 It thus seems natural  that the same considerations can be applied to  zero temperature quantum phase transitions.  
However, in the case of a 
QSMT, $L_{T}\to \infty$.  Consequently, the
 various electron response functions exhibit power-law decays at long distances 
and hence $S^{eff}[\Delta]$ 
 is non-local.
 
 In the cases we have discussed in which an anomalous metal phase is observed,  the single-particle states are presumably gapless.
 {\it Thus, no purely bosonic theory is adequate.}  

Currently there are no generally reliable methods to treat non-local  actions. 
This is not to say that it is never reasonable to approach the problem from this perspective.  
Studies of metallic criticality based on the 
    Herz-Millis \cite{Hertz1976,Millis1993} theory adopt such an approach.   
 In the context of the QSMT, there are a class of model problems, which correspond to a quantum version of the  phenomenologically defined RSJ model, 
  for which the theoretical solution is clear   as discussed in Appendix \ref{RSJ}.  
 
 A purely bosonic description may well be  possible in a system consisting of superconducting grains coupled by tunnel junctions system such that below a ``mean-field'' transition temperature there is a negligible density of low energy fermionic excitations.  
 In this case, a conventional  action describing  Josephson-coupled superconducting grains  supplemented with a  quantum capacitance term describing the quantum dynamics of the phase of the order parameter is appropriate.
 Typically, a proper treatment of such an action yields a quantum superconductor-insulator transition.\cite{Matthew}

  While we feel that the absence of gapless quasiparticle modes in purely bosonic models already disqualifies them as descriptions of the experiments discussed above, in  the interest of completeness, we conclude this subsection with a discussion of  a few of these exotic proposals.
  In some 
 circumstances, which are usually associated  with  strong correlation effects that can give rise to localized spins,
 even the sign of the Josephson coupling is a random variable. (See for example  \cite{Bulaevskii,KivelsonSpivakNeg}.) 
 It was further noted in 
 \cite{KivelsonSpivakNeg}  that  randomness in the signs of inter-grain Josephson couplings  
 can bring the system  into the universality class of a quantum XY spin-glass. Moreover, it was hypothesized in  \cite{Phillips2} that such a quantum superconducting glass has finite conductivity.  However, given that the anomalous metal regime has been observed in a broad range of systems, some of which are quite pure and with no other signs of strong correlations, it is difficult to believe that the random sign of the Josephson couplings can be a generic property of systems exhibiting anomalous metallic behavior. Not less importantly, transport properties of quantum spin and superconducting glasses 
 are almost totally uncharted territory, theoretically.   In particular, 
 it remains to be established whether or not the conductivity of the glass phase is finite.

 It was proposed in 
 \cite{dasanddoniach} that the Bose-Hubbard model can exhibit a ``Bose metal'' phase which is formally related to a spin-liquid with a spinon Fermi surface.  However, given that there is no intrinsic frustration in the model (the ground-state can be proven to be nodeless), and the fact that subsequent studies have clearly shown that spin-liquids arise only when any conventional ordering tendencies are strongly suppressed (extremely frustrated) it is now  clear this proposal is not correct.  
 In principle, extensions of this idea involving uniformly frustrated versions of the same model can give rise to spin-liquids with a spinon Fermi surface \cite{hongmaissamandme,hongzhangandme,fisherbosemetal} - however, in the absence of an emergent gauge field, such a spinon Fermi surface is inherently unstable.\cite{hongmaissamandme} 

 A more sophisticated version of a  Bose-metal proposal has been mooted \cite{srimetal,mikemetal} in the context of the magnetic field driven QSMT.  (Here the magnetic field implicitly introduces the requisite frustration.)  A related proposal - but one including gapless fermions - was made in \cite{senthil}.  In 2D there is a  precise correspondence~\cite{ZHK,lopezfradkin,HLR,son,sriandco} between charged bosons in a magnetic field, and charged fermions in a shifted average magnetic field and coupled to an emergent dynamical gauge field. It was argued in 
  \cite{srimetal} and \cite{mikemetal} that this mapping  provides a rational for metallic behavior in the neighborhood of a QSMT.
 This idea builds upon the earlier notions using similar theoretical technology, which establish an analogy between a field driven SIT and various quantum Hall plateau transitions.\cite{KLZ,chakravartykapitulniandme,KapKivDualityPaper}  
 
 While these  proposals are interesting in their own right, there are several additional reservations we have about their {\em application} to the QSMT: Firstly, 
  they do not offer any handle on the observed similarities between the field driven QSMT and the transition in the absence of a magnetic field.  
 Secondly, these theories treat the field-induced vortices as quantum mechanical point-particles; however, in the  systems of interest with  large normal state conductances  $G$, the vortices are quasi-macroscopic. Their quantum tunneling amplitude is controlled by a parameter $\exp(-G)\ll 1$  , so that they behave as essentially classical objects on all relevant energy and temperature scales.  
 
\subsubsection{How BCS theory implies the absence of quantum critical fluctuations  in systems without competing interactions}

It is natural to associate  the anomalous metal 
with growing ground-state superconducting correlations as a QSMT is approached from the metallic side.
 In Subsection \ref{granulartheory} we will discuss theoretically tractable circumstances in which the requisite quantum fluctuations indeed occur.
  First, however, we discuss why even the existence of  a quantum critical regime is an issue.  Specifically, because the uniform susceptibility of a Fermi liquid diverges (logarithmically) as $T\to 0$, even in the presence of weak disorder, any net attractive interaction  generally leads to a superconducting groundstate.  Conversely, weakly repulsive interactions are ``irrelevant'' and thus can be treated peturbatively.  According to this line of reasoning, the  QSMT occurs when the effective interactions vanish.

  Since this is an important point of perspective, let us explicitly consider 
the QSMT in the context of the 
 hamiltonian
 \begin{equation}\label{eq:BCS}
 H=H_{0}- \int d {\bf r} 
\ u({\bf r})\   \Psi^{\dagger}_{\sigma}({\bf r}) \Psi({\bf r})^\dagger_{-\sigma} \Psi_{-\sigma}({\bf r}) \Psi_{\sigma}({\bf r})
 \end{equation}
 where $\psi^\dagger_\sigma({\bf r})$ creates an electron with spin polarization $\sigma$ at position ${\bf r}$, and the sign convention is chosen such that $
 u({\bf r}) >0$ corresponds to a local attractive interaction between electrons.

 In generalized BCS mean-field theory, the local gap parameter is determined self-consistently in terms of the anomalous expectation value of the pair-field creation operator 
 according to
 \begin{equation}\label{Defin}
\langle \Delta({\bf r})  \rangle \equiv  -
u({\bf r})\  \langle\psi_{\uparrow}({\bf r})\psi_{\downarrow}({\bf r}) \rangle  
\label{BCS}
\end{equation}
where  $\langle \ \rangle$ represents the quantum mechanical   average.  Thus, the mean-field superconducting transition temperature $T_c$ (if it exists) is the temperature below which the largest eigenvalue, $\lambda(T)$, of the  the linearized gap equation
\be
\lambda(T) \Delta({\bf r}) = - u({\bf r})\int d{\bf r}^\prime\ K({\bf r},{\bf r}^\prime) \Delta({\bf r}^\prime) 
\ee
is larger than 1, i.e. $\lambda(T_c)=1$ and $\lambda(T) > 1$ for $T<T_c$.  Here $K({\bf r},{\bf r}')$ is the  nonlocal order parameter susceptibility, which for non-interacting electrons (or, more generally, for a Fermi liquid) can be expressed in terms of a convolution of single particle Matsubara Green functions,  
\be
K({\bf r},{\bf r}')=T\sum_{\omega} G_{\omega}({\bf r},{\bf r}')G_{-\omega}({\bf r},{\bf r}') 
\ee
 where $\omega=(2n+1)\pi T$. 
At finite temperature $K({\bf r},{\bf r}^\prime)\sim |{\bf r} -{\bf r}^\prime|^{-(d)}$ for $|{\bf r} -{\bf r}^\prime|\ll L_{T}$ and $K({\bf r},{\bf r}^\prime)\sim \sim e^{-|{\bf r} -{\bf r}^\prime|/L_T}$ for $R\gg L_{T}$. 
The essential feature is that $K$ is a  monotonically decreasing function of $|{\bf r} -{\bf r}^\prime| $ which falls sufficiently slowly with distance that its integral diverges as $T\to 0$:  
\be
\int d{\bf r}^\prime \ K({\bf r},{\bf r}^\prime) \sim \nu \log[E_F/T].
\label{BCSlog}
\ee
This is nothing more than a reflection of the Cooper instability of a Fermi liquid. That this relation is true even in the presence of disorder (at least out to distance scales comparable to the localization length, if the electronic states are weakly localized) is the essence of ``Anderson's theorem.'' \cite{Anderson1959}

To slightly belabor the point, notice that a variational lower bound to $\lambda(T)$ can be obtained 
 by considering a trial state, $\Delta({\bf r} )= \sqrt{u({\bf r})/\bar u}\ \Delta$,  which yields
\be
\lambda(T) \geq  \ \Omega^{-1} \int \ d{\bf r}\ d{\bf r}^\prime\ \sqrt{u({\bf r})u({\bf r}^\prime)}\ K({\bf r},{\bf r}^\prime)
\label{bound}
\ee
where $\Omega$ is the ``volume'' (area in 2D) of the system  and $\bar u$ is a suitable average of $u({\bf r})$.  This gives a lower-bound to the mean-field $T_c \geq E_F \exp[-1/(\bar u \nu)]$, which is manifestly non-vanishing for any $\bar u > 0$.  Further, since for small $\bar u$, the associated zero temperature coherence length is exponentially long, the  assumption that the pairing amplitude is uniform is self-consistently validated as all finite length-scale inhomogeneities are averaged out.  (In the literature, this is sometimes referred to as the ``Cooper limit.'')  Finally, the fact that the usual (thermal) Gilzburg-Levanyuk parameter diverges as $\bar u \to 0$ implies that the  mean-field estimate of $T_c$ becomes asymptotically exact. 

More realistic models, for instance those involving  low energy  attractive and high energy  repulsive interactions, 
when treated using the usual diagramatic approach give rise to the same conclusion:  
 the QSMT is driven by a change 
 in sign of the effective interaction. Consequently, the effective interaction
vanishes identically at the point of the quantum phase transition.

As already mentioned, in 2D even for $k_F\ell \gg 1$, all single-particle states are localized \cite{GangofFour,LarkinKhmelnitskiiGorkov,LeeRamakrishnan}, so the divergence of $\lambda_0(T)$ is cut off below $T\sim T^\star$, defined in Eq. \ref{Tstar}.  In principle, this could result in a non-vanishing interaction strength at criticality.  However,
   the corresponding critical regime is parametrically narrow and any critical effects would be confined to exponentially low temperatures, $T \lesssim T^\star$.  Thus, for clean metals, these considerations are of no practical importance. 
(We will return to the issue of localization in Sec. \ref{conclusion}.)

When localization physics can be neglected, it is not so much a question of why the metal is not an insulator as how can one understand the existence of a ``failed superconductor'' in which strong superconducting correlations develop below a non-zero crossover scale, but the ground-state fails to be globally phase coherent.  In the next subsection, we will show how this can arise in the case in which $u({\bf r})$ is attractive in some regions of space and repulsive in others. In Sec. \ref{conclusion} we discuss
other possible origins of  critical fluctuations near a QSMT.

\subsection{
Theory of the QSMT in granular systems}
\label{granulartheory}
 
In order to construct a theory of the quantum critical regime near a QSMT, it is necessary to identify loopholes in the  considerations outlined above that lead to a breakdown of BCS theory.  One route is to identify processes that cut off the  divergence of the superconducting susceptibility in the metallic state, Eq. \ref{BCSlog}.  Another is to consider the case in which there are competing attractive and repulsive interactions, or where a magnetic field cuts off the divergence of the susceptibility so that the QCP occurs at a point in the phase diagram at which interactions have non-negligible effects.  
   
In this section, following  on the analysis of Refs. \cite{LarkinFeigelman,OretoKivSp,Hruska}, we consider the QSMT   in a system with  a spatially non-uniform $
u({\bf r})$.   This  provides an important theoretical paradigm that explains how in principle at zero temperature the 
 conductivity can diverge upon approach to the point of the quantum phase transition.  
  As far as we know, these are  the only solvable {\em microscopically plausible} models of a QSMT with an observable quantum critical regime.

 \subsubsection{Strategy of solution}
 
 The system we will analyze consists of far separated superconducting puddles embedded in a normal metal background.  We consider the limit in which the distance between puddles is large compared to their size, as this separation of scales permits a controlled theoretical approach to the problem.  To begin with, we compute the zero temperature superconducting susceptibility of an isolated puddle, $\chi_j$.  Consistent with general expectations, 
 this susceptibility is always finite, but it 
 can depend exponentially on characteristic properties of the puddle, and so can be very large.  Then, we compute the Josephson coupling, $J_{ij}$, between pairs of puddles, $i$ and $j$.    Importantly, $J_{ij}$ reflects the quantum diffusion of Cooper pairs through the normal metal, and so falls relatively slowly with the separation between puddles, in sharp contrast to the behavior of the Josephson coupling through an insulating region.

 What enters thermodynamic considerations is the dimensionless coupling between puddles, 
 \begin{equation}
X_{ij}=\sqrt{|\chi_{i}J_{ij}\chi_{j}J_{ji}|}\ .
\end{equation}
Two puddles fluctuate essentially independently of each other if $|X_{ij}|\ll 1$, and they are phase locked to each other if  $|X_{ij}|\gg 1$.
The quantum transition to a globally phase-coherent state occurs at the point at which an infinite cluster of puddles is coupled together by links with $|X_{ij}|
\gtrsim 1$.
At slightly larger mean spacing between puddles, large clusters of puddles are still phase locked, which thus implies the existence of significant quantum critical effects.

 Note that for this procedure to be valid the sum  $\sum_{j}J_{ij}$ must be convergent.  At any non-zero $T$, $J$ falls exponentially with distance $R_{ij}$ between grains, so convergence is guaranteed.  
However, at $T=0$, and for repulsive $u_N>0$,
$J_{ij}\sim 1/|R_{ij}|^{-D}(1+u_{N} \ln^{2}R_{ij})$. 
In the special case in which there are no interactions in the normal metal ($u_{N}=0$) the sum is logarithmically divergent and the ground-state is thus always superconducting.
Thus the sum is convergent and the transition exists only because of the repulsion in the normal metal.

 Since most  experiments on the anomalous metal  are on 2D devices, we will consider this case.

 \subsubsection{ Model of superconducting puddles in a metal}
 
 Let us consider an s-wave superconducting grain  that is embedded in a normal metal.
 For simplicity we consider the following spatial structure of the  electron interaction 
\begin{equation}
u({\bf r})  = 
\left\{
\begin{array}{ccc}
u_{S} >0 &  {\rm for} &  |{\bf r}|<R   \\
u_{N}<0  &  {\rm for} &  |{\bf r}|<R 
\end{array}
\right.
  \label{decay}
\end{equation} 
 Here $
 u_{S}$ and $u_R$ are the interaction constants in the superconductor and in the normal metal respectively. 
  It follows from general statistical mechanical considerations that 
 the quantum mechanical average of 
 the order parameter of a zero-dimensional system
 $\langle \Delta \rangle=0$.  
 
 Theoretical investigation of the correlation function of the fluctuations of the order parameter has a long history. Here we briefly summarize the main results. 
 At mean-field level, there exists a critical puddle radius, $R_c$, such that for $R>R_c$ there is a non-zero solution of the mean-field equations (Eq. \ref{BCS}), while for $R<R_c$ no such solution exists.  
So long as there is no reflection at the puddle boundary we get $R_{c}\sim \xi$, where $\xi$ is the superconducting correlation length of a bulk superconductor 
with the interaction constant $u_{S}$.
The character of the superconducting quantum fluctuations are quite different depending on whether $R$ is less than or greater than $R_c$.

\subsubsection{Large puddles with $R\gg R_{c}$}

For puddles with $R\gg R_c$, there is a $T=0$ mean field solution for the order parameter with $\Delta_{MF}({\bf r})\approx \Delta_{0}$ for $|{\bf r}|<R$ and $\Delta_{MF}({\bf r})=0$ otherwise. 
In this case the quantum fluctuations of the modulus of the order parameter can be neglected,  the order parameter on an individual superconducting puddle can be parametrized as 
 $\Delta_i\equiv  |\Delta_{0} | e^{i\phi_i}$ and the quantum dynamics of the system                                          
  can be described in terms of phase variables alone. The corresponding phase fluctuations in 
  the  i-th puddle   can be described by the action introduced in 
\cite{Sudip} 

\begin{equation}\label{eq:phiAction}
S_{i}[\phi_{i}]= -\frac{G_i^{eff}}{(2\pi)^{2}}\int dt dt' \frac{\sin^{2}[\frac{1}{4}(\phi_{i}(t)-\phi_{i}(t'))]}{(t-t')^{2}} 
\end{equation}
Here $G_i^{eff}\gg 1$ is an effective conductance of the medium measured in units $e^{2}/\hbar$.
As a result 
\footnote{The fact that in both limits the correlation functions Eqs.~\ref{eq:correlationFunctionSmallR} and \ref{eq:correlationFunctionLargeR} decay at large times as $1/t^{2}$ is a manifestation of a more general principle: whenever the retarded Green function 
 decays exponentially with time, 
 the causal Green  function decays inversely proportional to time squared.}
\begin{equation}\label{eq:correlationFunctionLargeR}
\langle e^{-i(\phi_{i}(t)-\phi_{i}(0))} \rangle  = 
\left\{
\begin{array}{ccc}
\frac{1}{|\tau|^{1/G_i^{eff}}} & {\rm for} & \tau\ll \tau^{*}   \\
(\frac{\tau^{*}}{\tau})^{2} &  {\rm for} &  \tau \gg \tau_i^{*}
\end{array}
\right.
  \label{decay1}
\end{equation}
where 
\begin{equation}
\tau_i^{*}\sim \tau_{0}\exp(2\pi^{2}G_i^{eff})
\label{exponential}
\end{equation}
and 
\begin{equation}\label{eq:chi}
\chi_i 
\sim \tau_i^{*}.
\end{equation}

The definition of $G^{eff}$ in Eq.~\ref{eq:phiAction} requires clarification.  This expression is derived by considering the dynamical screening of charge fluctuations in the superconducting puddle by the surrounding metal.  In the 3D case, $G^{eff}\sim \hbar\nu D R_i$ is the two terminal conductance 
in units $e^{2}/\hbar$, which is obtained if one lead is put inside the superconducting puddle and the other 
is placed on a boundary at  infinity.   In the 2D case, the conductance defined in this way vanishes in the thermodynamic limit.  A more delicate analysis \cite{FiigelmanLarkinSkvortsov} shows that in this case
\begin{equation}
G^{eff}\sim \sqrt{\sigma_{D}^{(2d)}\hbar/e^{2}}
\end{equation}
 Note that in this case, $G^{eff}$ is independent of the puddle size.

The action describing a  collection of  large puddles embedded in a metal  has the form 
\be
S=\sum_iS_{i}[\phi_i]+S_{int}[\{\phi\}],
\ee
where $S_{int}=\int dt H_{int}$ and
\begin{equation}
H_{int} = - \frac{1}{2} \sum_{i\neq j} \tilde{J}_{ij}  \cos(\phi_{i}-\phi_{j})
\end{equation}
is the Josephson Hamiltonian. In  the case where the inter-grain distance is larger than the zero temperature superconducting coherence length
the inter-grain coupling is given by a conventional cooperon (ladder) diagram \cite{AGD,Sid} 
\begin{equation}\label{eq:Jlarge}
\tilde{J}_{ij}   = \frac{\hbar}{2e}\sigma_{2d}\frac{D}{R^{2}} \frac{R^{2}}{|r_{ij}|^{2}\left[1+2u_{N}\ln^{2}\left(|r_{ij}|/R\right)\right]}.
\end{equation}

Note that for $u_N >0$, the sum over couplings Eqs.~ 
\ref{eq:Jlarge}, between remote junctions converges, $\sum_j\tilde J_{ij} < \infty$, as required.  However, in the limit $u_N =0$, this sum diverges, consistent with our earlier observation that the existence of both attractive and repulsive interactions is a necessary condition for the existence of a critical regime.
  
\subsubsection{Near critical puddles with 
$|R-R_{c}|\ll R_{c}$:}

The value of $R_c$ can be obtained from the 
 linearized mean-field self-consistency equation 
\begin{equation}
F_{0}({\bf r})=-
u({\bf r})\int d {\bf r}_{1}K({\bf r}, {\bf r}_{1})
F_0({\bf r}_{1})
\end{equation}
which has a solution with non-zero 
$F_0$
when $R=R_c$.
We can always normalize $F_0$ such that  $\int |F_0|^{2} d {\bf r}=1$, and we chose a phase convention such that $F_0({\bf 0})$ is real and positive.

For $R< R_c$ so long as $R_c-R \ll R_c$, 
the important fluctuations  can be parameterized as $\Delta({\bf r},t)=\alpha(t) F_0({\bf r})$;
the fluctuations of the shape of $\Delta({\bf r},t)$ can be neglected and one need only calculate  the correlation function of the complex amplitude $\alpha(t)$. This  is given by the ladder diagrams (See for example  
\cite{AGD}) 

\begin{equation}\label{eq:correlationFunctionSmallR}
\langle \alpha_i(\omega) \alpha_i^{*}(-\omega) \rangle _{C}=\frac{1}{\nu\tau_{0}(-i|\omega|+\frac{1}{\tau_i^{*}})}
\end{equation}
which in $t$-representation gives us 
\begin{equation}
\langle \alpha_i^{*}(0) \alpha_i(t) \rangle_{C} = 
\left\{
\begin{array}{ccc}
\frac{1}{\nu \tau_{0}}\left(\frac{\tau_i^{*}}{t}\right)^{2} &  {\rm for} &  \tau_i^{*} \ll \tau_{0}   \\
\frac{1}{\nu \tau_{0}}i[-i\pi+2\ln (t/\tau_i^{*})] &  {\rm for} & \tau_i^{*} \gg \tau_{0}
\end{array}
\right.
  \label{eq:correltionalpha}
\end{equation}
Here the subscript $C$ refers to the casual (the time ordered) Green function,  $\tau_{0}=min[R_{c}/v_{F};R^{2}_{c}/D]$ is the time of electron flight through the grain, and 
\begin{equation}
\tau_i^{*}=\frac{\tau_{0}R}{R_{c}-R_i}
\end{equation}
As a result we get an estimate for the superconducting susceptibility of the grain
\begin{equation}\label{eq:chi1}
\chi_i
=\int dt \langle \alpha_i^{*}(0) \alpha_i(t) \rangle_{C} \sim \frac{\tau_i^{*}}{\nu\tau_{0}}
\end{equation}

Alternatively one can get the same results for the correlation function  using the non-local in time effective action obtained by integrating out the fermionic modes
which when  expanded up to quadratic terms in $\alpha_i(t)$ is
\begin{equation}
S_{i}[\alpha_i]=\nu \tau_{0}\int \frac{d \omega}{2\pi}\left(-i|\omega|+\frac{1}{\tau_i^\star}\right)|\tilde \alpha_i(\omega)|^{2}
\label{eq:actionsmall}
\end{equation}
where $\tilde \alpha_i$ is the Fourier transform of $\alpha_i$.

Eqs. ~\ref{eq:correltionalpha}, \ref{eq:chi1}, and  \ref{eq:actionsmall} are valid even for $R_i > R_c$ as long as the amplitude of the fluctuations of the order parameter is larger than its mean field value squared
$\langle |\delta\Delta_i|^2\rangle \approx \langle  |\Delta_i(0)- \Delta_i(t\sim\tau_i^{(*)})|^{2} \rangle \sim 1/\nu R^{2}_{c} \tau_{0} \sim \Delta_{0}^{2}/\sigma_{2d}\gg \Delta_{MF}^{2}\sim \Delta_{0}^{2}\left[(R_i-R_{c})/R_{c}\right]$, and the nonlinear terms in the action can be neglected.
Here $\Delta_{0}$ is the value of the
 order parameter in a bulk sample with  interaction constant
$u_{S}$. 

In the opposite limit,$\langle |\delta\Delta_i|^2\rangle \ll  \Delta_{MF}^{2}$ we have 
\begin{equation}
\tau_i^{*}\sim \exp \left[\frac{\Delta_{MF}^{2}}{\langle |\delta \Delta_i|^{2} \rangle}\right]
\end{equation}

To describe a system of superconducting  puddles embedded into a metalic host we can use the effective action $S[\{\alpha\}]=\sum_iS_{i}[\alpha_i]+S_{int}[\{\alpha\}]$,
where 
\begin{equation}
S_{int}=
-\sum_{ij}\int d t [J_{ij}\alpha_{i}^\star\alpha_{j}+ c.c]+
b\sum_{i} \int d t | \alpha_{i}|^{4}
\end{equation}
where $b\sim R_{c}^{2}\nu/aD^{2}$, and $a\ll \xi$ is the film thickness.
Since the quantum fluctuations of the phase of the order parameter are  slow compared to the inter-puddle electron propagation, the Josephson coupling energy can be calculated using the conventional cooperon (ladder) diagram 
\cite{AGD}
\begin{equation}\label{Jsmall}
J_{ij}=\frac{\nu R^{2}}{|r_{ij}|^{2}\left[1+2u_{N}\ln^{2}\left(|r_{ij}|/R\right)\right]}
\end{equation}
Again, the essential feature is the logarithmic correction to the long-distance fall-off, which makes the sum over $J_{ij}$ convergent.

\subsubsection{Effect of magnetic field on $J_{ij}$}

The expressions for $J_{ij}$ and $\tilde J_{ij}$ are somewhat more complicated 
in the presence of a magnetic field  and/or at finite temperature, in which cases they acquire  additional factors, $J_{ij} \to {\cal F}_{ij} J_{ij}$ and 
$\tilde J_{ij} \to {\cal F}_{ij} \tilde J_{ij}$ where
 \begin{equation}
{\cal F}_{ij} \sim \exp\left(-\frac {r_{ij}}{L_{T}}\right)\ \exp\left(-\frac{r_{ij}}{L_{B}}\right)\exp(i \gamma_{ij})
  \end{equation} 
 where 
 $L_{T}=min[v_{F}/T;\sqrt{D/T}]$ is the coherence length of the normal metal, $L_B = \sqrt{\hbar c/e B}$ is the magnetic length, and 
 $\gamma_{ij} = (hc/e)\int_i^j {\bf A}({\bf r})\cdot d{\bf r}$ is a gauge-dependent phase factor.

 \footnote{ If the metal is disordered, then the Josephson couplings, $J_{ij}$ exhibit sample specific  mesoscopic fluctuations, which  in the presence of the magnetic field and at $T=0$ decays only as a power low of the distance, $ \langle\langle J_{ij} ^{2}\rangle\rangle \sim 1/|r_{ij}|^{6}$. 
 \cite{SpivakZhou,ZhouSpivak}.  Here double-brackets stand for both the quantum mechanical averaging and the averaging over random scattering potential configurations.
However, in this article we are mainly interested in samples with good conductances where the relative amplitude of these fluctuations is small. Therefore in what follows we will  neglect mesoscopic effects of this sort. }
Obviously, in this case, the sum over $J_{ij}$ is convergent, even without the logarithmic correction due to non-zero $u_N$.

\subsubsection{Quantum critical region}
To illustrate the implications of the above analysis, let us consider the case in which there is a single characteristic puddle size, $R$, and the SMT is driven by changing the concentration of puddles.  The critical puddle concentration is thus determined according to $\chi J(r) \sim 1$ where $\chi$ is the single puddle susceptibility and $r$ is the typical distance between neighboring puddles.  
We are on secure theoretical grounds in all our considerations if the metallic portion of the system is relatively clean, so that $k_F\ell \gg 1$ (i.e. in 2D, when $\sigma^{(2d)}_{D} \gg e^2/h$).  

If in addition $R\gg R_c$, this means that $\chi$ is exponentially large, and hence that the Josephson coupling at criticality is exponentially small.   
 Therefore in the anomalous metallic regime proximate to the QCP, the temperature below which the T-independent regime can be reached is correspondingly small.
  If $|(R-R_{c})/R_{c}|\ll e^2/\hbar \sigma_D^{(2d)}$,  and if the  magnitude of 
  $u_{N}$ in the metal between the superconducting droplets is sufficiently large,   then the criterion $\chi
J\sim 1$ yields $(\tau^{(*)}/\tau_{0})^{2}(R^{2}N_{D})\sim 1$. Here $N_{D}$ is the concentration of puddles.  In this case, the critical regime is order $\mathcal{O}(1)$ both in concentration and in temperature.  However, satisfying this condition seemingly involves 
a certain amount of fine-tuning of the puddle geometry,  which 
appears to be at odds with the robustness of the observed phenomena.  This is a worrisome shortcoming of the model.

\subsubsection{Magnetic field driven QSM transition}

 The absence of a quantum-critical regime discussed in Sec. \ref{bosonictheories} above, ultimately reflects  the long-range (power law) falloff of the Cooperon correlation function in the normal metal at $T=0$.  However,   in the presence of a magnetic field, these correlations are cut off at the magnetic length.
Therefore , in principle, 
a quantum critical region may exist. 
 In the case of statistically uniform disorder the solution of the mean field equations has a form of Abrikosov lattice. (See for example Ref. ~\cite{Abrikosov}). If the magnetic field is close to $H_{c2}$ at $T=0$ one can 
write the mean field Gizburg-Landau equation for $\Delta({\bf r})$, which solution  gives us 
\begin{equation}\label{eq:meanfieldDeltaH}
\tilde {\Delta}^{2}(H)\sim \Delta_{0}^{2} \frac{H_{c2}-H}{H_{c2}}
\end{equation} 
where tilde indicates averaging of the modulus of the order parameter over a period of vortex lattice (See for example \cite{Abrikosov}). 

Beyond the mean field approximation 
 the classical transition sometimes turns out to be  the first order \cite{Herbut1995}.
However the first order is forbidden in disordered 2d samples \cite{ImryWortis,Aizenman,Berker,Chakrovarty}. 
Ultimately, in high conductance samples the transition is controlled by formation of rare droplets which are connected by Josephson junctions \cite{OretoKivSp,ZhouSpivak}.

 However, for the purpose of this article it is sufficient to say that 
in 2D case the  width of the critical region as a function of magnetic field is controlled by the conductance of the film $\sigma_{D}^{(2d)}$ (See for example Ref.~\cite{Blatter1994}).
Therefore it is narrow in samples with large conductance. 
Indeed, the amplitude of quantum fluctuations of the order parameter in metal, averaged over an area of order $L_{B_{c2}}^{2}$ is given by a standard Cooperon diagram \cite{AGD}
\begin{equation}\label{eq:deltaDeltaH}
\langle \delta \Delta^{2} \rangle \sim \frac{\Delta_{0}}{\nu L_{Bc2}^{2}}\sim \frac{e^{2} \Delta_{0}^{2}}{\hbar \sigma_{D}^{(2d)}}
\end{equation}

Thus we arrive to a conclusion that the interval of magnetic field where quantum fluctuations of the order parameter are important is
\begin{equation}\label{MagneticFieldInterval}
\frac{|H-H_{c2}|}{H_{c2}}\sim \frac{e^{2}}{\hbar \sigma_{D}^{(2d)}} \ll 1
\end{equation}

In other words Eqs.~\ref{eq:meanfieldDeltaH},\ref{eq:deltaDeltaH}, and \ref{MagneticFieldInterval} imply that in highly conductive samples not very close to $H_{c2}$, vortices remain macroscopic objects, and their quantum tunneling probability is negligible.

In Section \ref{conclusion} we will discuss significance of this result for interpretation of experiments.

\subsubsection{Role of disorder and Griffiths phenomenon}
\label{griffiths}

We conclude this section with a short remark about some possible consequences for the QSMT of rare events associated with certain types of disorder.
Note, however, that the wide variations in  the character of the disorder of the experimental platforms  that exhibit an anomalous metallic phase already suggests that  such effects are unlikely to be of central importance.  

In classical phase transitions the role of disorder has been discussed in detail in the past. 
For example, 
a degree of randomness in the local $T_c$'s can shift the ordering temperature, and in some cases can change the universal critical exponents at a transition, but leaves the character of the phases and the general nature of critical scaling intact.
In addition, in the neighborhood of any critical point in a disordered system, there are universal ``Griffiths phenomena:''  Even for statistically uniform disorder, due to mesoscopic variations in the local disorder configuration there always exist rare regions that are effectively on one side of the critical point even though the bulk of the system is on the other.  In classical systems ({\it i.e.} for finite $T$ transitions), this is largely an academic issue, as these rare regions lead to extremely subtle and  difficult to detect effects.  

The effects of rare regions can be much more important in the neighborhood of a quantum critical point, and especially  in a metallic system. 
In the present context, the essential point is that according to Eq.~\ref{exponential}, the susceptibility of an individual grain  depends exponentially on its properties ({\it i.e.} on $G_{eff}$).
Consequently  the existence of  dilute rare droplets in which the electronic properties differ from average  ({\it e.g.} in which $G^{eff}$ is anomalously large)
can have enormously amplified effect on the physics close to 
the QSMT.   This can lead to significant spatial inhomogeneities in the electronic properties and the relevance of a percolation analysis even in otherwise highly homogeneous materials.   

The superconducting puddles discussed above can arise from such considerations. 
However, as is shown in  
\cite{OretoKivSp}, this amplification is typically not sufficient to lead to a true quantum Griffith phase\cite{DSFisher1,DSFisher2} in which the superconducting susceptibility would diverge in a range of parameters proximate to the quantum critical point. In order for this to happen, the susceptibility of an anomalously large puddle would need to diverge exponentially with its ``volume,'' (its area in 2D).  Conversely, in most situations relevant to the present discussion, in any puddle that is large compared to the superconducting coherence length, the relevant value of $G^{eff}$ grows at most in proportion to the surface area (the circumference in 2D) of the puddle.\footnote{ It was correctly  shown in Refs. \cite{DelMaestro1,Vojta,DelMaestro2} that a true quantum Griffith phase occurs in a disordered version of  a phenomenological model of an array of resistively shunted Josephson junctions discussed in Appendix \ref{RSJ};  while this may well capture correct intermediate scale physics, we argue in that Appendix that ultimately, the quantum Griffith phase} is a consequence of an unphysical aspect of the model itself.

\subsection{
The  QSMT of  a d-wave superconductor}

One might expect that the symmetry of the superconducting state could have important consequences for the nature of the quantum critical phenomena associated with its demise.  
For instance, 
the sign of the Josephson coupling between two d-wave grains can be 
either positive or negative depending on the orientation and shape of the grains and the character of the intervening metal.   Random Josephson couplings 
can lead to the existence of a superconducting glass phase with all the complexity so implied.  Thus, at least at zero magnetic field, 
 one might expect that the nature   non-superconducting state 
 is different 
 proximate to an s-wave or a d-wave superconductor.  
 
However, while many aspects of the more general theory remain to be developed,  
it is possible to argue that in some range of parameters the nature of the anomalous metal phase originating from a d-wave superconductor
is similar to the s-wave case.
 In particular, the 
 phase diagram of a system of d-wave superconducting grains of  random shape  with $R\gg R_{c}$  in a metallic matrix has been discussed in 
  Refs.~ \cite{OretoKivSp,KivelsonSpivakD,KivelsonSpivakD1}. 
  While each grain has dominantly d-wave pairing, an admixture of an  s-wave component is implied by the (generically) assymetric character of each grain - effectively each grain is either a d+s or a d-s wave superconductor.  When the grains are close to one another, the Josephson couplings have a sign structure that generically leads to frustration and, presumably, the complex physics associated with an XY-spin glass.  However,    the longest-range portion of the Josephson couplings connect the subdominant s-wave components.    This results in a situation analagous to that which arises in the Mattis-model, in which the  couplings between the d-wave components are random in sign, but in a way that is thermodynamically equivalent to a ferromagnet (i.e. no frustration).  In effect, while the superconducting order is locally d-wave, on distance scales of order the distance between grains and longer, it is only the s-wave component of the order that matters.  Thus, the physics of the QSMT transition is identical in the two cases.

\section{Conclusions and Further Directions}
\label{conclusion}

In view of the number and diversity of systems in which very similar phenomena have been seen in experiment, we feel that the case for the existence of an      anomalous metallic phase proximate to a QSMT is  compelling.    This conclusion contradicts the widespread  ``belief,'' based on perturbative considerations, that no metallic phase can exist in 2d.

The theory of conventional superconducting grains embedded in a weakly interacting metal establishes  that such a metallic phase is a valid theoretical possibility.
The anomalous metal regime 
 close to the point of the 
 QSMT is very different from  a normal Fermi liquid. 
It is characterized by values of conductivity that diverge upon approach to   the transition.  
 However, for reasons we will analyze below, it seems to us difficult from this approach to account for the robustness of the anomalous metal to variation of circumstances (e.g. whether the QSMT occurs 
 in disordered films, crystalline flakes, or engineered Josephson junction arrays) and for its extension over a relatively broad range of temperatures and quantum tuning parameters.  We therefore conclude with the  {\em suggestion} that the set of observations described in this review may reflect a still more fundamental and far reaching failure of the perturbative (Fermi-liquid based) approach to the theory of metals.

\subsubsection{Is there a satisfactory theory of anomalous metals?}

On a phenomenological level,  fluctuational  contributions to  the conductivity near classical and quantum phase transitions are similar  (See for example, \cite{AslamazovLarkin,Maki,Thompson,LarkinVarlamov,Blatter1994,Hartnoll}).
However, in our opinion the  central problem in the area is the microscopic origins of the QSMT.

The theory in which  the QSMT is driven entirely by  quantum fluctuations of the  order parameter of superconducting grains embedded in a metal
can qualitatively reproduce the salient experimentally observed features of the anomalous metal regime. 
Note that such grains could arise directly as a consequence of sample inhomogeneities, or arise as an intrinsic feature of the electronic structure.  For instance, the interaction between SC and another form of order generally enhances the effects of even weak structural inhomogeneities.  In many unconventional superconductors, the superconducting state occurs in close association with  charge-density-wave (CDW) 
 phases.  While the coupling between the SC order parameter and quenched disorder is constrained by gauge invariance, disorder necessarily destroys long-range incommensurate CDW order, leading to an inhomogeneous state with Larkin-Ovchinikov-Lee-Rice-Imry-Ma domains.\cite{Larkin1979,LeeRice,ImryMa}
 Consequently, if there is strong coupling
between the CDW order and the SC order, 
this can lead to an intrinsically granular SC state. 

However, 
there are reasons to question whether any granular picture can provide an adequate (semi) quantitative account.
The central reason is that the underlying Ginzburg-Levanyuk parameter  
is typically 
small. In the case of large grains  $R>R_{c}$ and large $
\sigma_{D}^{(2d)}$,
it follows from Eqs.~\ref{exponential}, \ref{eq:chi}, and \ref{eq:Jlarge}  that
the critical concentration
of superconducting grains at the point of the QSMT is exponentially small;  this situation is realized, for example,  in the experiments in 
\cite{Bottcher2017}.
Consequently, 
genuinely quantum critical effects should be observable only at exponentially low temperatures.  
 
 The situation may be somewhat better if the grain's  radius is close to the critical one 
 $|R-R_{c}|/R_c \ll \hbar 
 \sigma_{D}^{(2d)}/e^{2}$. This, however,  requires   fine-tuning of the puddle geometry which 
 seems to be at odds with the robustness of the experimentally observed phenomena.  Moreover, even here, the temperature range is small in proportion to $\hbar 
 \sigma^{(2d)}_{D}/e^{2}$.  In principle,  in the case  
of a magnetic field driven QSMT, 
  it is also possible to consider a theoretical approach based on the quantum melting of the vortex lattice. 
  However, as we have discussed in Section \ref{granulartheory},  in addition to providing no insight into the anomalous metal observed in zero magnetic field, we 
think it is likely that this type of theory will  
encounter similar quantitative problems in accounting for the breadth of the anomalous metallic regime.
 For example, the magnetic field driven anomalous metal regime has been observed in samples with 
 $
 \sigma_{D}^{(2d)}$ as large as $\sim 300 e^{2}/\hbar$ in 
 \cite{Tsen2016}, and  $\sim 40 e^{2}/\hbar$ in 
 \cite{MasonKapitulnik1,MasonKapitulnik2}. In the absence of a magnetic field, the gate voltage driven anomalous metal regime  has been observed 
in samples with $
\sigma_{D}^{(2d)}\sim 26 e^{2}/\hbar$ in 
\cite{Bottcher2017,Chen2017}.

This leads us to 
speculate about ways in which the quantum transition can be driven by mechanisms 
that involve more than simply the quantum fluctuations of the superconducting order parameter itself.
Below we list some not entirely logically  distinct possibilities:

{\bf a)} The interaction constant $u({\bf r},t)$ even in pure systems can be a dynamical variable 
that fluctuates as a function of (imaginary) time. 
At an intuitive level, many of the same considerations we have discussed in the context of the effect of purely spatial variations of $u({\bf r})$ due to quenched disorder apply to this case, as well.  
As we will discuss next, this might occur in the neighborhood of another (non superconducting) quantum critical point.
Along the same lines we can consider the situation in which the lattice contains many low-lying excitations in the form of two-level systems.
It is known that the nature of  glassy two-level systems is related to soft local potentials. Thus, in places where they are located the amplitude of quantum oscillations of atoms is large, while the frequency of the oscillations is low.  This could induce the requisite sort of variations of $\lambda$.

{\bf b)} The nature of the superconducting state near a metallic QCP associated with another form of order is a subject that has attracted considerable interest for many reasons.  In the present context, one can certainly imagine that the quantum critical fluctuations of a non-superconducting order parameter will produce the sort of large amplitude slowly fluctuating variations in  the effective $\lambda$ that can enhance the critical regime associated with a QSMT.  
  Indeed, an inspiring study of the superconducting instability in the neighborhood of a class of QCPs\cite{raghuQSMT} found an associated QSMT that occurs at a finite mean BCS coupling - just the sort of situation that can lead to a large quantum critical regime.  However, beyond the point of principle, it is not plausible  that a previously undetected  quantum phase transition can account for the experiments reviewed above.

{\bf c)}  Finally, it is not impossible that the perturbation theory based  conventional theory  of pure uniform superconductors 
  misses  an important aspect of the problem.  It has long been noted that the condensation energy density, $\Delta U = (1/2) \nu |\Delta_0|^2$, is enormously small compared to even high order Fermi-liquid corrections.  Thus, one can wonder whether non-perturbative effects that are of negligible importance for normal state properties, might play a larger role in superconductivity than has been previously appreciated.  For instance, an interacting Fermi system 
exhibits quantum fluctuations of the spin magnetization.  
Moreover, in systems with several bands there are 
low lying excitations related with 
fluctuations  of relative electron densities in 
different bands.  At small $r_{s}$ these oscillations have small amplitude and 
relax rapidly.  
However, in most 
real metals $r_{s}>1$.  In this regime, 
the above 
fluctuations can have 
large amplitude, large correlation radius, and long 
lifetimes.  This implies that it is possible to have a scenario similar to that discussed above in the case of frozen non-uniformity of $u({\bf r})$.

\subsubsection{Is there  a distinct anomalous metallic phase?}

The question arises whether the ``anomalous metal phase'' is a {\em phase} that is distinct from the Fermi liquid phase,
or  if instead there is a smooth cross-over between an anomalous metal and a Femi liquid metal {\em regime}?
 We believe that it is the latter. Indeed, in the anomalous metal regime the electron transport is dominated by 
bosonic quantum fluctuations of the order parameter. However, since the situation is gapless there are also generally fermionic excitations in the system that contribute the  transport.
Conversely,  in  conventional  Fermi liquid metals, there are also  bosonic excitations, such as  for example,  zero spin sound,  
that can carry current. 
While at low temperatures, the relative
contribution to  transport from 
bosonic modes is small in the Fermi liquid and dominant in the anomalous metal, one can envisage a continuous crossover between these two extremes.

\subsubsection{Is there ever a direct SIT?}

In a purely bosonic system with no magnetic field and no other forms of frustration, 
  there is a direct SIT.  Moreover, since the theoretical considerations above 
  imply  that the existence of the anomalous metal depends on the 
presence of repulsive interactions, it seems likely that in an artificial model with {\em purely attractive} interactions between electrons, the superconducting 
state can only be destroyed by localization, implying again the existence of a SIT.  

However, both of these situations are fine-tuned.  Even in a granular superconductor, it is likely that there is some concentration of gapless fermionic excitations, 
and surely there are always some remaining repulsive interactions between electrons.  
Therefore, 
while it is possible that a direct SIT occurs in some cases, it is also possible that 
 even where there appears to be such a transition, there might in actuality be  an extremely narrow intermediate metallic phase.

\subsubsection{Related problems and future directions}
\label{future}
We end with a few observations concerning related problems and future directions.

{\bf 1)}  We are not aware of experimental evidence of the existence of  an anomalous metallic regime in 3d samples.  However, to the extent that the theoretical considerations we have discussed are pertinent, they apply in 3D as well as in 2D.  The existence of a QSMT in 3D is not, in itself, surprising as the stability of the metallic state is not in question.  However, it is important to determine to what extent the metallic state proximate to this transition is anomalous, and and to delineate the extent of the quantum critical regime.
 Magnetic field can always tune such a transition, and in the case of unconventional superconductors, pressure, chemical doping, and the amount of disorder can all be used to tune $T_c$ to 0. 
  To mention just one ``classic'' example, there is a 
  superconductor-metal transition as a function of doping in the low-carrier density compound La$_{3-x}$[v]$_x$Se$_4$, where [v] denote vacancies on the La site \cite{Holtzberg1968,Seiden1968}.

Already there is evidence of strange anomalies near the doping-tuned QSMT in the cuprate superconductor LSCO\cite{bozovic,wen,taillefer,hussey}, as well as near the field-tuned transition in 
YBCO.\cite{brad,suchitra,ong}  
Here the large temperature scales involved make these materials attractive 
  for studying these phenomena.
 QSMTs also occur in numerous heavy fermion materials, but the low temperatures involved make may make it more difficult to access regimes in which quantum fluctuations 
 are significant.

{\bf 2)} Strictly speaking, at $T=0$ a disordered superconductor in the presence of a magnetic field is a superconducting glass.
    It is possible that in the anomalous metal regime  the system inherits 
  the orbital glassy properties of the underlying superconductor, in which case it should exhibit  hysteresis, and long time relaxations.

{\bf 3}) Finally, experimental and theoretical studies of mesoscopic samples may provide  additional information about physical properties of anomalous metals.  In particular,  the fact that the role of rare events is enhanced near a quantum phase  transition may manifest itself in large amplitude of mesoscopic fluctuations.

\acknowledgements
We would like to acknowledge helpful discussions with  A. Andreev, S. Chakravarty, M. Gershenson, S. Raghu,  Xiaodong Xu, C. Marcus, H. Hwang, D. Cobden, J. Folk, N. Mason, O. Agam, D. J. Scalapino,  E. Berg and especially M. Feigelman S. Hartnoll, and D. Shahar.  This work was supported in part by  the Department of Energy, Office of Basic Energy Sciences, under contract no. DE-AC02- 76SF00515. (AK and SAK).   B. Spivak and S.A. Kivelson acknowledge the hospitality of the KITP and UCSB where some of this work was carried out.
\appendix

\section{Phenomenological models with dissipative heat baths}

There is a considerable body of theoretical analysis that has been carried out on models in which the quantum fluctuations of  local superconducting phases (defined, for instance, on the superconducting nodes of a Josephson junction array) are coupled locally to a phenomenological ``heat bath.''  These models have somewhat similar structure to the problem considered in Sec. \ref{theory}, but are simpler to the extent that the heat-bath and superconducting degrees of freedom are treated separately.  On the plus side, this means that aspects of the solution of these models, some of which we summarize,  can be obtained with a greater level of certainty.  On the negative side, there is an unphysical aspect of all these models - which we will hightlight as well - in that the presence of distinct heat baths on each site of the system corresponds to the assumed existence of an infinite number of degrees of freedom per unit volume.

 \subsection{QSMT in the Quantum RSJ model}
 \label{RSJ}
  
 The quantum fluctuations of a superconducting order parameter in the presence of a dissipative heat bath is a problem with a long history - it was the subject of intense study as the focus of early work on ``macroscopic quantum tunnelling,'' a precursor of the studies that led to the study of superconducting Q-bits.  (For a review, see Ref. \cite{leggettrmp}.)  In the context of the QSMT, this problem was studied in the context of an array of resistively shunted Josephson junctions  \cite{ingold,chakravartyprb,chakravartykapitulniandme,zimanyi,Matthew,chaktoner,troyer,sudbo}.  The same model has been revisited recently in the context of quantum criticality in a dissipative  XY model.\cite{varma3,varma2,varma1}  This model is simple and explicit, and has a number of features that capture  aspects of the phenomena characteristic of the QSMT, as discussed in Ref. \cite{chakravartykapitulniandme}.  We will summarize some aspects of the solution here. However, it has some physical shortcomings that we will also discuss.
 
 The classical RSJ model gives an extremely useful phenomenological description of a resistively shunted Josephson junction.  Here, there are two contributions to the current across the junction, a supercurrent $I_{sc} = J \sin[\theta]$, and a normal current, $I_{norm} = V/R$, where $\theta$ is the difference in phase across the junction, $V$ is the voltage across the junction, and $J$ and $R$ are, respectively, the Josephson coupling and the resistance across the junction.  To obtain dynamical equations for the superconducting phase, one invokes the Josephson relation, $V=2e\dot\theta$ between the voltage and the phase.  It is also sometimes important to account for the capacitance of the junction, according to $\dot Q = C\dot V$ where $\dot Q = I_{sc} + I_{norm}$ is the time derivative of the charge on the two sides of the junction (treated as the two sides of a capacitor).  Combining these consideration leads to the classical equation of motion for the phase across the junction,
 \be
 C\ddot \theta +2e J \sin(\theta) + (1/R)\dot\theta = 0.
 \label{classical}
 \ee
 
Given the success of this description of the collective properties of Josephson junctions at non-zero $T$, it was natural to ask about its properties thought of as an example of dissipative quantum mechanics.  In order to quantize this problem, a representation of the ``heat-bath'' must be introduced.  The key assertion\cite{caldera} is that the details of the heat-bath do not matter -- what matters is that it consists of a large number of degrees of freedom, each weakly coupled to the ``macroscopic quantum variable'' $\theta$, so that the heat-bath can be treated in linear response approximation.  Then, since it is going to be treated in linear response in any case, the heat bath can always be represented as a collection of harmonic oscillators, with a spectral distribution designed to yield the frictional  term in Eq. \ref{classical}.  One particular representation that is useful is to couple $\dot\theta$ to a 1+1 D acoustic boson - a representation that was introduced originally to model the effect of coupling the Josephson junction to an open transmission line.\cite{schmid,zwerger} 
The advantage of this representation is that it maps the problem of the dissipative Josephson junction onto a boundary conformal field theory, for which many exact results exist.

Although the effective action obtained by integrating out the gapless heat-bath degrees of freedom is non-local in time, it is still of a simple enough form that it can be analyzed.   
The result is an effective imaginary time action for a quantum resistively shunted Josephson junction:
\ba
S^{RSJ}[\theta] =&& \int d\tau \left\{ \frac { |\dot \theta|^2}{2E_C} - J \cos[\theta] \right\} 
\label{RSJmodel}
\\
&&+\frac{ \alpha}{4\pi} \int d\tau d\tau^\prime  \left|\frac{\theta(\tau) - \theta(\tau^\prime)}{\tau-\tau^\prime}\right|^2
\nonumber
\ea
where the ``charging energy'' $E_C = 1/(4e^2C)$, while $\alpha \propto 1/R$ is the single coupling that reflects the strength of the coupling to the heat bath.

Eq. \ref{RSJmodel} describes a 0+1 dimensional system.  In general, neither finite size quantum systems, nor infinite 1D classical (finite $T$) systems  can  exhibit phase transitions.  However, 1D classical systems with $1/r^2$ interactions are a notable counter-example to this general expectation, which carries over to the 0+1 dimensional quantum system for the   special case  in which interactions fall as  $1/\tau^2$ in imaginary time.  (A heat bath that produces such an interaction (which in Fourier transform is linear in $|\omega|$) is also known as an ``ohmic heat bath,''\cite{caldera}. 
)  In the case of the RSJ model at $T=0$, this system exhibits a phase transition\cite{Sudip,chakravartyKondo,A.Schmid}  as a function of $\alpha$ from an ordered (``superconducting'') phase for $\alpha>\alpha_c$, in which $\langle [\theta(\tau) - \theta(0)]^2\rangle$ is bounded as $\tau \to \infty$, to a phase disordered (``metallic'') phase in which $\langle [\theta(\tau) - \theta(0)]^2\rangle\to \infty$ as $\tau \to \infty$.  ($\alpha_c$ is probably not  universal, but $\alpha_c(J/E_c) \to1$ as $J/E_c \to 0$.)  As the names suggest, in the superconducting phase the junction can support a dissipationless supercurrent across the junction in the $T\to 0$ limit.  In contrast, in the metallic phase, the junction resistance is finite; there are additive contributions to the conductivity (parallel resistors) from the shunt resistor and the Cooper pair tunnelling.\cite{halperin}   

Thus, the single quantum RSJ junction undergoes a non-trivial QSMT, although obviously in this case the superconducting phase only exists at $T= 0$.  The same considerations have been extended to higher dimensional arrays of resistively shunted Josephson junctions.\cite{ingold,chakravartyprb,Matthew,zimanyi,varma3,varma2,varma1} 
\footnote{There are various ways to imagine generalizing the RSJ model to an array.  For instance, one can imagine the heat-bath is coupled to the phase difference, $\theta_i-\theta_j$, across each junction, directly generalizing the model of the single junction, or one could imagine a circumstance in which each node of the array, i.e. each superconducting grain, is capacitively coupled to a heat-bath,\cite{emeryandme} leading to a dissipative term that depends separately on the phase, $\theta_j$, of each grain separately.}
 Perhaps the most interesting thing about this model is that in the limit of  small $J/E_c$, there is a QSMT that occurs as a function of $\alpha$;  for $\alpha>\alpha_c$, the ground-state exhibits   long-range phase coherence even in the limit $J/E_C \to 0$, while for $\alpha < \alpha_c$, quantum fluctuations destroy long-range phase coherence.  Any phase with long-range phase coherence has a non-vanishing helicity modulous (and hence is superconducting);  manifestly, because current can always be carried by normal electrons through the shunt resistors, the phase without long-range phase coherence is metallic, despite the fact that it is referred to as
``insulating'' in much of the  theoretical literature on the subject.  In fact, for small enough $J/E_C$, the parallel contribution to the conductivity can be computed perturbatively in powers of $J/E_C$, from which it can be seen that the $T\to 0$ conductivity diverges continuously as $\alpha \to \alpha_c^-$.  

Clearly this model has many attractive features in the present context:  It has a QSMT.  The metallic phase proximate to the transition is anomalous in very similar ways to that observed in experiment.  Moreover, the model can be (and has been)  solved using well controlled perturbative RG methods and quantum Monte Carlo methods in various limits.  However, there are  some very peculiar features of the model:  It has an effective dynamical exponent $z\to \infty$. Indeed in the metallic phase it has power-law correlations in time while exhibiting exponential fall of correlations in space - something known\cite{nayak} as ``sliding in time.''  Along with this, it has non-universal critical exponents - corresponding to a line of fixed points rather than a usual fixed point.

Thus, while we view the solution of this model as extremely illuminating, and extremely useful as a charicature of the QSMT, we now focus on some of the unphysical features that prevent it from being considered an entirely satisfactory model.  Firstly, the heat-bath that has been integrated out has, by construction, an infinite number of degrees of freedom.  When thought of as a description of a macroscopic object, this is a reasonable abstraction, but when we come to think of this as representing a local mesoscopic degree of freedom in an extended array of such junctions, this assumption must ultimately break down. 
 It is this feature of the model that is responsible for the 0+1 dimensional character of the critical phenomena.  This is a fundamental criticism of all models which assume coupling to a local heat bath.
 Secondly, in all microscopic derivations of which we know, superconducting fluctuations in a metallic environment couple to the dissipative degrees of freedom through the sine of the phase, as in Eqs. \ref{eq:phiAction} and \ref{AESmodel}, rather than through the phase itself, as in Eq. \ref{RSJmodel}.  
 This observation applies both in the Josephson-junction context\cite{Sudip,ambegoakar} and in the context of superconducting grains embedded in a metal\cite{LarkinFeigelman,OretoKivSp}.  Clearly, as far as the dynamics of small amplitude phase fluctuations are concerned, the two forms of the effective action are identical, but for large amplitude fluctuations the effects are very different, as shown in Appendix \ref{griffith}.

 \section{Josephson junction arrays and quantum Griffith phases }
 \label{griffith}
 A microscopic derivation of the heat-bath associated with the normal electrons in a macroscopic Josephson junction was first  obtained Ambegoakar, Eckern, and Schoern\cite{ambegoakar}. 
 This leads to an effective action of the same form as in  Eq. \ref{eq:phiAction}, where as mentioned above, the ohmic heat bath is coupled to $e^{i\theta}$.  Again at the phenomenological level, it seems reasonable to extend this model to Josephson junction arrays (JJA), leading to an effective action 
 \ba
S^{
JJA} =&&\sum_j \int d\tau \left\{ \frac { |\dot \theta_j|^2}{2E_j} - \sum_{i\neq j} \frac {J_{ij}}2\cos(\theta_i-\theta_j \right\} 
\label{AESmodel}
\\
&&+\sum_j\frac{ \alpha_j}{4\pi} \int d\tau d\tau^\prime  \left|\frac{e^{i\theta_j(\tau)} - e^{i\theta_j(\tau^\prime)}}{\tau-\tau^\prime}\right|^2
\nonumber
\ea
where $\theta_j$ is the superconducting phase on node $j$ and the various couplings are typically taken to be random variables reflecting the degree of disorder.  In most places where such a model is considered, the Josephson coupling $J_{ij}$ is assumed to be short-ranged, although as we have discussed, for superconducting grains embedded in a metal,  this assumption is unphysical - at least in the absence of a magnetic field.  In defining the explicit model, we have taken  a heat-bath coupled separately to the phase on each grain, but similar considerations apply to the model in which we associate a heat-bath with each junction, $(i,j)$. 

Firstly, to illustrate the difference between the 
JJA and RSJ models, we consider the solution of this model problem for a single grain (i.e. $J_{ij}=0$).  Again, the action is that of a 0+1d system, but now one that is recognizable;  if we think of imaginary time as being a spatial dimension, this problem is a version of the classical 1D $1/r^2$ XY ferromagnet at an effective temperature $T \sim 1/\alpha$.  1D is the lower critical dimension for this problem (or if we consider models with interactions of the form $1/r^a$ in 1D, then $a=2$ is the critical range);  while there is no phase transition in this model at any finite $T$, the correlation length diverges exponentially at low $T$.  In other words, for the quantum model, the superconducting susceptibility and the corresponding correlation time diverge as
\be
\tau_0 \sim \chi_{sc} \sim \exp[ Z\pi \alpha] \ \ {\rm as} \ \ \alpha \to \infty
\ee
where $Z \approx 1$. On the one hand, this stands in contrast with the RSJ model in which there is a true superconducting phase ($\chi_{sc}=\infty$) for $\alpha>\alpha_c$.  On the otherhand, for $\alpha$ large, the exponentially large value of $\chi_{sc}$ reflects a similar suppression of quantum fluctuations produced by the coupling to the heat-bath. 

The existence of a susceptibility that depends exponentially on a local parameter is an essential feature needed to obtain a quantum Griffith phase.  Indeed, it has been shown in Refs. \cite{DelMaestro1,Vojta,DelMaestro2} that such a model does indeed support a quantum Griffith phase.  Loosely, the line of argumentation goes as follows:  Consider the case in which $J_{ij}$ is short-raged and the grains are far separated, so that globally the system is not superconducting.  However, there will be rare regions in which there is a group of $N$ grains that are strongly coupled by unusually large Josephson couplings, $J_{ij} \gg E_i$ and $E_j$.  In such a cluster of grains, the phases are effectively locked leaving only one low-energy phase variable, an average phase for the cluster, $\theta$.  This phase, is in turn, coupled to a dissipative heat bath with an effective value of $\alpha =\sum_j \alpha_j \sim N\bar \alpha$, where the sum runs over grains in the cluster.  From the above, this implies that the superconducting susceptibility of the cluster grows exponentially with the size of the cluster, $\chi_{sc} \sim \exp[ Z \pi \bar \alpha N]$.  Such a cluster is rare for large $N$ - under general circumstances the probability of finding such a cluster is  exponentially small in proportion to $N$:  $P(N) \sim \exp[-\gamma N]$.  So long as $\gamma > Z\pi \bar \alpha$, such rare grains are entirely negligible.  However, when $\gamma < Z\pi \bar \alpha$, even though large clusters are rare, they make a divergent contribution to the average superconducting susceptibility.  Since presumably $\gamma \to 0$ as one approaches the point of a superconducting transition, the susceptibility necessarily diverges before this transition is reached - the defining feature of a quantum Griffith phase.

Appealing as this result is, it ultimately depends on the same unphysical feature of the heat-bath already noted.  (See discussion in Refs. \cite{OretoKivSp} and \cite{nogriffith} for further details.)  For a large superconducting cluster, there is a well developed superconducting gap and correspondingly a well-defined superconducting coherence length, $\xi_0$.  Gapless degrees of freedom associated with the surrounding metallic state can only penetrate at most a distance $\xi_0$ into the cluster.  Therefore, in the end, the coupling to the heat-bath can at most grow in proportion to the perimeter (surface area in 3D) of the cluster. In addition, as already mentioned, this model treats $J_{ij}$ as short-ranged, whereas in fact in a metal it falls with a slow power law with distance.
 Thus, while it is an extremely attractive possibility that quantum-Griffith-like phenomena may occur in real materials over an interesting intermediate range of scales, we consider the sharp existence of such a phase to be an artifact of the model.
 
\bibliographystyle{apsrmp4-1}
 \bibliography{metal}

\begin{thebibliography}{137}%
\makeatletter
\providecommand \@ifxundefined [1]{%
 \@ifx{#1\undefined}
}%
\providecommand \@ifnum [1]{%
 \ifnum #1\expandafter \@firstoftwo
 \else \expandafter \@secondoftwo
 \fi
}%
\providecommand \@ifx [1]{%
 \ifx #1\expandafter \@firstoftwo
 \else \expandafter \@secondoftwo
 \fi
}%
\providecommand \natexlab [1]{#1}%
\providecommand \enquote  [1]{``#1''}%
\providecommand \bibnamefont  [1]{#1}%
\providecommand \bibfnamefont [1]{#1}%
\providecommand \citenamefont [1]{#1}%
\providecommand \href@noop [0]{\@secondoftwo}%
\providecommand \href [0]{\begingroup \@sanitize@url \@href}%
\providecommand \@href[1]{\@@startlink{#1}\@@href}%
\providecommand \@@href[1]{\endgroup#1\@@endlink}%
\providecommand \@sanitize@url [0]{\catcode `\\12\catcode `\$12\catcode
  `\&12\catcode `\#12\catcode `\^12\catcode `\_12\catcode `\%12\relax}%
\providecommand \@@startlink[1]{}%
\providecommand \@@endlink[0]{}%
\providecommand \url  [0]{\begingroup\@sanitize@url \@url }%
\providecommand \@url [1]{\endgroup\@href {#1}{\urlprefix }}%
\providecommand \urlprefix  [0]{URL }%
\providecommand \Eprint [0]{\href }%
\providecommand \doibase [0]{http://dx.doi.org/}%
\providecommand \selectlanguage [0]{\@gobble}%
\providecommand \bibinfo  [0]{\@secondoftwo}%
\providecommand \bibfield  [0]{\@secondoftwo}%
\providecommand \translation [1]{[#1]}%
\providecommand \BibitemOpen [0]{}%
\providecommand \bibitemStop [0]{}%
\providecommand \bibitemNoStop [0]{.\EOS\space}%
\providecommand \EOS [0]{\spacefactor3000\relax}%
\providecommand \BibitemShut  [1]{\csname bibitem#1\endcsname}%
\let\auto@bib@innerbib\@empty
\bibitem [{\citenamefont {Abeles}\ \emph {et~al.}(1975)\citenamefont {Abeles},
  \citenamefont {Sheng}, \citenamefont {Coutts},\ and\ \citenamefont
  {Arie}}]{Abeles1975}%
  \BibitemOpen
  \bibfield  {author} {\bibinfo {author} {\bibnamefont {Abeles}, \bibfnamefont
  {B.}}, \bibinfo {author} {\bibfnamefont {P.}~\bibnamefont {Sheng}}, \bibinfo
  {author} {\bibfnamefont {M.}~\bibnamefont {Coutts}}, \ and\ \bibinfo {author}
  {\bibfnamefont {Y.}~\bibnamefont {Arie}}} (\bibinfo {year} {1975}),\ \href
  {\doibase 10.1080/00018737500101431} {\bibfield  {journal} {\bibinfo
  {journal} {Advances in Physics}\ }\textbf {\bibinfo {volume} {24}}~(\bibinfo
  {number} {3}),\ \bibinfo {pages} {407}}\BibitemShut {NoStop}%
\bibitem [{\citenamefont {Abrahams}\ \emph {et~al.}(1979)\citenamefont
  {Abrahams}, \citenamefont {Anderson}, \citenamefont {Licciardello},\ and\
  \citenamefont {Ramakrishnan}}]{GangofFour}%
  \BibitemOpen
  \bibfield  {author} {\bibinfo {author} {\bibnamefont {Abrahams},
  \bibfnamefont {E.}}, \bibinfo {author} {\bibfnamefont {P.~W.}\ \bibnamefont
  {Anderson}}, \bibinfo {author} {\bibfnamefont {D.~C.}\ \bibnamefont
  {Licciardello}}, \ and\ \bibinfo {author} {\bibfnamefont {T.~V.}\
  \bibnamefont {Ramakrishnan}}} (\bibinfo {year} {1979}),\ \href {\doibase
  10.1103/PhysRevLett.42.673} {\bibfield  {journal} {\bibinfo  {journal} {Phys.
  Rev. Lett.}\ }\textbf {\bibinfo {volume} {42}},\ \bibinfo {pages}
  {673}}\BibitemShut {NoStop}%
\bibitem [{\citenamefont {Abrikosov}(1988)}]{Abrikosov}%
  \BibitemOpen
  \bibfield  {author} {\bibinfo {author} {\bibnamefont {Abrikosov},
  \bibfnamefont {A.~A.}}} (\bibinfo {year} {1988}),\ \href@noop {} {\emph
  {\bibinfo {title} {Fundamentals of the theory of metals}}}\BibitemShut
  {NoStop}%
\bibitem [{\citenamefont {Abrikosov}\ \emph {et~al.}(1975)\citenamefont
  {Abrikosov}, \citenamefont {Gor'kov},\ and\ \citenamefont
  {Dzyaloshinski}}]{AGD}%
  \BibitemOpen
  \bibfield  {author} {\bibinfo {author} {\bibnamefont {Abrikosov},
  \bibfnamefont {A.~A.}}, \bibinfo {author} {\bibfnamefont {L.}~\bibnamefont
  {Gor'kov}}, \ and\ \bibinfo {author} {\bibfnamefont {I.~E.}\ \bibnamefont
  {Dzyaloshinski}}} (\bibinfo {year} {1975}),\ \href@noop {} {\emph {\bibinfo
  {title} {Methods of Quantum Field Theory in Statistical Physics}}},\ \bibinfo
  {edition} {revised edition}\ ed.,\ Dover Books on Physics\BibitemShut
  {NoStop}%
\bibitem [{\citenamefont {Aizenman}\ and\ \citenamefont
  {Wehr}(1989)}]{Aizenman}%
  \BibitemOpen
  \bibfield  {author} {\bibinfo {author} {\bibnamefont {Aizenman},
  \bibfnamefont {M.}}, \ and\ \bibinfo {author} {\bibfnamefont
  {J.}~\bibnamefont {Wehr}}} (\bibinfo {year} {1989}),\ \href {\doibase
  10.1103/PhysRevLett.62.2503} {\bibfield  {journal} {\bibinfo  {journal}
  {Phys. Rev. Lett.}\ }\textbf {\bibinfo {volume} {62}},\ \bibinfo {pages}
  {2503}}\BibitemShut {NoStop}%
\bibitem [{\citenamefont {Altshuler}\ \emph {et~al.}(1980)\citenamefont
  {Altshuler}, \citenamefont {Aronov},\ and\ \citenamefont
  {Lee}}]{AronovAltshuler}%
  \BibitemOpen
  \bibfield  {author} {\bibinfo {author} {\bibnamefont {Altshuler},
  \bibfnamefont {B.}}, \bibinfo {author} {\bibfnamefont {A.}~\bibnamefont
  {Aronov}}, \ and\ \bibinfo {author} {\bibfnamefont {P.}~\bibnamefont {Lee}}}
  (\bibinfo {year} {1980}),\ \href {\doibase 10.1103/PhysRevLett.44.1288}
  {\bibfield  {journal} {\bibinfo  {journal} {Physical Review Letters}\
  }\textbf {\bibinfo {volume} {44}}~(\bibinfo {number} {19}),\ \bibinfo {pages}
  {1288}}\BibitemShut {NoStop}%
\bibitem [{\citenamefont {Anderson}(1959)}]{Anderson1959}%
  \BibitemOpen
  \bibfield  {author} {\bibinfo {author} {\bibnamefont {Anderson},
  \bibfnamefont {P.}}} (\bibinfo {year} {1959}),\ \href {\doibase
  https://doi.org/10.1016/0022-3697(59)90036-8} {\bibfield  {journal} {\bibinfo
   {journal} {Journal of Physics and Chemistry of Solids}\ }\textbf {\bibinfo
  {volume} {11}}~(\bibinfo {number} {1}),\ \bibinfo {pages} {26 }}\BibitemShut
  {NoStop}%
\bibitem [{\citenamefont {Aslamazov}\ and\ \citenamefont
  {Larkin}(1968)}]{AslamazovLarkin}%
  \BibitemOpen
  \bibfield  {author} {\bibinfo {author} {\bibnamefont {Aslamazov},
  \bibfnamefont {L.}}, \ and\ \bibinfo {author} {\bibfnamefont
  {A.}~\bibnamefont {Larkin}}} (\bibinfo {year} {{1968}}),\ \href@noop {}
  {\bibfield  {journal} {\bibinfo  {journal} {Soviet Physics Solid State,
  USSR}\ }\textbf {\bibinfo {volume} {{10}}}~(\bibinfo {number} {{4}}),\
  \bibinfo {pages} {{875}}}\BibitemShut {NoStop}%
\bibitem [{\citenamefont {Barkeshli}\ \emph {et~al.}(2013)\citenamefont
  {Barkeshli}, \citenamefont {Yao},\ and\ \citenamefont
  {Kivelson}}]{hongmaissamandme}%
  \BibitemOpen
  \bibfield  {author} {\bibinfo {author} {\bibnamefont {Barkeshli},
  \bibfnamefont {M.}}, \bibinfo {author} {\bibfnamefont {H.}~\bibnamefont
  {Yao}}, \ and\ \bibinfo {author} {\bibfnamefont {S.~A.}\ \bibnamefont
  {Kivelson}}} (\bibinfo {year} {2013}),\ \href {\doibase
  10.1103/PhysRevB.87.140402} {\bibfield  {journal} {\bibinfo  {journal} {Phys.
  Rev. B}\ }\textbf {\bibinfo {volume} {87}},\ \bibinfo {pages}
  {140402}}\BibitemShut {NoStop}%
\bibitem [{\citenamefont {Bilbro}\ \emph {et~al.}(2011)\citenamefont {Bilbro},
  \citenamefont {ValdŽs}, \citenamefont {Aguilar}, \citenamefont {Logvenov},
  \citenamefont {Pelleg}, \citenamefont {Bozovic},\ and\ \citenamefont
  {Armitage}}]{Bilbro11}%
  \BibitemOpen
  \bibfield  {author} {\bibinfo {author} {\bibnamefont {Bilbro}, \bibfnamefont
  {L.~S.}}, \bibinfo {author} {\bibfnamefont {R.}~\bibnamefont {ValdŽs}},
  \bibinfo {author} {\bibfnamefont {R.}~\bibnamefont {Aguilar}}, \bibinfo
  {author} {\bibfnamefont {G.}~\bibnamefont {Logvenov}}, \bibinfo {author}
  {\bibfnamefont {O.}~\bibnamefont {Pelleg}}, \bibinfo {author} {\bibfnamefont
  {I.}~\bibnamefont {Bozovic}}, \ and\ \bibinfo {author} {\bibfnamefont
  {N.~P.}\ \bibnamefont {Armitage}}} (\bibinfo {year} {2011}),\ \href {\doibase
  10.1103/PhysRevB.96.075148} {\bibfield  {journal} {\bibinfo  {journal}
  {Nature Physics}\ }\textbf {\bibinfo {volume} {7}},\ \bibinfo {pages}
  {298}}\BibitemShut {NoStop}%
\bibitem [{\citenamefont {Blatter}\ \emph {et~al.}(1994)\citenamefont
  {Blatter}, \citenamefont {Feigel'man}, \citenamefont {Geshkenbein},
  \citenamefont {Larkin},\ and\ \citenamefont {Vinokur}}]{Blatter1994}%
  \BibitemOpen
  \bibfield  {author} {\bibinfo {author} {\bibnamefont {Blatter}, \bibfnamefont
  {G.}}, \bibinfo {author} {\bibfnamefont {M.~V.}\ \bibnamefont {Feigel'man}},
  \bibinfo {author} {\bibfnamefont {V.~B.}\ \bibnamefont {Geshkenbein}},
  \bibinfo {author} {\bibfnamefont {A.~I.}\ \bibnamefont {Larkin}}, \ and\
  \bibinfo {author} {\bibfnamefont {V.~M.}\ \bibnamefont {Vinokur}}} (\bibinfo
  {year} {1994}),\ \href {\doibase 10.1103/RevModPhys.66.1125} {\bibfield
  {journal} {\bibinfo  {journal} {Rev. Mod. Phys.}\ }\textbf {\bibinfo {volume}
  {66}},\ \bibinfo {pages} {1125}}\BibitemShut {NoStop}%
\bibitem [{\citenamefont {B{\o}ttcher}\ \emph {et~al.}(2017)\citenamefont
  {B{\o}ttcher}, \citenamefont {F.~Nichele}, \citenamefont {Suominen},
  \citenamefont {Shabani}, \citenamefont {Palmstrom},\ and\ \citenamefont
  {Marcus}}]{Bottcher2017}%
  \BibitemOpen
  \bibfield  {author} {\bibinfo {author} {\bibnamefont {B{\o}ttcher},
  \bibfnamefont {C.~G.~L.}}, \bibinfo {author} {\bibfnamefont {a.~M.~K.}\
  \bibnamefont {F.~Nichele}}, \bibinfo {author} {\bibfnamefont {H.~J.}\
  \bibnamefont {Suominen}}, \bibinfo {author} {\bibfnamefont {J.}~\bibnamefont
  {Shabani}}, \bibinfo {author} {\bibfnamefont {C.~J.}\ \bibnamefont
  {Palmstrom}}, \ and\ \bibinfo {author} {\bibfnamefont {C.~M.}\ \bibnamefont
  {Marcus}}} (\bibinfo {year} {2017}),\ \href@noop {} {\ }\BibitemShut
  {NoStop}%
\bibitem [{\citenamefont {Bozovic}\ \emph {et~al.}(2016)\citenamefont
  {Bozovic}, \citenamefont {He}, \citenamefont {Wu},\ and\ \citenamefont
  {Bollinger}}]{bozovic}%
  \BibitemOpen
  \bibfield  {author} {\bibinfo {author} {\bibnamefont {Bozovic}, \bibfnamefont
  {I.}}, \bibinfo {author} {\bibfnamefont {X.}~\bibnamefont {He}}, \bibinfo
  {author} {\bibfnamefont {J.}~\bibnamefont {Wu}}, \ and\ \bibinfo {author}
  {\bibfnamefont {A.~T.}\ \bibnamefont {Bollinger}}} (\bibinfo {year}
  {{2016}}),\ \href {\doibase {10.1038/nature19061}} {\bibfield  {journal}
  {\bibinfo  {journal} {{Nature}}\ }\textbf {\bibinfo {volume}
  {{536}}}~(\bibinfo {number} {{7616}}),\ \bibinfo {pages}
  {{309+}}}\BibitemShut {NoStop}%
\bibitem [{\citenamefont {Brazovskii}\ \emph {et~al.}(1993)\citenamefont
  {Brazovskii}, \citenamefont {Matveenko},\ and\ \citenamefont
  {Nozi\'eres}}]{BrazovskiiNozieres}%
  \BibitemOpen
  \bibfield  {author} {\bibinfo {author} {\bibnamefont {Brazovskii},
  \bibfnamefont {S.}}, \bibinfo {author} {\bibfnamefont {F.}~\bibnamefont
  {Matveenko}}, \ and\ \bibinfo {author} {\bibfnamefont {P.}~\bibnamefont
  {Nozi\'eres}}} (\bibinfo {year} {1993}),\ \href@noop {} {\bibfield  {journal}
  {\bibinfo  {journal} {JETP Lett.}\ }\textbf {\bibinfo {volume}
  {58}}~(\bibinfo {number} {10}),\ \bibinfo {pages} {796}}\BibitemShut
  {NoStop}%
\bibitem [{\citenamefont {Breznay}\ and\ \citenamefont
  {Kapitulnik}(2017)}]{Breznay2017}%
  \BibitemOpen
  \bibfield  {author} {\bibinfo {author} {\bibnamefont {Breznay}, \bibfnamefont
  {N.~P.}}, \ and\ \bibinfo {author} {\bibfnamefont {A.}~\bibnamefont
  {Kapitulnik}}} (\bibinfo {year} {2017}),\ \href {\doibase
  10.1126/sciadv.1700612} {\bibfield  {journal} {\bibinfo  {journal} {Science
  Advances}\ }\textbf {\bibinfo {volume} {3}}~(\bibinfo {number} {9}),\
  \bibinfo {pages} {e1700612}}\BibitemShut {NoStop}%
\bibitem [{\citenamefont {Breznay}\ \emph {et~al.}(2016)\citenamefont
  {Breznay}, \citenamefont {Steiner}, \citenamefont {Kivelson},\ and\
  \citenamefont {Kapitulnik}}]{KapKivDualityPaper}%
  \BibitemOpen
  \bibfield  {author} {\bibinfo {author} {\bibnamefont {Breznay}, \bibfnamefont
  {N.~P.}}, \bibinfo {author} {\bibfnamefont {M.~A.}\ \bibnamefont {Steiner}},
  \bibinfo {author} {\bibfnamefont {S.~A.}\ \bibnamefont {Kivelson}}, \ and\
  \bibinfo {author} {\bibfnamefont {A.}~\bibnamefont {Kapitulnik}}} (\bibinfo
  {year} {2016}),\ \href {\doibase 10.1073/pnas.1522435113} {\bibfield
  {journal} {\bibinfo  {journal} {Proceedings of the National Academy of
  Sciences of the United States of America}\ }\textbf {\bibinfo {volume}
  {113}}~(\bibinfo {number} {2}),\ \bibinfo {pages} {280 }}\BibitemShut
  {NoStop}%
\bibitem [{\citenamefont {Bulaevskii}\ \emph {et~al.}(1977)\citenamefont
  {Bulaevskii}, \citenamefont {Kuzii},\ and\ \citenamefont
  {Sobyanin}}]{Bulaevskii}%
  \BibitemOpen
  \bibfield  {author} {\bibinfo {author} {\bibnamefont {Bulaevskii},
  \bibfnamefont {L.}}, \bibinfo {author} {\bibfnamefont {V.}~\bibnamefont
  {Kuzii}}, \ and\ \bibinfo {author} {\bibfnamefont {A.}~\bibnamefont
  {Sobyanin}}} (\bibinfo {year} {1977}),\ \href@noop {} {\bibfield  {journal}
  {\bibinfo  {journal} {JETP Lett.}\ }\textbf {\bibinfo {volume}
  {25}}~(\bibinfo {number} {7}),\ \bibinfo {pages} {290}}\BibitemShut {NoStop}%
\bibitem [{\citenamefont {Caldeira}\ and\ \citenamefont
  {Leggett}(1981)}]{caldera}%
  \BibitemOpen
  \bibfield  {author} {\bibinfo {author} {\bibnamefont {Caldeira},
  \bibfnamefont {A.~O.}}, \ and\ \bibinfo {author} {\bibfnamefont {A.~J.}\
  \bibnamefont {Leggett}}} (\bibinfo {year} {1981}),\ \href {\doibase
  10.1103/PhysRevLett.46.211} {\bibfield  {journal} {\bibinfo  {journal} {Phys.
  Rev. Lett.}\ }\textbf {\bibinfo {volume} {46}},\ \bibinfo {pages}
  {211}}\BibitemShut {NoStop}%
\bibitem [{\citenamefont {Chakravarty}(1982)}]{Sudip}%
  \BibitemOpen
  \bibfield  {author} {\bibinfo {author} {\bibnamefont {Chakravarty},
  \bibfnamefont {S.}}} (\bibinfo {year} {1982}),\ \href {\doibase
  10.1103/PhysRevLett.49.681} {\bibfield  {journal} {\bibinfo  {journal} {Phys.
  Rev. Lett.}\ }\textbf {\bibinfo {volume} {49}},\ \bibinfo {pages}
  {681}}\BibitemShut {NoStop}%
\bibitem [{\citenamefont {Chakravarty}\ \emph {et~al.}(1986)\citenamefont
  {Chakravarty}, \citenamefont {Ingold}, \citenamefont {Kivelson},\ and\
  \citenamefont {Luther}}]{ingold}%
  \BibitemOpen
  \bibfield  {author} {\bibinfo {author} {\bibnamefont {Chakravarty},
  \bibfnamefont {S.}}, \bibinfo {author} {\bibfnamefont {G.-L.}\ \bibnamefont
  {Ingold}}, \bibinfo {author} {\bibfnamefont {S.}~\bibnamefont {Kivelson}}, \
  and\ \bibinfo {author} {\bibfnamefont {A.}~\bibnamefont {Luther}}} (\bibinfo
  {year} {1986}),\ \href {\doibase 10.1103/PhysRevLett.56.2303} {\bibfield
  {journal} {\bibinfo  {journal} {Phys. Rev. Lett.}\ }\textbf {\bibinfo
  {volume} {56}},\ \bibinfo {pages} {2303}}\BibitemShut {NoStop}%
\bibitem [{\citenamefont {Chakravarty}\ \emph {et~al.}(1988)\citenamefont
  {Chakravarty}, \citenamefont {Ingold}, \citenamefont {Kivelson},\ and\
  \citenamefont {Zimanyi}}]{chakravartyprb}%
  \BibitemOpen
  \bibfield  {author} {\bibinfo {author} {\bibnamefont {Chakravarty},
  \bibfnamefont {S.}}, \bibinfo {author} {\bibfnamefont {G.-L.}\ \bibnamefont
  {Ingold}}, \bibinfo {author} {\bibfnamefont {S.}~\bibnamefont {Kivelson}}, \
  and\ \bibinfo {author} {\bibfnamefont {G.}~\bibnamefont {Zimanyi}}} (\bibinfo
  {year} {1988}),\ \href {\doibase 10.1103/PhysRevB.37.3283} {\bibfield
  {journal} {\bibinfo  {journal} {Phys. Rev. B}\ }\textbf {\bibinfo {volume}
  {37}},\ \bibinfo {pages} {3283}}\BibitemShut {NoStop}%
\bibitem [{\citenamefont {Chakravarty}\ and\ \citenamefont
  {Rudnick}(1995)}]{chakravartyKondo}%
  \BibitemOpen
  \bibfield  {author} {\bibinfo {author} {\bibnamefont {Chakravarty},
  \bibfnamefont {S.}}, \ and\ \bibinfo {author} {\bibfnamefont
  {J.}~\bibnamefont {Rudnick}}} (\bibinfo {year} {1995}),\ \href {\doibase
  10.1103/PhysRevLett.75.501} {\bibfield  {journal} {\bibinfo  {journal} {Phys.
  Rev. Lett.}\ }\textbf {\bibinfo {volume} {75}},\ \bibinfo {pages}
  {501}}\BibitemShut {NoStop}%
\bibitem [{\citenamefont {Chakravarty}\ and\ \citenamefont
  {Schmid}(1986)}]{schmid}%
  \BibitemOpen
  \bibfield  {author} {\bibinfo {author} {\bibnamefont {Chakravarty},
  \bibfnamefont {S.}}, \ and\ \bibinfo {author} {\bibfnamefont
  {A.}~\bibnamefont {Schmid}}} (\bibinfo {year} {1986}),\ \href {\doibase
  10.1103/PhysRevB.33.2000} {\bibfield  {journal} {\bibinfo  {journal} {Phys.
  Rev. B}\ }\textbf {\bibinfo {volume} {33}},\ \bibinfo {pages}
  {2000}}\BibitemShut {NoStop}%
\bibitem [{\citenamefont {Chen}\ \emph {et~al.}(2017)\citenamefont {Chen},
  \citenamefont {Swartz}, \citenamefont {Yoon}, \citenamefont {Inoue},
  \citenamefont {Merz}, \citenamefont {Di~Lu}, \citenamefont {Yuan},
  \citenamefont {Hikita}, \citenamefont {Raghu},\ and\ \citenamefont
  {Hwang}}]{Chen2017}%
  \BibitemOpen
  \bibfield  {author} {\bibinfo {author} {\bibnamefont {Chen}, \bibfnamefont
  {Z.}}, \bibinfo {author} {\bibfnamefont {A.~G.}\ \bibnamefont {Swartz}},
  \bibinfo {author} {\bibfnamefont {H.}~\bibnamefont {Yoon}}, \bibinfo {author}
  {\bibfnamefont {H.}~\bibnamefont {Inoue}}, \bibinfo {author} {\bibfnamefont
  {T.}~\bibnamefont {Merz}}, \bibinfo {author} {\bibfnamefont {Y.~a.~X.}\
  \bibnamefont {Di~Lu}}, \bibinfo {author} {\bibfnamefont {H.}~\bibnamefont
  {Yuan}}, \bibinfo {author} {\bibfnamefont {Y.}~\bibnamefont {Hikita}},
  \bibinfo {author} {\bibfnamefont {S.}~\bibnamefont {Raghu}}, \ and\ \bibinfo
  {author} {\bibfnamefont {H.~Y.}\ \bibnamefont {Hwang}}} (\bibinfo {year}
  {2017}),\ \href@noop {} {\ }\BibitemShut {NoStop}%
\bibitem [{\citenamefont {Cheraghchi}(2006)}]{Schreiber}%
  \BibitemOpen
  \bibfield  {author} {\bibinfo {author} {\bibnamefont {Cheraghchi},
  \bibfnamefont {H.}}} (\bibinfo {year} {2006}),\ \href
  {http://stacks.iop.org/1742-5468/2006/i=11/a=P11006} {\bibfield  {journal}
  {\bibinfo  {journal} {journal of Statistical Mechanics: Theory and
  Experiment}\ }\textbf {\bibinfo {volume} {2006}}~(\bibinfo {number} {11}),\
  \bibinfo {pages} {P11006}}\BibitemShut {NoStop}%
\bibitem [{\citenamefont {Cooper}\ \emph {et~al.}(2009)\citenamefont {Cooper},
  \citenamefont {Wang}, \citenamefont {Vignolle}, \citenamefont {Lipscombe},
  \citenamefont {Hayden}, \citenamefont {Tanabe}, \citenamefont {Adachi},
  \citenamefont {Koike}, \citenamefont {Nohara}, \citenamefont {Takagi},
  \citenamefont {Proust},\ and\ \citenamefont {Hussey}}]{hussey}%
  \BibitemOpen
  \bibfield  {author} {\bibinfo {author} {\bibnamefont {Cooper}, \bibfnamefont
  {R.~A.}}, \bibinfo {author} {\bibfnamefont {Y.}~\bibnamefont {Wang}},
  \bibinfo {author} {\bibfnamefont {B.}~\bibnamefont {Vignolle}}, \bibinfo
  {author} {\bibfnamefont {O.~J.}\ \bibnamefont {Lipscombe}}, \bibinfo {author}
  {\bibfnamefont {S.~M.}\ \bibnamefont {Hayden}}, \bibinfo {author}
  {\bibfnamefont {Y.}~\bibnamefont {Tanabe}}, \bibinfo {author} {\bibfnamefont
  {T.}~\bibnamefont {Adachi}}, \bibinfo {author} {\bibfnamefont
  {Y.}~\bibnamefont {Koike}}, \bibinfo {author} {\bibfnamefont
  {M.}~\bibnamefont {Nohara}}, \bibinfo {author} {\bibfnamefont
  {H.}~\bibnamefont {Takagi}}, \bibinfo {author} {\bibfnamefont
  {C.}~\bibnamefont {Proust}}, \ and\ \bibinfo {author} {\bibfnamefont {N.~E.}\
  \bibnamefont {Hussey}}} (\bibinfo {year} {2009}),\ \href {\doibase
  10.1126/science.1165015} {\bibfield  {journal} {\bibinfo  {journal}
  {Science}\ }\textbf {\bibinfo {volume} {323}}~(\bibinfo {number} {5914}),\
  \bibinfo {pages} {603}}\BibitemShut {NoStop}%
\bibitem [{\citenamefont {Corson}\ \emph {et~al.}(1999)\citenamefont {Corson},
  \citenamefont {Mallozzi}, \citenamefont {Orenstein}, \citenamefont
  {Eckstein},\ and\ \citenamefont {Bozovic}}]{Corson1999}%
  \BibitemOpen
  \bibfield  {author} {\bibinfo {author} {\bibnamefont {Corson}, \bibfnamefont
  {J.}}, \bibinfo {author} {\bibfnamefont {R.}~\bibnamefont {Mallozzi}},
  \bibinfo {author} {\bibfnamefont {J.}~\bibnamefont {Orenstein}}, \bibinfo
  {author} {\bibfnamefont {J.~N.}\ \bibnamefont {Eckstein}}, \ and\ \bibinfo
  {author} {\bibfnamefont {I.}~\bibnamefont {Bozovic}}} (\bibinfo {year}
  {1999}),\ \href@noop {} {\bibfield  {journal} {\bibinfo  {journal} {Nature}\
  }\textbf {\bibinfo {volume} {398}},\ \bibinfo {pages} {221}}\BibitemShut
  {NoStop}%
\bibitem [{\citenamefont {Crane}\ \emph {et~al.}(2007)\citenamefont {Crane},
  \citenamefont {Armitage}, \citenamefont {Johansson}, \citenamefont
  {Sambandamurthy}, \citenamefont {Shahar},\ and\ \citenamefont
  {Gruner}}]{Crane2007}%
  \BibitemOpen
  \bibfield  {author} {\bibinfo {author} {\bibnamefont {Crane}, \bibfnamefont
  {R.~W.}}, \bibinfo {author} {\bibfnamefont {N.~P.}\ \bibnamefont {Armitage}},
  \bibinfo {author} {\bibfnamefont {A.}~\bibnamefont {Johansson}}, \bibinfo
  {author} {\bibfnamefont {G.}~\bibnamefont {Sambandamurthy}}, \bibinfo
  {author} {\bibfnamefont {D.}~\bibnamefont {Shahar}}, \ and\ \bibinfo {author}
  {\bibfnamefont {G.}~\bibnamefont {Gruner}}} (\bibinfo {year} {2007}),\ \href
  {\doibase 10.1103/PhysRevB.96.075148} {\bibfield  {journal} {\bibinfo
  {journal} {Phys. Rev. B}\ }\textbf {\bibinfo {volume} {75}},\ \bibinfo
  {pages} {184530}}\BibitemShut {NoStop}%
\bibitem [{\citenamefont {Crauste}\ \emph {et~al.}(2009)\citenamefont
  {Crauste}, \citenamefont {Marrache-Kikuchi}, \citenamefont {Berge},
  \citenamefont {Stanescu},\ and\ \citenamefont {Dumoulin}}]{Crauste2009}%
  \BibitemOpen
  \bibfield  {author} {\bibinfo {author} {\bibnamefont {Crauste}, \bibfnamefont
  {O.}}, \bibinfo {author} {\bibfnamefont {C.~A.}\ \bibnamefont
  {Marrache-Kikuchi}}, \bibinfo {author} {\bibfnamefont {L.}~\bibnamefont
  {Berge}}, \bibinfo {author} {\bibfnamefont {D.}~\bibnamefont {Stanescu}}, \
  and\ \bibinfo {author} {\bibfnamefont {L.}~\bibnamefont {Dumoulin}}}
  (\bibinfo {year} {{2009}}),\ \href {\doibase
  {10.1088/1742-6596/150/4/042019}} {\enquote {\bibinfo {title} {{Thickness
  dependence of the superconductivity in thin disordered NbSi films}},}\
  }\bibinfo {note} {{25th International Conference on Low Temperature Physics
  (LT25), Leiden Inst Phys, Kamerlingh Onnes Lab, Amsterdam, Netherlands, AUG
  06-13, 2008}}\BibitemShut {NoStop}%
\bibitem [{\citenamefont {Das}\ and\ \citenamefont
  {Doniach}(1999)}]{dasanddoniach}%
  \BibitemOpen
  \bibfield  {author} {\bibinfo {author} {\bibnamefont {Das}, \bibfnamefont
  {D.}}, \ and\ \bibinfo {author} {\bibfnamefont {S.}~\bibnamefont {Doniach}}}
  (\bibinfo {year} {1999}),\ \href {\doibase 10.1103/PhysRevB.60.1261}
  {\bibfield  {journal} {\bibinfo  {journal} {Phys. Rev. B}\ }\textbf {\bibinfo
  {volume} {60}},\ \bibinfo {pages} {1261}}\BibitemShut {NoStop}%
\bibitem [{\citenamefont {Davison}\ \emph {et~al.}(2016)\citenamefont
  {Davison}, \citenamefont {Delacr\'etaz}, \citenamefont {Gout\'eraux},\ and\
  \citenamefont {Hartnoll}}]{Hartnoll}%
  \BibitemOpen
  \bibfield  {author} {\bibinfo {author} {\bibnamefont {Davison}, \bibfnamefont
  {R.~A.}}, \bibinfo {author} {\bibfnamefont {L.~V.}\ \bibnamefont
  {Delacr\'etaz}}, \bibinfo {author} {\bibfnamefont {B.}~\bibnamefont
  {Gout\'eraux}}, \ and\ \bibinfo {author} {\bibfnamefont {S.~A.}\ \bibnamefont
  {Hartnoll}}} (\bibinfo {year} {2016}),\ \href {\doibase
  10.1103/PhysRevB.94.054502} {\bibfield  {journal} {\bibinfo  {journal} {Phys.
  Rev. B}\ }\textbf {\bibinfo {volume} {94}},\ \bibinfo {pages}
  {054502}}\BibitemShut {NoStop}%
\bibitem [{\citenamefont {Del~Maestro}\ \emph {et~al.}(2010)\citenamefont
  {Del~Maestro}, \citenamefont {Rosenow}, \citenamefont {Hoyos},\ and\
  \citenamefont {Vojta}}]{DelMaestro2}%
  \BibitemOpen
  \bibfield  {author} {\bibinfo {author} {\bibnamefont {Del~Maestro},
  \bibfnamefont {A.}}, \bibinfo {author} {\bibfnamefont {B.}~\bibnamefont
  {Rosenow}}, \bibinfo {author} {\bibfnamefont {J.~A.}\ \bibnamefont {Hoyos}},
  \ and\ \bibinfo {author} {\bibfnamefont {T.}~\bibnamefont {Vojta}}} (\bibinfo
  {year} {2010}),\ \href {\doibase 10.1103/PhysRevLett.105.145702} {\bibfield
  {journal} {\bibinfo  {journal} {Phys. Rev. Lett.}\ }\textbf {\bibinfo
  {volume} {105}},\ \bibinfo {pages} {145702}}\BibitemShut {NoStop}%
\bibitem [{\citenamefont {Del~Maestro}\ \emph {et~al.}(2008)\citenamefont
  {Del~Maestro}, \citenamefont {Rosenow}, \citenamefont {M\"uller},\ and\
  \citenamefont {Sachdev}}]{DelMaestro1}%
  \BibitemOpen
  \bibfield  {author} {\bibinfo {author} {\bibnamefont {Del~Maestro},
  \bibfnamefont {A.}}, \bibinfo {author} {\bibfnamefont {B.}~\bibnamefont
  {Rosenow}}, \bibinfo {author} {\bibfnamefont {M.}~\bibnamefont {M\"uller}}, \
  and\ \bibinfo {author} {\bibfnamefont {S.}~\bibnamefont {Sachdev}}} (\bibinfo
  {year} {2008}),\ \href {\doibase 10.1103/PhysRevLett.101.035701} {\bibfield
  {journal} {\bibinfo  {journal} {Phys. Rev. Lett.}\ }\textbf {\bibinfo
  {volume} {101}},\ \bibinfo {pages} {035701}}\BibitemShut {NoStop}%
\bibitem [{\citenamefont {Deutcher}\ \emph {et~al.}(1983)\citenamefont
  {Deutcher}, \citenamefont {Zallen},\ and\ \citenamefont {J}}]{Deutscher1983}%
  \BibitemOpen
  \bibinfo {editor} {\bibnamefont {Deutcher}, \bibfnamefont {G.}}, \bibinfo
  {editor} {\bibfnamefont {R.}~\bibnamefont {Zallen}}, \ and\ \bibinfo {editor}
  {\bibfnamefont {A.}~\bibnamefont {J}},\ Eds. (\bibinfo {year} {1983}),\
  \href@noop {} {\emph {\bibinfo {title} {Percolation, Structures, and
  Processes}}},\ \bibinfo {series} {Ann. of the Israel Physical Society},
  Vol.~\bibinfo {volume} {5},\ \bibinfo {organization} {Israel Physical
  Society}\ (\bibinfo  {publisher} {Adam Hilger},\ \bibinfo {address}
  {Bristol})\BibitemShut {NoStop}%
\bibitem [{\citenamefont {Eckern}\ \emph {et~al.}(1984)\citenamefont {Eckern},
  \citenamefont {Sch\"on},\ and\ \citenamefont {Ambegaokar}}]{ambegoakar}%
  \BibitemOpen
  \bibfield  {author} {\bibinfo {author} {\bibnamefont {Eckern}, \bibfnamefont
  {U.}}, \bibinfo {author} {\bibfnamefont {G.}~\bibnamefont {Sch\"on}}, \ and\
  \bibinfo {author} {\bibfnamefont {V.}~\bibnamefont {Ambegaokar}}} (\bibinfo
  {year} {1984}),\ \href {\doibase 10.1103/PhysRevB.30.6419} {\bibfield
  {journal} {\bibinfo  {journal} {Phys. Rev. B}\ }\textbf {\bibinfo {volume}
  {30}},\ \bibinfo {pages} {6419}}\BibitemShut {NoStop}%
\bibitem [{\citenamefont {Eley}\ \emph {et~al.}(2012)\citenamefont {Eley},
  \citenamefont {Gopalakrishnan}, \citenamefont {Goldbart},\ and\ \citenamefont
  {Mason}}]{Eley2013}%
  \BibitemOpen
  \bibfield  {author} {\bibinfo {author} {\bibnamefont {Eley}, \bibfnamefont
  {S.}}, \bibinfo {author} {\bibfnamefont {S.}~\bibnamefont {Gopalakrishnan}},
  \bibinfo {author} {\bibfnamefont {P.~M.}\ \bibnamefont {Goldbart}}, \ and\
  \bibinfo {author} {\bibfnamefont {N.}~\bibnamefont {Mason}}} (\bibinfo {year}
  {2012}),\ \href@noop {} {\bibfield  {journal} {\bibinfo  {journal} {Nature
  Physics}\ }\textbf {\bibinfo {volume} {8}}~(\bibinfo {number} {1}),\ \bibinfo
  {pages} {59}}\BibitemShut {NoStop}%
\bibitem [{\citenamefont {Emery}\ and\ \citenamefont
  {Kivelson}(1995)}]{emeryandme}%
  \BibitemOpen
  \bibfield  {author} {\bibinfo {author} {\bibnamefont {Emery}, \bibfnamefont
  {V.~J.}}, \ and\ \bibinfo {author} {\bibfnamefont {S.~A.}\ \bibnamefont
  {Kivelson}}} (\bibinfo {year} {1995}),\ \href {\doibase
  10.1103/PhysRevLett.74.3253} {\bibfield  {journal} {\bibinfo  {journal}
  {Phys. Rev. Lett.}\ }\textbf {\bibinfo {volume} {74}},\ \bibinfo {pages}
  {3253}}\BibitemShut {NoStop}%
\bibitem [{\citenamefont {Entin-Wohlman}\ \emph {et~al.}(1981)\citenamefont
  {Entin-Wohlman}, \citenamefont {Kapitulnik},\ and\ \citenamefont
  {Shapira}}]{Entin1981}%
  \BibitemOpen
  \bibfield  {author} {\bibinfo {author} {\bibnamefont {Entin-Wohlman},
  \bibfnamefont {O.}}, \bibinfo {author} {\bibfnamefont {A.}~\bibnamefont
  {Kapitulnik}}, \ and\ \bibinfo {author} {\bibfnamefont {Y.}~\bibnamefont
  {Shapira}}} (\bibinfo {year} {1981}),\ \href {\doibase
  10.1103/PhysRevB.24.6464} {\bibfield  {journal} {\bibinfo  {journal} {Phys.
  Rev. B}\ }\textbf {\bibinfo {volume} {24}},\ \bibinfo {pages}
  {6464}}\BibitemShut {NoStop}%
\bibitem [{\citenamefont {Ephron}(1996)}]{EphronThesis}%
  \BibitemOpen
  \bibfield  {author} {\bibinfo {author} {\bibnamefont {Ephron}, \bibfnamefont
  {D.}}} (\bibinfo {year} {1996}),\ \emph {\bibinfo {title} {Correlated
  Electron Tunneling and Quantum Motion of Vortices in Disordered Model
  Systems}},\ \href@noop {} {Ph.D. thesis}\ (\bibinfo  {school} {Stanford
  University}, \bibinfo {address} {Stanford, CA 94305})\BibitemShut {NoStop}%
\bibitem [{\citenamefont {Ephron}\ \emph {et~al.}(1996)\citenamefont {Ephron},
  \citenamefont {Yazdani}, \citenamefont {Kapitulnik},\ and\ \citenamefont
  {Beasley}}]{Ephron}%
  \BibitemOpen
  \bibfield  {author} {\bibinfo {author} {\bibnamefont {Ephron}, \bibfnamefont
  {D.}}, \bibinfo {author} {\bibfnamefont {A.}~\bibnamefont {Yazdani}},
  \bibinfo {author} {\bibfnamefont {A.}~\bibnamefont {Kapitulnik}}, \ and\
  \bibinfo {author} {\bibfnamefont {M.~R.}\ \bibnamefont {Beasley}}} (\bibinfo
  {year} {1996}),\ \href {\doibase 10.1103/PhysRevLett.76.1529} {\bibfield
  {journal} {\bibinfo  {journal} {Phys. Rev. Lett.}\ }\textbf {\bibinfo
  {volume} {76}},\ \bibinfo {pages} {1529}}\BibitemShut {NoStop}%
\bibitem [{\citenamefont {Feigel'man}\ and\ \citenamefont
  {Larkin}(1998)}]{LarkinFeigelman}%
  \BibitemOpen
  \bibfield  {author} {\bibinfo {author} {\bibnamefont {Feigel'man},
  \bibfnamefont {M.}}, \ and\ \bibinfo {author} {\bibfnamefont
  {A.}~\bibnamefont {Larkin}}} (\bibinfo {year} {1998}),\ \href {\doibase
  https://doi.org/10.1016/S0301-0104(98)00075-5} {\bibfield  {journal}
  {\bibinfo  {journal} {Chemical Physics}\ }\textbf {\bibinfo {volume}
  {235}}~(\bibinfo {number} {1}),\ \bibinfo {pages} {107 }}\BibitemShut
  {NoStop}%
\bibitem [{\citenamefont {Feigel'man}\ \emph {et~al.}(2001)\citenamefont
  {Feigel'man}, \citenamefont {Larkin},\ and\ \citenamefont
  {Skvortsov}}]{FiigelmanLarkinSkvortsov}%
  \BibitemOpen
  \bibfield  {author} {\bibinfo {author} {\bibnamefont {Feigel'man},
  \bibfnamefont {M.~V.}}, \bibinfo {author} {\bibfnamefont {A.~I.}\
  \bibnamefont {Larkin}}, \ and\ \bibinfo {author} {\bibfnamefont {M.~A.}\
  \bibnamefont {Skvortsov}}} (\bibinfo {year} {2001}),\ \href {\doibase
  10.1103/PhysRevLett.86.1869} {\bibfield  {journal} {\bibinfo  {journal}
  {Phys. Rev. Lett.}\ }\textbf {\bibinfo {volume} {86}},\ \bibinfo {pages}
  {1869}}\BibitemShut {NoStop}%
\bibitem [{\citenamefont {Finkelshtein}(1987)}]{Finkelstein2}%
  \BibitemOpen
  \bibfield  {author} {\bibinfo {author} {\bibnamefont {Finkelshtein},
  \bibfnamefont {A.}}} (\bibinfo {year} {1987}),\ \href@noop {} {\bibfield
  {journal} {\bibinfo  {journal} {JETP Lett.}\ }\textbf {\bibinfo {volume}
  {45}}~(\bibinfo {number} {1}),\ \bibinfo {pages} {46}}\BibitemShut {NoStop}%
\bibitem [{\citenamefont {Fisher}(1992)}]{DSFisher1}%
  \BibitemOpen
  \bibfield  {author} {\bibinfo {author} {\bibnamefont {Fisher}, \bibfnamefont
  {D.~S.}}} (\bibinfo {year} {1992}),\ \href {\doibase
  10.1103/PhysRevLett.69.534} {\bibfield  {journal} {\bibinfo  {journal} {Phys.
  Rev. Lett.}\ }\textbf {\bibinfo {volume} {69}},\ \bibinfo {pages}
  {534}}\BibitemShut {NoStop}%
\bibitem [{\citenamefont {Fisher}(1995)}]{DSFisher2}%
  \BibitemOpen
  \bibfield  {author} {\bibinfo {author} {\bibnamefont {Fisher}, \bibfnamefont
  {D.~S.}}} (\bibinfo {year} {1995}),\ \href {\doibase
  10.1103/PhysRevB.51.6411} {\bibfield  {journal} {\bibinfo  {journal} {Phys.
  Rev. B}\ }\textbf {\bibinfo {volume} {51}},\ \bibinfo {pages}
  {6411}}\BibitemShut {NoStop}%
\bibitem [{\citenamefont {Fisher}(1986)}]{Matthew}%
  \BibitemOpen
  \bibfield  {author} {\bibinfo {author} {\bibnamefont {Fisher}, \bibfnamefont
  {M.~P.~A.}}} (\bibinfo {year} {1986}),\ \href {\doibase
  10.1103/PhysRevLett.57.885} {\bibfield  {journal} {\bibinfo  {journal} {Phys.
  Rev. Lett.}\ }\textbf {\bibinfo {volume} {57}},\ \bibinfo {pages}
  {885}}\BibitemShut {NoStop}%
\bibitem [{\citenamefont {Fisher}(1990)}]{MPAFisher1990}%
  \BibitemOpen
  \bibfield  {author} {\bibinfo {author} {\bibnamefont {Fisher}, \bibfnamefont
  {M.~P.~A.}}} (\bibinfo {year} {1990}),\ \href {\doibase
  10.1103/PhysRevLett.65.923} {\bibfield  {journal} {\bibinfo  {journal} {Phys.
  Rev. Lett.}\ }\textbf {\bibinfo {volume} {65}},\ \bibinfo {pages}
  {923}}\BibitemShut {NoStop}%
\bibitem [{\citenamefont {Galitski}\ \emph {et~al.}(2005)\citenamefont
  {Galitski}, \citenamefont {Refael}, \citenamefont {Fisher},\ and\
  \citenamefont {Senthil}}]{senthil}%
  \BibitemOpen
  \bibfield  {author} {\bibinfo {author} {\bibnamefont {Galitski},
  \bibfnamefont {V.~M.}}, \bibinfo {author} {\bibfnamefont {G.}~\bibnamefont
  {Refael}}, \bibinfo {author} {\bibfnamefont {M.~P.~A.}\ \bibnamefont
  {Fisher}}, \ and\ \bibinfo {author} {\bibfnamefont {T.}~\bibnamefont
  {Senthil}}} (\bibinfo {year} {2005}),\ \href {\doibase
  10.1103/PhysRevLett.95.077002} {\bibfield  {journal} {\bibinfo  {journal}
  {Phys. Rev. Lett.}\ }\textbf {\bibinfo {volume} {95}},\ \bibinfo {pages}
  {077002}}\BibitemShut {NoStop}%
\bibitem [{\citenamefont {Garcia-Barriocanal}\ \emph
  {et~al.}(2013)\citenamefont {Garcia-Barriocanal}, \citenamefont {Kobrinskii},
  \citenamefont {Leng}, \citenamefont {Kinney}, \citenamefont {Yang},
  \citenamefont {Snyder},\ and\ \citenamefont {Goldman}}]{GoldmanLSCO}%
  \BibitemOpen
  \bibfield  {author} {\bibinfo {author} {\bibnamefont {Garcia-Barriocanal},
  \bibfnamefont {J.}}, \bibinfo {author} {\bibfnamefont {A.}~\bibnamefont
  {Kobrinskii}}, \bibinfo {author} {\bibfnamefont {X.}~\bibnamefont {Leng}},
  \bibinfo {author} {\bibfnamefont {J.}~\bibnamefont {Kinney}}, \bibinfo
  {author} {\bibfnamefont {B.}~\bibnamefont {Yang}}, \bibinfo {author}
  {\bibfnamefont {S.}~\bibnamefont {Snyder}}, \ and\ \bibinfo {author}
  {\bibfnamefont {A.~M.}\ \bibnamefont {Goldman}}} (\bibinfo {year} {2013}),\
  \href {\doibase 10.1103/PhysRevB.87.024509} {\bibfield  {journal} {\bibinfo
  {journal} {Phys. Rev. B}\ }\textbf {\bibinfo {volume} {87}},\ \bibinfo
  {pages} {024509}}\BibitemShut {NoStop}%
\bibitem [{\citenamefont {Ginzburg}(1961)}]{Ginzburg1961}%
  \BibitemOpen
  \bibfield  {author} {\bibinfo {author} {\bibnamefont {Ginzburg},
  \bibfnamefont {V.}}} (\bibinfo {year} {1961}),\ \href@noop {} {\bibfield
  {journal} {\bibinfo  {journal} {Soviet Physics-Solid State}\ }\textbf
  {\bibinfo {volume} {2}}~(\bibinfo {number} {9}),\ \bibinfo {pages} {1824}},\
  \bibinfo {note} {fiz. Tverd. Tela 2 (9), 2031-2043, 1960}\BibitemShut
  {NoStop}%
\bibitem [{\citenamefont {Gorkov}\ \emph {et~al.}(1979)\citenamefont {Gorkov},
  \citenamefont {Larkin},\ and\ \citenamefont
  {Khmelnitskii}}]{LarkinKhmelnitskiiGorkov}%
  \BibitemOpen
  \bibfield  {author} {\bibinfo {author} {\bibnamefont {Gorkov}, \bibfnamefont
  {L.}}, \bibinfo {author} {\bibfnamefont {A.}~\bibnamefont {Larkin}}, \ and\
  \bibinfo {author} {\bibfnamefont {D.}~\bibnamefont {Khmelnitskii}}} (\bibinfo
  {year} {1979}),\ \href@noop {} {\bibfield  {journal} {\bibinfo  {journal}
  {JETP Lett.}\ }\textbf {\bibinfo {volume} {30}}~(\bibinfo {number} {4}),\
  \bibinfo {pages} {228}}\BibitemShut {NoStop}%
\bibitem [{\citenamefont {Goswami}\ \emph {et~al.}(2008)\citenamefont
  {Goswami}, \citenamefont {Schwab},\ and\ \citenamefont
  {Chakravarty}}]{Chakrovarty}%
  \BibitemOpen
  \bibfield  {author} {\bibinfo {author} {\bibnamefont {Goswami}, \bibfnamefont
  {P.}}, \bibinfo {author} {\bibfnamefont {D.}~\bibnamefont {Schwab}}, \ and\
  \bibinfo {author} {\bibfnamefont {S.}~\bibnamefont {Chakravarty}}} (\bibinfo
  {year} {2008}),\ \href {\doibase 10.1103/PhysRevLett.100.015703} {\bibfield
  {journal} {\bibinfo  {journal} {Phys. Rev. Lett.}\ }\textbf {\bibinfo
  {volume} {100}},\ \bibinfo {pages} {015703}}\BibitemShut {NoStop}%
\bibitem [{\citenamefont {Halperin}\ \emph {et~al.}(1993)\citenamefont
  {Halperin}, \citenamefont {Lee},\ and\ \citenamefont {Read}}]{HLR}%
  \BibitemOpen
  \bibfield  {author} {\bibinfo {author} {\bibnamefont {Halperin},
  \bibfnamefont {B.~I.}}, \bibinfo {author} {\bibfnamefont {P.~A.}\
  \bibnamefont {Lee}}, \ and\ \bibinfo {author} {\bibfnamefont
  {N.}~\bibnamefont {Read}}} (\bibinfo {year} {1993}),\ \href {\doibase
  10.1103/PhysRevB.47.7312} {\bibfield  {journal} {\bibinfo  {journal} {Phys.
  Rev. B}\ }\textbf {\bibinfo {volume} {47}},\ \bibinfo {pages}
  {7312}}\BibitemShut {NoStop}%
\bibitem [{\citenamefont {Halperin}\ \emph {et~al.}(2010)\citenamefont
  {Halperin}, \citenamefont {Refael},\ and\ \citenamefont {Demler}}]{halperin}%
  \BibitemOpen
  \bibfield  {author} {\bibinfo {author} {\bibnamefont {Halperin},
  \bibfnamefont {B.~I.}}, \bibinfo {author} {\bibfnamefont {G.}~\bibnamefont
  {Refael}}, \ and\ \bibinfo {author} {\bibfnamefont {E.}~\bibnamefont
  {Demler}}} (\bibinfo {year} {2010}),\ \href {\doibase
  10.1142/S021797921005644X} {\bibfield  {journal} {\bibinfo  {journal} {Int.
  J. Mod. Phys. B}\ }\textbf {\bibinfo {volume} {24}}~(\bibinfo {number}
  {20-21}),\ \bibinfo {pages} {4039}}\BibitemShut {NoStop}%
\bibitem [{\citenamefont {Han}\ \emph {et~al.}(2014)\citenamefont {Han},
  \citenamefont {Allain}, \citenamefont {Arjmandi-Tash}, \citenamefont
  {Tikhonov}, \citenamefont {FeigelÕMan}, \citenamefont {Sac{\'e}p{\'e}},\ and\
  \citenamefont {Bouchiat}}]{BouchiatFeigelman}%
  \BibitemOpen
  \bibfield  {author} {\bibinfo {author} {\bibnamefont {Han}, \bibfnamefont
  {Z.}}, \bibinfo {author} {\bibfnamefont {A.}~\bibnamefont {Allain}}, \bibinfo
  {author} {\bibfnamefont {H.}~\bibnamefont {Arjmandi-Tash}}, \bibinfo {author}
  {\bibfnamefont {K.}~\bibnamefont {Tikhonov}}, \bibinfo {author}
  {\bibfnamefont {M.}~\bibnamefont {FeigelÕMan}}, \bibinfo {author}
  {\bibfnamefont {B.}~\bibnamefont {Sac{\'e}p{\'e}}}, \ and\ \bibinfo {author}
  {\bibfnamefont {V.}~\bibnamefont {Bouchiat}}} (\bibinfo {year} {2014}),\
  \href@noop {} {\bibfield  {journal} {\bibinfo  {journal} {Nature Physics}\
  }\textbf {\bibinfo {volume} {10}}~(\bibinfo {number} {5}),\ \bibinfo {pages}
  {380}}\BibitemShut {NoStop}%
\bibitem [{\citenamefont {Haviland}\ \emph {et~al.}(1989)\citenamefont
  {Haviland}, \citenamefont {Liu},\ and\ \citenamefont
  {Goldman}}]{GoldmanIconic}%
  \BibitemOpen
  \bibfield  {author} {\bibinfo {author} {\bibnamefont {Haviland},
  \bibfnamefont {D.~B.}}, \bibinfo {author} {\bibfnamefont {Y.}~\bibnamefont
  {Liu}}, \ and\ \bibinfo {author} {\bibfnamefont {A.~M.}\ \bibnamefont
  {Goldman}}} (\bibinfo {year} {1989}),\ \href {\doibase
  10.1103/PhysRevLett.62.2180} {\bibfield  {journal} {\bibinfo  {journal}
  {Phys. Rev. Lett.}\ }\textbf {\bibinfo {volume} {62}},\ \bibinfo {pages}
  {2180}}\BibitemShut {NoStop}%
\bibitem [{\citenamefont {Hebard}\ and\ \citenamefont
  {Paalanen}(1990)}]{HebardPaalanen}%
  \BibitemOpen
  \bibfield  {author} {\bibinfo {author} {\bibnamefont {Hebard}, \bibfnamefont
  {A.~F.}}, \ and\ \bibinfo {author} {\bibfnamefont {M.~A.}\ \bibnamefont
  {Paalanen}}} (\bibinfo {year} {1990}),\ \href {\doibase
  10.1103/PhysRevLett.65.927} {\bibfield  {journal} {\bibinfo  {journal} {Phys.
  Rev. Lett.}\ }\textbf {\bibinfo {volume} {65}},\ \bibinfo {pages}
  {927}}\BibitemShut {NoStop}%
\bibitem [{\citenamefont {Herbut}\ and\ \citenamefont
  {Te\v{s}anovi\'c‡}(1995)}]{Herbut1995}%
  \BibitemOpen
  \bibfield  {author} {\bibinfo {author} {\bibnamefont {Herbut}, \bibfnamefont
  {I.~F.}}, \ and\ \bibinfo {author} {\bibfnamefont {Z.}~\bibnamefont
  {Te\v{s}anovi\'c‡}}} (\bibinfo {year} {1995}),\ \href {\doibase
  https://doi.org/10.1016/0921-4534(95)00613-3} {\bibfield  {journal} {\bibinfo
   {journal} {Physica C: Superconductivity}\ }\textbf {\bibinfo {volume}
  {255}}~(\bibinfo {number} {3}),\ \bibinfo {pages} {324 }}\BibitemShut
  {NoStop}%
\bibitem [{\citenamefont {Hertz}(1976)}]{Hertz1976}%
  \BibitemOpen
  \bibfield  {author} {\bibinfo {author} {\bibnamefont {Hertz}, \bibfnamefont
  {J.~A.}}} (\bibinfo {year} {1976}),\ \href {\doibase
  10.1103/PhysRevB.14.1165} {\bibfield  {journal} {\bibinfo  {journal} {Phys.
  Rev. B}\ }\textbf {\bibinfo {volume} {14}},\ \bibinfo {pages}
  {1165}}\BibitemShut {NoStop}%
\bibitem [{\citenamefont {Holtzberg}\ \emph {et~al.}(1968)\citenamefont
  {Holtzberg}, \citenamefont {Seiden},\ and\ \citenamefont {von
  Molnar}}]{Holtzberg1968}%
  \BibitemOpen
  \bibfield  {author} {\bibinfo {author} {\bibnamefont {Holtzberg},
  \bibfnamefont {F.}}, \bibinfo {author} {\bibfnamefont {P.~E.}\ \bibnamefont
  {Seiden}}, \ and\ \bibinfo {author} {\bibfnamefont {S.}~\bibnamefont {von
  Molnar}}} (\bibinfo {year} {1968}),\ \href {\doibase 10.1103/PhysRev.168.408}
  {\bibfield  {journal} {\bibinfo  {journal} {Phys. Rev.}\ }\textbf {\bibinfo
  {volume} {168}},\ \bibinfo {pages} {408}}\BibitemShut {NoStop}%
\bibitem [{\citenamefont {Hou}\ and\ \citenamefont {Varma}(2016)}]{varma2}%
  \BibitemOpen
  \bibfield  {author} {\bibinfo {author} {\bibnamefont {Hou}, \bibfnamefont
  {C.}}, \ and\ \bibinfo {author} {\bibfnamefont {C.~M.}\ \bibnamefont
  {Varma}}} (\bibinfo {year} {2016}),\ \href {\doibase
  10.1103/PhysRevB.94.201101} {\bibfield  {journal} {\bibinfo  {journal} {Phys.
  Rev. B}\ }\textbf {\bibinfo {volume} {94}},\ \bibinfo {pages}
  {201101}}\BibitemShut {NoStop}%
\bibitem [{\citenamefont {Hsu}\ \emph {et~al.}(2017)\citenamefont {Hsu},
  \citenamefont {M}, \citenamefont {Davies}, \citenamefont {Chan},
  \citenamefont {Porras}, \citenamefont {Loew}, \citenamefont {Taylor},
  \citenamefont {Liu}, \citenamefont {Tacon}, \citenamefont {Zuo},
  \citenamefont {Wang}, \citenamefont {Zhu}, \citenamefont {Lonzarich},
  \citenamefont {Keimer}, \citenamefont {Harrison},\ and\ \citenamefont
  {Sebastian}}]{suchitra}%
  \BibitemOpen
  \bibfield  {author} {\bibinfo {author} {\bibnamefont {Hsu}, \bibfnamefont
  {Y.-T.}}, \bibinfo {author} {\bibfnamefont {H.}~\bibnamefont {M}}, \bibinfo
  {author} {\bibfnamefont {A.~J.}\ \bibnamefont {Davies}}, \bibinfo {author}
  {\bibfnamefont {M.~K.}\ \bibnamefont {Chan}}, \bibinfo {author}
  {\bibfnamefont {J.}~\bibnamefont {Porras}}, \bibinfo {author} {\bibfnamefont
  {T.}~\bibnamefont {Loew}}, \bibinfo {author} {\bibfnamefont {S.~V.}\
  \bibnamefont {Taylor}}, \bibinfo {author} {\bibfnamefont {H.}~\bibnamefont
  {Liu}}, \bibinfo {author} {\bibfnamefont {M.~L.}\ \bibnamefont {Tacon}},
  \bibinfo {author} {\bibfnamefont {H.}~\bibnamefont {Zuo}}, \bibinfo {author}
  {\bibfnamefont {J.}~\bibnamefont {Wang}}, \bibinfo {author} {\bibfnamefont
  {Z.}~\bibnamefont {Zhu}}, \bibinfo {author} {\bibfnamefont {G.~G.}\
  \bibnamefont {Lonzarich}}, \bibinfo {author} {\bibfnamefont {B.}~\bibnamefont
  {Keimer}}, \bibinfo {author} {\bibfnamefont {N.}~\bibnamefont {Harrison}}, \
  and\ \bibinfo {author} {\bibfnamefont {S.~E.}\ \bibnamefont {Sebastian}}}
  (\bibinfo {year} {2017}),\ \href@noop {} {\ }\bibinfo {note}
  {Preprint}\BibitemShut {NoStop}%
\bibitem [{\citenamefont {Hui}\ and\ \citenamefont {Berker}(1989)}]{Berker}%
  \BibitemOpen
  \bibfield  {author} {\bibinfo {author} {\bibnamefont {Hui}, \bibfnamefont
  {K.}}, \ and\ \bibinfo {author} {\bibfnamefont {A.~N.}\ \bibnamefont
  {Berker}}} (\bibinfo {year} {1989}),\ \href {\doibase
  10.1103/PhysRevLett.62.2507} {\bibfield  {journal} {\bibinfo  {journal}
  {Phys. Rev. Lett.}\ }\textbf {\bibinfo {volume} {62}},\ \bibinfo {pages}
  {2507}}\BibitemShut {NoStop}%
\bibitem [{\citenamefont {Imry}\ and\ \citenamefont {Ma}(1975)}]{ImryMa}%
  \BibitemOpen
  \bibfield  {author} {\bibinfo {author} {\bibnamefont {Imry}, \bibfnamefont
  {Y.}}, \ and\ \bibinfo {author} {\bibfnamefont {S.-k.}\ \bibnamefont {Ma}}}
  (\bibinfo {year} {1975}),\ \href {\doibase 10.1103/PhysRevLett.35.1399}
  {\bibfield  {journal} {\bibinfo  {journal} {Phys. Rev. Lett.}\ }\textbf
  {\bibinfo {volume} {35}},\ \bibinfo {pages} {1399}}\BibitemShut {NoStop}%
\bibitem [{\citenamefont {Imry}\ and\ \citenamefont
  {Strongin}(1981)}]{ImryStrongin}%
  \BibitemOpen
  \bibfield  {author} {\bibinfo {author} {\bibnamefont {Imry}, \bibfnamefont
  {Y.}}, \ and\ \bibinfo {author} {\bibfnamefont {M.}~\bibnamefont {Strongin}}}
  (\bibinfo {year} {1981}),\ \href {\doibase 10.1103/PhysRevB.24.6353}
  {\bibfield  {journal} {\bibinfo  {journal} {Phys. Rev. B}\ }\textbf {\bibinfo
  {volume} {24}},\ \bibinfo {pages} {6353}}\BibitemShut {NoStop}%
\bibitem [{\citenamefont {Imry}\ and\ \citenamefont
  {Wortis}(1979)}]{ImryWortis}%
  \BibitemOpen
  \bibfield  {author} {\bibinfo {author} {\bibnamefont {Imry}, \bibfnamefont
  {Y.}}, \ and\ \bibinfo {author} {\bibfnamefont {M.}~\bibnamefont {Wortis}}}
  (\bibinfo {year} {1979}),\ \href {\doibase 10.1103/PhysRevB.19.3580}
  {\bibfield  {journal} {\bibinfo  {journal} {Phys. Rev. B}\ }\textbf {\bibinfo
  {volume} {19}},\ \bibinfo {pages} {3580}}\BibitemShut {NoStop}%
\bibitem [{\citenamefont {Ioffe}\ and\ \citenamefont
  {Larkin}(1981)}]{Ioffe1981}%
  \BibitemOpen
  \bibfield  {author} {\bibinfo {author} {\bibnamefont {Ioffe}, \bibfnamefont
  {L.}}, \ and\ \bibinfo {author} {\bibfnamefont {A.}~\bibnamefont {Larkin}}}
  (\bibinfo {year} {1981}),\ \href@noop {} {\bibfield  {journal} {\bibinfo
  {journal} {Zh. \'Eksp. Teor. Fiz}\ }\textbf {\bibinfo {volume}
  {81}}~(\bibinfo {number} {2}),\ \bibinfo {pages} {707}},\ \bibinfo {note}
  {[Sov. Phys. JETP, 54 (1981), p. 378]}\BibitemShut {NoStop}%
\bibitem [{\citenamefont {Jaeger}\ \emph {et~al.}(1989)\citenamefont {Jaeger},
  \citenamefont {Haviland}, \citenamefont {Orr},\ and\ \citenamefont
  {Goldman}}]{Goldman2}%
  \BibitemOpen
  \bibfield  {author} {\bibinfo {author} {\bibnamefont {Jaeger}, \bibfnamefont
  {H.~M.}}, \bibinfo {author} {\bibfnamefont {D.~B.}\ \bibnamefont {Haviland}},
  \bibinfo {author} {\bibfnamefont {B.~G.}\ \bibnamefont {Orr}}, \ and\
  \bibinfo {author} {\bibfnamefont {A.~M.}\ \bibnamefont {Goldman}}} (\bibinfo
  {year} {1989}),\ \href {\doibase 10.1103/PhysRevB.40.182} {\bibfield
  {journal} {\bibinfo  {journal} {Phys. Rev. B}\ }\textbf {\bibinfo {volume}
  {40}},\ \bibinfo {pages} {182}}\BibitemShut {NoStop}%
\bibitem [{\citenamefont {Kapitulnik}\ and\ \citenamefont
  {Deutscher}(1982)}]{Kapitulnik1982}%
  \BibitemOpen
  \bibfield  {author} {\bibinfo {author} {\bibnamefont {Kapitulnik},
  \bibfnamefont {A.}}, \ and\ \bibinfo {author} {\bibfnamefont
  {G.}~\bibnamefont {Deutscher}}} (\bibinfo {year} {1982}),\ \href {\doibase
  10.1103/PhysRevLett.49.1444} {\bibfield  {journal} {\bibinfo  {journal}
  {Phys. Rev. Lett.}\ }\textbf {\bibinfo {volume} {49}},\ \bibinfo {pages}
  {1444}}\BibitemShut {NoStop}%
\bibitem [{\citenamefont {Kapitulnik}\ \emph {et~al.}(2001)\citenamefont
  {Kapitulnik}, \citenamefont {Mason}, \citenamefont {Kivelson},\ and\
  \citenamefont {Chakravarty}}]{chakravartykapitulniandme}%
  \BibitemOpen
  \bibfield  {author} {\bibinfo {author} {\bibnamefont {Kapitulnik},
  \bibfnamefont {A.}}, \bibinfo {author} {\bibfnamefont {N.}~\bibnamefont
  {Mason}}, \bibinfo {author} {\bibfnamefont {S.~A.}\ \bibnamefont {Kivelson}},
  \ and\ \bibinfo {author} {\bibfnamefont {S.}~\bibnamefont {Chakravarty}}}
  (\bibinfo {year} {2001}),\ \href {\doibase 10.1103/PhysRevB.63.125322}
  {\bibfield  {journal} {\bibinfo  {journal} {Phys. Rev. B}\ }\textbf {\bibinfo
  {volume} {63}},\ \bibinfo {pages} {125322}}\BibitemShut {NoStop}%
\bibitem [{\citenamefont {Kivelson}\ \emph {et~al.}(1992)\citenamefont
  {Kivelson}, \citenamefont {Lee},\ and\ \citenamefont {Zhang}}]{KLZ}%
  \BibitemOpen
  \bibfield  {author} {\bibinfo {author} {\bibnamefont {Kivelson},
  \bibfnamefont {S.}}, \bibinfo {author} {\bibfnamefont {D.-H.}\ \bibnamefont
  {Lee}}, \ and\ \bibinfo {author} {\bibfnamefont {S.-C.}\ \bibnamefont
  {Zhang}}} (\bibinfo {year} {1992}),\ \href {\doibase
  10.1103/PhysRevB.46.2223} {\bibfield  {journal} {\bibinfo  {journal} {Phys.
  Rev. B}\ }\textbf {\bibinfo {volume} {46}},\ \bibinfo {pages}
  {2223}}\BibitemShut {NoStop}%
\bibitem [{\citenamefont {Kivelson}\ and\ \citenamefont
  {Spivak}(2015)}]{KivelsonSpivakD1}%
  \BibitemOpen
  \bibfield  {author} {\bibinfo {author} {\bibnamefont {Kivelson},
  \bibfnamefont {S.~A.}}, \ and\ \bibinfo {author} {\bibfnamefont
  {B.}~\bibnamefont {Spivak}}} (\bibinfo {year} {2015}),\ \href {\doibase
  10.1103/PhysRevB.92.184502} {\bibfield  {journal} {\bibinfo  {journal} {Phys.
  Rev. B}\ }\textbf {\bibinfo {volume} {92}},\ \bibinfo {pages}
  {184502}}\BibitemShut {NoStop}%
\bibitem [{\citenamefont {Kivelson}\ and\ \citenamefont
  {Spivak}(1992)}]{KivelsonSpivakNeg}%
  \BibitemOpen
  \bibfield  {author} {\bibinfo {author} {\bibnamefont {Kivelson},
  \bibfnamefont {S.~A.}}, \ and\ \bibinfo {author} {\bibfnamefont {B.~Z.}\
  \bibnamefont {Spivak}}} (\bibinfo {year} {1992}),\ \href {\doibase
  10.1103/PhysRevB.45.10490} {\bibfield  {journal} {\bibinfo  {journal} {Phys.
  Rev. B}\ }\textbf {\bibinfo {volume} {45}},\ \bibinfo {pages}
  {10490}}\BibitemShut {NoStop}%
\bibitem [{\citenamefont {Larkin}\ and\ \citenamefont
  {Varlamov}(2005)}]{LarkinVarlamov}%
  \BibitemOpen
  \bibfield  {author} {\bibinfo {author} {\bibnamefont {Larkin}, \bibfnamefont
  {A.}}, \ and\ \bibinfo {author} {\bibfnamefont {A.}~\bibnamefont {Varlamov}}}
  (\bibinfo {year} {2005}),\ \href@noop {} {\emph {\bibinfo {title} {Theory of
  Fluctuations in Superconductors}}},\ \bibinfo {edition} {1st}\ ed.\ (\bibinfo
   {publisher} {Oxford University Press})\BibitemShut {NoStop}%
\bibitem [{\citenamefont {Larkin}\ and\ \citenamefont
  {Ovchinnikov}(1979)}]{Larkin1979}%
  \BibitemOpen
  \bibfield  {author} {\bibinfo {author} {\bibnamefont {Larkin}, \bibfnamefont
  {A.~I.}}, \ and\ \bibinfo {author} {\bibfnamefont {Y.~N.}\ \bibnamefont
  {Ovchinnikov}}} (\bibinfo {year} {1979}),\ \href {\doibase
  10.1007/BF00117160} {\bibfield  {journal} {\bibinfo  {journal} {Journal of
  Low Temperature Physics}\ }\textbf {\bibinfo {volume} {34}}~(\bibinfo
  {number} {3}),\ \bibinfo {pages} {409}}\BibitemShut {NoStop}%
\bibitem [{\citenamefont {Lee}\ and\ \citenamefont
  {Ramakrishnan}(1985)}]{LeeRamakrishnan}%
  \BibitemOpen
  \bibfield  {author} {\bibinfo {author} {\bibnamefont {Lee}, \bibfnamefont
  {P.~A.}}, \ and\ \bibinfo {author} {\bibfnamefont {T.~V.}\ \bibnamefont
  {Ramakrishnan}}} (\bibinfo {year} {1985}),\ \href {\doibase
  10.1103/RevModPhys.57.287} {\bibfield  {journal} {\bibinfo  {journal} {Rev.
  Mod. Phys.}\ }\textbf {\bibinfo {volume} {57}},\ \bibinfo {pages}
  {287}}\BibitemShut {NoStop}%
\bibitem [{\citenamefont {Lee}\ and\ \citenamefont {Rice}(1979)}]{LeeRice}%
  \BibitemOpen
  \bibfield  {author} {\bibinfo {author} {\bibnamefont {Lee}, \bibfnamefont
  {P.~A.}}, \ and\ \bibinfo {author} {\bibfnamefont {T.~M.}\ \bibnamefont
  {Rice}}} (\bibinfo {year} {1979}),\ \href {\doibase 10.1103/PhysRevB.19.3970}
  {\bibfield  {journal} {\bibinfo  {journal} {Phys. Rev. B}\ }\textbf {\bibinfo
  {volume} {19}},\ \bibinfo {pages} {3970}}\BibitemShut {NoStop}%
\bibitem [{\citenamefont {Leggett}\ \emph {et~al.}(1987)\citenamefont
  {Leggett}, \citenamefont {Chakravarty}, \citenamefont {Dorsey}, \citenamefont
  {Fisher}, \citenamefont {Garg},\ and\ \citenamefont {Zwerger}}]{leggettrmp}%
  \BibitemOpen
  \bibfield  {author} {\bibinfo {author} {\bibnamefont {Leggett}, \bibfnamefont
  {A.~J.}}, \bibinfo {author} {\bibfnamefont {S.}~\bibnamefont {Chakravarty}},
  \bibinfo {author} {\bibfnamefont {A.~T.}\ \bibnamefont {Dorsey}}, \bibinfo
  {author} {\bibfnamefont {M.~P.~A.}\ \bibnamefont {Fisher}}, \bibinfo {author}
  {\bibfnamefont {A.}~\bibnamefont {Garg}}, \ and\ \bibinfo {author}
  {\bibfnamefont {W.}~\bibnamefont {Zwerger}}} (\bibinfo {year} {1987}),\ \href
  {\doibase 10.1103/RevModPhys.59.1} {\bibfield  {journal} {\bibinfo  {journal}
  {Rev. Mod. Phys.}\ }\textbf {\bibinfo {volume} {59}},\ \bibinfo {pages}
  {1}}\BibitemShut {NoStop}%
\bibitem [{\citenamefont {Levanyuk}(1959)}]{Levanyuk1959}%
  \BibitemOpen
  \bibfield  {author} {\bibinfo {author} {\bibnamefont {Levanyuk},
  \bibfnamefont {A.}}} (\bibinfo {year} {1959}),\ \href@noop {} {\bibfield
  {journal} {\bibinfo  {journal} {Soviet Phys. JETP-USSR}\ }\textbf {\bibinfo
  {volume} {9}}~(\bibinfo {number} {3}),\ \bibinfo {pages} {571}}\BibitemShut
  {NoStop}%
\bibitem [{\citenamefont {Liu}\ \emph {et~al.}(2013)\citenamefont {Liu},
  \citenamefont {Pan}, \citenamefont {Wen}, \citenamefont {Kim}, \citenamefont
  {Sambandamurthy},\ and\ \citenamefont {Armitage}}]{Liu2013}%
  \BibitemOpen
  \bibfield  {author} {\bibinfo {author} {\bibnamefont {Liu}, \bibfnamefont
  {W.}}, \bibinfo {author} {\bibfnamefont {L.}~\bibnamefont {Pan}}, \bibinfo
  {author} {\bibfnamefont {J.}~\bibnamefont {Wen}}, \bibinfo {author}
  {\bibfnamefont {M.}~\bibnamefont {Kim}}, \bibinfo {author} {\bibfnamefont
  {G.}~\bibnamefont {Sambandamurthy}}, \ and\ \bibinfo {author} {\bibfnamefont
  {N.~P.}\ \bibnamefont {Armitage}}} (\bibinfo {year} {2013}),\ \href {\doibase
  10.1103/PhysRevLett.111.067003} {\bibfield  {journal} {\bibinfo  {journal}
  {Phys. Rev. Lett.}\ }\textbf {\bibinfo {volume} {111}},\ \bibinfo {pages}
  {067003}}\BibitemShut {NoStop}%
\bibitem [{\citenamefont {Lopez}\ and\ \citenamefont
  {Fradkin}(1991)}]{lopezfradkin}%
  \BibitemOpen
  \bibfield  {author} {\bibinfo {author} {\bibnamefont {Lopez}, \bibfnamefont
  {A.}}, \ and\ \bibinfo {author} {\bibfnamefont {E.}~\bibnamefont {Fradkin}}}
  (\bibinfo {year} {1991}),\ \href {\doibase 10.1103/PhysRevB.44.5246}
  {\bibfield  {journal} {\bibinfo  {journal} {Phys. Rev. B}\ }\textbf {\bibinfo
  {volume} {44}},\ \bibinfo {pages} {5246}}\BibitemShut {NoStop}%
\bibitem [{\citenamefont {Maki}(1968)}]{Maki}%
  \BibitemOpen
  \bibfield  {author} {\bibinfo {author} {\bibnamefont {Maki}, \bibfnamefont
  {K.}}} (\bibinfo {year} {1968}),\ \href {\doibase 10.1143/PTP.39.897}
  {\bibfield  {journal} {\bibinfo  {journal} {Progress of Theoretical Physics}\
  }\textbf {\bibinfo {volume} {39}}~(\bibinfo {number} {4}),\ \bibinfo {pages}
  {897}}\BibitemShut {NoStop}%
\bibitem [{\citenamefont {Markos}(2006)}]{Markos06}%
  \BibitemOpen
  \bibfield  {author} {\bibinfo {author} {\bibnamefont {Markos}, \bibfnamefont
  {P.}}} (\bibinfo {year} {2006}),\ \href {\doibase 10.2478/v10155-010-0081-0}
  {\bibfield  {journal} {\bibinfo  {journal} {Acta Physica Slovaca}\ }\textbf
  {\bibinfo {volume} {56}}~(\bibinfo {number} {5}),\ \bibinfo {pages}
  {561}}\BibitemShut {NoStop}%
\bibitem [{\citenamefont {Masker}\ \emph {et~al.}(1969)\citenamefont {Masker},
  \citenamefont {Mar\ifmmode~\check{c}\else \v{c}\fi{}elja},\ and\
  \citenamefont {Parks}}]{Masker1969}%
  \BibitemOpen
  \bibfield  {author} {\bibinfo {author} {\bibnamefont {Masker}, \bibfnamefont
  {W.~E.}}, \bibinfo {author} {\bibfnamefont {S.}~\bibnamefont
  {Mar\ifmmode~\check{c}\else \v{c}\fi{}elja}}, \ and\ \bibinfo {author}
  {\bibfnamefont {R.~D.}\ \bibnamefont {Parks}}} (\bibinfo {year} {1969}),\
  \href {\doibase 10.1103/PhysRev.188.745} {\bibfield  {journal} {\bibinfo
  {journal} {Phys. Rev.}\ }\textbf {\bibinfo {volume} {188}},\ \bibinfo {pages}
  {745}}\BibitemShut {NoStop}%
\bibitem [{\citenamefont {Mason}(2001)}]{MasonThesis}%
  \BibitemOpen
  \bibfield  {author} {\bibinfo {author} {\bibnamefont {Mason}, \bibfnamefont
  {N.}}} (\bibinfo {year} {2001}),\ \emph {\bibinfo {title}
  {Superconductor-Metal-Insulator Transitions in Two Dimensional}},\ \href@noop
  {} {Ph.D. thesis}\ (\bibinfo  {school} {Stanford University}, \bibinfo
  {address} {Stanford, CA 94305})\BibitemShut {NoStop}%
\bibitem [{\citenamefont {Mason}\ and\ \citenamefont
  {Kapitulnik}(1999)}]{MasonKapitulnik1}%
  \BibitemOpen
  \bibfield  {author} {\bibinfo {author} {\bibnamefont {Mason}, \bibfnamefont
  {N.}}, \ and\ \bibinfo {author} {\bibfnamefont {A.}~\bibnamefont
  {Kapitulnik}}} (\bibinfo {year} {1999}),\ \href {\doibase
  10.1103/PhysRevLett.82.5341} {\bibfield  {journal} {\bibinfo  {journal}
  {Phys. Rev. Lett.}\ }\textbf {\bibinfo {volume} {82}},\ \bibinfo {pages}
  {5341}}\BibitemShut {NoStop}%
\bibitem [{\citenamefont {Mason}\ and\ \citenamefont
  {Kapitulnik}(2001)}]{MasonKapitulnik2}%
  \BibitemOpen
  \bibfield  {author} {\bibinfo {author} {\bibnamefont {Mason}, \bibfnamefont
  {N.}}, \ and\ \bibinfo {author} {\bibfnamefont {A.}~\bibnamefont
  {Kapitulnik}}} (\bibinfo {year} {2001}),\ \href {\doibase
  10.1103/PhysRevB.64.060504} {\bibfield  {journal} {\bibinfo  {journal} {Phys.
  Rev. B}\ }\textbf {\bibinfo {volume} {64}},\ \bibinfo {pages}
  {060504}}\BibitemShut {NoStop}%
\bibitem [{\citenamefont {Merchant}\ \emph {et~al.}(2001)\citenamefont
  {Merchant}, \citenamefont {Ostrick}, \citenamefont {Barber},\ and\
  \citenamefont {Dynes}}]{Merchant}%
  \BibitemOpen
  \bibfield  {author} {\bibinfo {author} {\bibnamefont {Merchant},
  \bibfnamefont {L.}}, \bibinfo {author} {\bibfnamefont {J.}~\bibnamefont
  {Ostrick}}, \bibinfo {author} {\bibfnamefont {R.~P.}\ \bibnamefont {Barber}},
  \ and\ \bibinfo {author} {\bibfnamefont {R.~C.}\ \bibnamefont {Dynes}}}
  (\bibinfo {year} {2001}),\ \href {\doibase 10.1103/PhysRevB.63.134508}
  {\bibfield  {journal} {\bibinfo  {journal} {Phys. Rev. B}\ }\textbf {\bibinfo
  {volume} {63}},\ \bibinfo {pages} {134508}}\BibitemShut {NoStop}%
\bibitem [{\citenamefont {Millis}(1993)}]{Millis1993}%
  \BibitemOpen
  \bibfield  {author} {\bibinfo {author} {\bibnamefont {Millis}, \bibfnamefont
  {A.~J.}}} (\bibinfo {year} {1993}),\ \href {\doibase
  10.1103/PhysRevB.48.7183} {\bibfield  {journal} {\bibinfo  {journal} {Phys.
  Rev. B}\ }\textbf {\bibinfo {volume} {48}},\ \bibinfo {pages}
  {7183}}\BibitemShut {NoStop}%
\bibitem [{\citenamefont {Millis}\ \emph {et~al.}(2002)\citenamefont {Millis},
  \citenamefont {Morr},\ and\ \citenamefont {Schmalian}}]{nogriffith}%
  \BibitemOpen
  \bibfield  {author} {\bibinfo {author} {\bibnamefont {Millis}, \bibfnamefont
  {A.~J.}}, \bibinfo {author} {\bibfnamefont {D.~K.}\ \bibnamefont {Morr}}, \
  and\ \bibinfo {author} {\bibfnamefont {J.}~\bibnamefont {Schmalian}}}
  (\bibinfo {year} {2002}),\ \href {\doibase 10.1103/PhysRevB.66.174433}
  {\bibfield  {journal} {\bibinfo  {journal} {Phys. Rev. B}\ }\textbf {\bibinfo
  {volume} {66}},\ \bibinfo {pages} {174433}}\BibitemShut {NoStop}%
\bibitem [{\citenamefont {Mulligan}(2017)}]{mikemetal}%
  \BibitemOpen
  \bibfield  {author} {\bibinfo {author} {\bibnamefont {Mulligan},
  \bibfnamefont {M.}}} (\bibinfo {year} {2017}),\ \href {\doibase
  10.1103/PhysRevB.95.045118} {\bibfield  {journal} {\bibinfo  {journal} {Phys.
  Rev. B}\ }\textbf {\bibinfo {volume} {95}},\ \bibinfo {pages}
  {045118}}\BibitemShut {NoStop}%
\bibitem [{\citenamefont {Mulligan}\ and\ \citenamefont
  {Raghu}(2016)}]{sriandco}%
  \BibitemOpen
  \bibfield  {author} {\bibinfo {author} {\bibnamefont {Mulligan},
  \bibfnamefont {M.}}, \ and\ \bibinfo {author} {\bibfnamefont
  {S.}~\bibnamefont {Raghu}}} (\bibinfo {year} {2016}),\ \href {\doibase
  10.1103/PhysRevB.93.205116} {\bibfield  {journal} {\bibinfo  {journal} {Phys.
  Rev. B}\ }\textbf {\bibinfo {volume} {93}},\ \bibinfo {pages}
  {205116}}\BibitemShut {NoStop}%
\bibitem [{\citenamefont {Nayak}\ \emph {et~al.}(2001)\citenamefont {Nayak},
  \citenamefont {Shtengel}, \citenamefont {Orgad}, \citenamefont {Fisher},\
  and\ \citenamefont {Girvin}}]{nayakorgad}%
  \BibitemOpen
  \bibfield  {author} {\bibinfo {author} {\bibnamefont {Nayak}, \bibfnamefont
  {C.}}, \bibinfo {author} {\bibfnamefont {K.}~\bibnamefont {Shtengel}},
  \bibinfo {author} {\bibfnamefont {D.}~\bibnamefont {Orgad}}, \bibinfo
  {author} {\bibfnamefont {M.~P.~A.}\ \bibnamefont {Fisher}}, \ and\ \bibinfo
  {author} {\bibfnamefont {S.~M.}\ \bibnamefont {Girvin}}} (\bibinfo {year}
  {2001}),\ \href {\doibase 10.1103/PhysRevB.64.235113} {\bibfield  {journal}
  {\bibinfo  {journal} {Phys. Rev. B}\ }\textbf {\bibinfo {volume} {64}},\
  \bibinfo {pages} {235113}}\BibitemShut {NoStop}%
\bibitem [{\citenamefont {Parameswaran}\ \emph {et~al.}(2010)\citenamefont
  {Parameswaran}, \citenamefont {Shankar},\ and\ \citenamefont {Sondhi}}]{Sid}%
  \BibitemOpen
  \bibfield  {author} {\bibinfo {author} {\bibnamefont {Parameswaran},
  \bibfnamefont {S.~A.}}, \bibinfo {author} {\bibfnamefont {R.}~\bibnamefont
  {Shankar}}, \ and\ \bibinfo {author} {\bibfnamefont {S.~L.}\ \bibnamefont
  {Sondhi}}} (\bibinfo {year} {2010}),\ \href {\doibase
  10.1103/PhysRevB.82.195104} {\bibfield  {journal} {\bibinfo  {journal} {Phys.
  Rev. B}\ }\textbf {\bibinfo {volume} {82}},\ \bibinfo {pages}
  {195104}}\BibitemShut {NoStop}%
\bibitem [{\citenamefont {Phillips}\ and\ \citenamefont
  {Dalidovich}(2003)}]{Phillips2}%
  \BibitemOpen
  \bibfield  {author} {\bibinfo {author} {\bibnamefont {Phillips},
  \bibfnamefont {P.}}, \ and\ \bibinfo {author} {\bibfnamefont
  {D.}~\bibnamefont {Dalidovich}}} (\bibinfo {year} {2003}),\ \href {\doibase
  10.1126/science.1088253} {\bibfield  {journal} {\bibinfo  {journal}
  {Science}\ }\textbf {\bibinfo {volume} {302}}~(\bibinfo {number} {5643}),\
  \bibinfo {pages} {243}}\BibitemShut {NoStop}%
\bibitem [{\citenamefont {Qin}\ \emph {et~al.}(2006)\citenamefont {Qin},
  \citenamefont {Vicente},\ and\ \citenamefont {Yoon}}]{Qin2006}%
  \BibitemOpen
  \bibfield  {author} {\bibinfo {author} {\bibnamefont {Qin}, \bibfnamefont
  {Y.}}, \bibinfo {author} {\bibfnamefont {C.~L.}\ \bibnamefont {Vicente}}, \
  and\ \bibinfo {author} {\bibfnamefont {J.}~\bibnamefont {Yoon}}} (\bibinfo
  {year} {2006}),\ \href {\doibase 10.1103/PhysRevB.73.100505} {\bibfield
  {journal} {\bibinfo  {journal} {Phys. Rev. B}\ }\textbf {\bibinfo {volume}
  {73}},\ \bibinfo {pages} {100505}}\BibitemShut {NoStop}%
\bibitem [{\citenamefont {Raghu}\ \emph
  {et~al.}(2015{\natexlab{a}})\citenamefont {Raghu}, \citenamefont {Torroba},\
  and\ \citenamefont {Wang}}]{srimetal}%
  \BibitemOpen
  \bibfield  {author} {\bibinfo {author} {\bibnamefont {Raghu}, \bibfnamefont
  {S.}}, \bibinfo {author} {\bibfnamefont {G.}~\bibnamefont {Torroba}}, \ and\
  \bibinfo {author} {\bibfnamefont {H.}~\bibnamefont {Wang}}} (\bibinfo {year}
  {2015}{\natexlab{a}}),\ \href {\doibase 10.1103/PhysRevB.92.205104}
  {\bibfield  {journal} {\bibinfo  {journal} {Phys. Rev. B}\ }\textbf {\bibinfo
  {volume} {92}},\ \bibinfo {pages} {205104}}\BibitemShut {NoStop}%
\bibitem [{\citenamefont {Raghu}\ \emph
  {et~al.}(2015{\natexlab{b}})\citenamefont {Raghu}, \citenamefont {Torroba},\
  and\ \citenamefont {Wang}}]{raghuQSMT}%
  \BibitemOpen
  \bibfield  {author} {\bibinfo {author} {\bibnamefont {Raghu}, \bibfnamefont
  {S.}}, \bibinfo {author} {\bibfnamefont {G.}~\bibnamefont {Torroba}}, \ and\
  \bibinfo {author} {\bibfnamefont {H.}~\bibnamefont {Wang}}} (\bibinfo {year}
  {2015}{\natexlab{b}}),\ \href {\doibase 10.1103/PhysRevB.92.205104}
  {\bibfield  {journal} {\bibinfo  {journal} {Phys. Rev. B}\ }\textbf {\bibinfo
  {volume} {92}},\ \bibinfo {pages} {205104}}\BibitemShut {NoStop}%
\bibitem [{\citenamefont {Ramshaw}\ \emph {et~al.}(2012)\citenamefont
  {Ramshaw}, \citenamefont {Day}, \citenamefont {Vignolle}, \citenamefont
  {LeBoeuf}, \citenamefont {Dosanjh}, \citenamefont {Proust}, \citenamefont
  {Taillefer}, \citenamefont {Liang}, \citenamefont {Hardy},\ and\
  \citenamefont {Bonn}}]{brad}%
  \BibitemOpen
  \bibfield  {author} {\bibinfo {author} {\bibnamefont {Ramshaw}, \bibfnamefont
  {B.~J.}}, \bibinfo {author} {\bibfnamefont {J.}~\bibnamefont {Day}}, \bibinfo
  {author} {\bibfnamefont {B.}~\bibnamefont {Vignolle}}, \bibinfo {author}
  {\bibfnamefont {D.}~\bibnamefont {LeBoeuf}}, \bibinfo {author} {\bibfnamefont
  {P.}~\bibnamefont {Dosanjh}}, \bibinfo {author} {\bibfnamefont
  {C.}~\bibnamefont {Proust}}, \bibinfo {author} {\bibfnamefont
  {L.}~\bibnamefont {Taillefer}}, \bibinfo {author} {\bibfnamefont
  {R.}~\bibnamefont {Liang}}, \bibinfo {author} {\bibfnamefont {W.~N.}\
  \bibnamefont {Hardy}}, \ and\ \bibinfo {author} {\bibfnamefont {D.~A.}\
  \bibnamefont {Bonn}}} (\bibinfo {year} {2012}),\ \href {\doibase
  10.1103/PhysRevB.86.174501} {\bibfield  {journal} {\bibinfo  {journal} {Phys.
  Rev. B}\ }\textbf {\bibinfo {volume} {86}},\ \bibinfo {pages}
  {174501}}\BibitemShut {NoStop}%
\bibitem [{\citenamefont {Rimberg}\ \emph {et~al.}(1997)\citenamefont
  {Rimberg}, \citenamefont {Ho}, \citenamefont {Kurdak}, \citenamefont
  {Clarke}, \citenamefont {Campman},\ and\ \citenamefont
  {Gossard}}]{RimbergClarke}%
  \BibitemOpen
  \bibfield  {author} {\bibinfo {author} {\bibnamefont {Rimberg}, \bibfnamefont
  {A.~J.}}, \bibinfo {author} {\bibfnamefont {T.~R.}\ \bibnamefont {Ho}},
  \bibinfo {author} {\bibfnamefont {i.~m.~c.}\ \bibnamefont {Kurdak}}, \bibinfo
  {author} {\bibfnamefont {J.}~\bibnamefont {Clarke}}, \bibinfo {author}
  {\bibfnamefont {K.~L.}\ \bibnamefont {Campman}}, \ and\ \bibinfo {author}
  {\bibfnamefont {A.~C.}\ \bibnamefont {Gossard}}} (\bibinfo {year} {1997}),\
  \href {\doibase 10.1103/PhysRevLett.78.2632} {\bibfield  {journal} {\bibinfo
  {journal} {Phys. Rev. Lett.}\ }\textbf {\bibinfo {volume} {78}},\ \bibinfo
  {pages} {2632}}\BibitemShut {NoStop}%
\bibitem [{\citenamefont {Saito}\ \emph {et~al.}(2015)\citenamefont {Saito},
  \citenamefont {Kasahara}, \citenamefont {Ye}, \citenamefont {Iwasa},\ and\
  \citenamefont {Nojima}}]{Saito2015}%
  \BibitemOpen
  \bibfield  {author} {\bibinfo {author} {\bibnamefont {Saito}, \bibfnamefont
  {Y.}}, \bibinfo {author} {\bibfnamefont {Y.}~\bibnamefont {Kasahara}},
  \bibinfo {author} {\bibfnamefont {J.}~\bibnamefont {Ye}}, \bibinfo {author}
  {\bibfnamefont {Y.}~\bibnamefont {Iwasa}}, \ and\ \bibinfo {author}
  {\bibfnamefont {T.}~\bibnamefont {Nojima}}} (\bibinfo {year} {2015}),\ \href
  {\doibase 10.1126/science.1259440} {\bibfield  {journal} {\bibinfo  {journal}
  {Science}\ }\textbf {\bibinfo {volume} {350}}~(\bibinfo {number} {6259}),\
  \bibinfo {pages} {409}}\BibitemShut {NoStop}%
\bibitem [{\citenamefont {Sajadi}\ \emph {et~al.}(2017)\citenamefont {Sajadi},
  \citenamefont {Palomaki}, \citenamefont {Fei}, \citenamefont {Zhao},
  \citenamefont {Bement}, \citenamefont {Olsen}, \citenamefont {Xu},
  \citenamefont {Folk},\ and\ \citenamefont {Cobden}}]{JoshFolk}%
  \BibitemOpen
  \bibfield  {author} {\bibinfo {author} {\bibnamefont {Sajadi}, \bibfnamefont
  {E.}}, \bibinfo {author} {\bibfnamefont {T.}~\bibnamefont {Palomaki}},
  \bibinfo {author} {\bibfnamefont {Z.}~\bibnamefont {Fei}}, \bibinfo {author}
  {\bibfnamefont {W.}~\bibnamefont {Zhao}}, \bibinfo {author} {\bibfnamefont
  {P.}~\bibnamefont {Bement}}, \bibinfo {author} {\bibfnamefont
  {C.}~\bibnamefont {Olsen}}, \bibinfo {author} {\bibfnamefont
  {X.}~\bibnamefont {Xu}}, \bibinfo {author} {\bibfnamefont {J.}~\bibnamefont
  {Folk}}, \ and\ \bibinfo {author} {\bibfnamefont {D.~H.}\ \bibnamefont
  {Cobden}}} (\bibinfo {year} {2017}),\ \href@noop {} {\ }\BibitemShut
  {NoStop}%
\bibitem [{\citenamefont {Schmid}(1983)}]{A.Schmid}%
  \BibitemOpen
  \bibfield  {author} {\bibinfo {author} {\bibnamefont {Schmid}, \bibfnamefont
  {A.}}} (\bibinfo {year} {1983}),\ \href {\doibase
  10.1103/PhysRevLett.51.1506} {\bibfield  {journal} {\bibinfo  {journal}
  {Phys. Rev. Lett.}\ }\textbf {\bibinfo {volume} {51}},\ \bibinfo {pages}
  {1506}}\BibitemShut {NoStop}%
\bibitem [{\citenamefont {Seiden}(1968)}]{Seiden1968}%
  \BibitemOpen
  \bibfield  {author} {\bibinfo {author} {\bibnamefont {Seiden}, \bibfnamefont
  {P.~E.}}} (\bibinfo {year} {1968}),\ \href {\doibase 10.1103/PhysRev.168.403}
  {\bibfield  {journal} {\bibinfo  {journal} {Phys. Rev.}\ }\textbf {\bibinfo
  {volume} {168}},\ \bibinfo {pages} {403}}\BibitemShut {NoStop}%
\bibitem [{\citenamefont {Sheng}\ \emph {et~al.}(2009)\citenamefont {Sheng},
  \citenamefont {Motrunich},\ and\ \citenamefont {Fisher}}]{fisherbosemetal}%
  \BibitemOpen
  \bibfield  {author} {\bibinfo {author} {\bibnamefont {Sheng}, \bibfnamefont
  {D.~N.}}, \bibinfo {author} {\bibfnamefont {O.~I.}\ \bibnamefont
  {Motrunich}}, \ and\ \bibinfo {author} {\bibfnamefont {M.~P.~A.}\
  \bibnamefont {Fisher}}} (\bibinfo {year} {2009}),\ \href {\doibase
  10.1103/PhysRevB.79.205112} {\bibfield  {journal} {\bibinfo  {journal} {Phys.
  Rev. B}\ }\textbf {\bibinfo {volume} {79}},\ \bibinfo {pages}
  {205112}}\BibitemShut {NoStop}%
\bibitem [{\citenamefont {Shtengel}\ \emph {et~al.}(2005)\citenamefont
  {Shtengel}, \citenamefont {Nayak}, \citenamefont {Bishara},\ and\
  \citenamefont {Chamon}}]{nayak}%
  \BibitemOpen
  \bibfield  {author} {\bibinfo {author} {\bibnamefont {Shtengel},
  \bibfnamefont {K.}}, \bibinfo {author} {\bibfnamefont {C.}~\bibnamefont
  {Nayak}}, \bibinfo {author} {\bibfnamefont {W.}~\bibnamefont {Bishara}}, \
  and\ \bibinfo {author} {\bibfnamefont {C.}~\bibnamefont {Chamon}}} (\bibinfo
  {year} {2005}),\ \href {\doibase 10.1088/0305-4470/38/36/L01} {\bibfield
  {journal} {\bibinfo  {journal} {J. Phys. A: Math. Theor.}\ }\textbf {\bibinfo
  {volume} {38}}~(\bibinfo {number} {36}),\ \bibinfo {pages}
  {L589}}\BibitemShut {NoStop}%
\bibitem [{\citenamefont {Siemons}\ \emph {et~al.}(2008)\citenamefont
  {Siemons}, \citenamefont {Steiner}, \citenamefont {Koster}, \citenamefont
  {Blank}, \citenamefont {Beasley},\ and\ \citenamefont
  {Kapitulnik}}]{Siemons2008}%
  \BibitemOpen
  \bibfield  {author} {\bibinfo {author} {\bibnamefont {Siemons}, \bibfnamefont
  {W.}}, \bibinfo {author} {\bibfnamefont {M.~A.}\ \bibnamefont {Steiner}},
  \bibinfo {author} {\bibfnamefont {G.}~\bibnamefont {Koster}}, \bibinfo
  {author} {\bibfnamefont {D.~H.~A.}\ \bibnamefont {Blank}}, \bibinfo {author}
  {\bibfnamefont {M.~R.}\ \bibnamefont {Beasley}}, \ and\ \bibinfo {author}
  {\bibfnamefont {A.}~\bibnamefont {Kapitulnik}}} (\bibinfo {year} {2008}),\
  \href {\doibase 10.1103/PhysRevB.77.174506} {\bibfield  {journal} {\bibinfo
  {journal} {Phys. Rev. B}\ }\textbf {\bibinfo {volume} {77}},\ \bibinfo
  {pages} {174506}}\BibitemShut {NoStop}%
\bibitem [{\citenamefont {Son}(2015)}]{son}%
  \BibitemOpen
  \bibfield  {author} {\bibinfo {author} {\bibnamefont {Son}, \bibfnamefont
  {D.~T.}}} (\bibinfo {year} {2015}),\ \href {\doibase
  10.1103/PhysRevX.5.031027} {\bibfield  {journal} {\bibinfo  {journal} {Phys.
  Rev. X}\ }\textbf {\bibinfo {volume} {5}},\ \bibinfo {pages}
  {031027}}\BibitemShut {NoStop}%
\bibitem [{\citenamefont {Spivak}\ \emph {et~al.}(2009)\citenamefont {Spivak},
  \citenamefont {Oreto},\ and\ \citenamefont {Kivelson}}]{KivelsonSpivakD}%
  \BibitemOpen
  \bibfield  {author} {\bibinfo {author} {\bibnamefont {Spivak}, \bibfnamefont
  {B.}}, \bibinfo {author} {\bibfnamefont {P.}~\bibnamefont {Oreto}}, \ and\
  \bibinfo {author} {\bibfnamefont {S.}~\bibnamefont {Kivelson}}} (\bibinfo
  {year} {2009}),\ \href {\doibase https://doi.org/10.1016/j.physb.2008.11.062}
  {\bibfield  {journal} {\bibinfo  {journal} {Physica B: Condensed Matter}\
  }\textbf {\bibinfo {volume} {404}}~(\bibinfo {number} {3}),\ \bibinfo {pages}
  {462 }}\BibitemShut {NoStop}%
\bibitem [{\citenamefont {Spivak}\ \emph {et~al.}(2008)\citenamefont {Spivak},
  \citenamefont {Oreto},\ and\ \citenamefont {Kivelson}}]{OretoKivSp}%
  \BibitemOpen
  \bibfield  {author} {\bibinfo {author} {\bibnamefont {Spivak}, \bibfnamefont
  {B.}}, \bibinfo {author} {\bibfnamefont {P.}~\bibnamefont {Oreto}}, \ and\
  \bibinfo {author} {\bibfnamefont {S.~A.}\ \bibnamefont {Kivelson}}} (\bibinfo
  {year} {2008}),\ \href {\doibase 10.1103/PhysRevB.77.214523} {\bibfield
  {journal} {\bibinfo  {journal} {Phys. Rev. B}\ }\textbf {\bibinfo {volume}
  {77}},\ \bibinfo {pages} {214523}}\BibitemShut {NoStop}%
\bibitem [{\citenamefont {Spivak}\ and\ \citenamefont
  {Zhou}(1995)}]{SpivakZhou}%
  \BibitemOpen
  \bibfield  {author} {\bibinfo {author} {\bibnamefont {Spivak}, \bibfnamefont
  {B.}}, \ and\ \bibinfo {author} {\bibfnamefont {F.}~\bibnamefont {Zhou}}}
  (\bibinfo {year} {1995}),\ \href {\doibase 10.1103/PhysRevLett.74.2800}
  {\bibfield  {journal} {\bibinfo  {journal} {Phys. Rev. Lett.}\ }\textbf
  {\bibinfo {volume} {74}},\ \bibinfo {pages} {2800}}\BibitemShut {NoStop}%
\bibitem [{\citenamefont {Spivak}\ \emph {et~al.}(2001)\citenamefont {Spivak},
  \citenamefont {Zyuzin},\ and\ \citenamefont {Hruska}}]{Hruska}%
  \BibitemOpen
  \bibfield  {author} {\bibinfo {author} {\bibnamefont {Spivak}, \bibfnamefont
  {B.}}, \bibinfo {author} {\bibfnamefont {A.}~\bibnamefont {Zyuzin}}, \ and\
  \bibinfo {author} {\bibfnamefont {M.}~\bibnamefont {Hruska}}} (\bibinfo
  {year} {2001}),\ \href {\doibase 10.1103/PhysRevB.64.132502} {\bibfield
  {journal} {\bibinfo  {journal} {Phys. Rev. B}\ }\textbf {\bibinfo {volume}
  {64}},\ \bibinfo {pages} {132502}}\BibitemShut {NoStop}%
\bibitem [{\citenamefont {Stauffer}\ and\ \citenamefont
  {Aharony}(1994)}]{percolation}%
  \BibitemOpen
  \bibfield  {author} {\bibinfo {author} {\bibnamefont {Stauffer},
  \bibfnamefont {D.}}, \ and\ \bibinfo {author} {\bibfnamefont
  {A.}~\bibnamefont {Aharony}}} (\bibinfo {year} {1994}),\ \href@noop {} {\emph
  {\bibinfo {title} {Introduction to percolation theory}}},\ \bibinfo {edition}
  {2nd}\ ed.\ (\bibinfo  {publisher} {Taylor $\&$ Francis})\BibitemShut
  {NoStop}%
\bibitem [{\citenamefont {Steiner}\ \emph {et~al.}(2008)\citenamefont
  {Steiner}, \citenamefont {Breznay},\ and\ \citenamefont
  {Kapitulnik}}]{Steiner}%
  \BibitemOpen
  \bibfield  {author} {\bibinfo {author} {\bibnamefont {Steiner}, \bibfnamefont
  {M.~A.}}, \bibinfo {author} {\bibfnamefont {N.~P.}\ \bibnamefont {Breznay}},
  \ and\ \bibinfo {author} {\bibfnamefont {A.}~\bibnamefont {Kapitulnik}}}
  (\bibinfo {year} {2008}),\ \href {\doibase 10.1103/PhysRevB.77.212501}
  {\bibfield  {journal} {\bibinfo  {journal} {Phys. Rev. B}\ }\textbf {\bibinfo
  {volume} {77}},\ \bibinfo {pages} {212501}}\BibitemShut {NoStop}%
\bibitem [{\citenamefont {Stiansen}\ \emph {et~al.}(2012)\citenamefont
  {Stiansen}, \citenamefont {Sperstad},\ and\ \citenamefont
  {Sudb\o{}}}]{sudbo}%
  \BibitemOpen
  \bibfield  {author} {\bibinfo {author} {\bibnamefont {Stiansen},
  \bibfnamefont {E.~B.}}, \bibinfo {author} {\bibfnamefont {I.~B.}\
  \bibnamefont {Sperstad}}, \ and\ \bibinfo {author} {\bibfnamefont
  {A.}~\bibnamefont {Sudb\o{}}}} (\bibinfo {year} {2012}),\ \href {\doibase
  10.1103/PhysRevB.85.224531} {\bibfield  {journal} {\bibinfo  {journal} {Phys.
  Rev. B}\ }\textbf {\bibinfo {volume} {85}},\ \bibinfo {pages}
  {224531}}\BibitemShut {NoStop}%
\bibitem [{\citenamefont {Stroud}\ and\ \citenamefont
  {Bergman}(1984)}]{Stroud1984}%
  \BibitemOpen
  \bibfield  {author} {\bibinfo {author} {\bibnamefont {Stroud}, \bibfnamefont
  {D.}}, \ and\ \bibinfo {author} {\bibfnamefont {D.~J.}\ \bibnamefont
  {Bergman}}} (\bibinfo {year} {1984}),\ \href {\doibase
  10.1103/PhysRevB.30.447} {\bibfield  {journal} {\bibinfo  {journal} {Phys.
  Rev. B}\ }\textbf {\bibinfo {volume} {30}},\ \bibinfo {pages}
  {447}}\BibitemShut {NoStop}%
\bibitem [{\citenamefont {Taillefer}(2010)}]{taillefer}%
  \BibitemOpen
  \bibfield  {author} {\bibinfo {author} {\bibnamefont {Taillefer},
  \bibfnamefont {L.}}} (\bibinfo {year} {2010}),\ \href {\doibase
  10.1146/annurev-conmatphys-070909-104117} {\bibfield  {journal} {\bibinfo
  {journal} {Annual Review of Condensed Matter Physics}\ }\textbf {\bibinfo
  {volume} {1}}~(\bibinfo {number} {1}),\ \bibinfo {pages} {51}},\ \Eprint
  {http://arxiv.org/abs/https://doi.org/10.1146/annurev-conmatphys-070909-104117}
  {https://doi.org/10.1146/annurev-conmatphys-070909-104117} \BibitemShut
  {NoStop}%
\bibitem [{\citenamefont {Tewari}\ \emph {et~al.}(2005)\citenamefont {Tewari},
  \citenamefont {Toner},\ and\ \citenamefont {Chakravarty}}]{chaktoner}%
  \BibitemOpen
  \bibfield  {author} {\bibinfo {author} {\bibnamefont {Tewari}, \bibfnamefont
  {S.}}, \bibinfo {author} {\bibfnamefont {J.}~\bibnamefont {Toner}}, \ and\
  \bibinfo {author} {\bibfnamefont {S.}~\bibnamefont {Chakravarty}}} (\bibinfo
  {year} {2005}),\ \href {\doibase 10.1103/PhysRevB.72.060505} {\bibfield
  {journal} {\bibinfo  {journal} {Phys. Rev. B}\ }\textbf {\bibinfo {volume}
  {72}},\ \bibinfo {pages} {060505}}\BibitemShut {NoStop}%
\bibitem [{\citenamefont {Thompson}(1970)}]{Thompson}%
  \BibitemOpen
  \bibfield  {author} {\bibinfo {author} {\bibnamefont {Thompson},
  \bibfnamefont {R.~S.}}} (\bibinfo {year} {1970}),\ \href {\doibase
  10.1103/PhysRevB.1.327} {\bibfield  {journal} {\bibinfo  {journal} {Phys.
  Rev. B}\ }\textbf {\bibinfo {volume} {1}},\ \bibinfo {pages}
  {327}}\BibitemShut {NoStop}%
\bibitem [{\citenamefont {Tsen}\ \emph {et~al.}(2016)\citenamefont {Tsen},
  \citenamefont {Hunt}, \citenamefont {Kim}, \citenamefont {Yuan},
  \citenamefont {Jia}, \citenamefont {Cava}, \citenamefont {Hone},
  \citenamefont {Kim}, \citenamefont {Dean},\ and\ \citenamefont
  {Pasupathy}}]{Tsen2016}%
  \BibitemOpen
  \bibfield  {author} {\bibinfo {author} {\bibnamefont {Tsen}, \bibfnamefont
  {A.~W.}}, \bibinfo {author} {\bibfnamefont {B.}~\bibnamefont {Hunt}},
  \bibinfo {author} {\bibfnamefont {Y.~D.}\ \bibnamefont {Kim}}, \bibinfo
  {author} {\bibfnamefont {Z.~J.}\ \bibnamefont {Yuan}}, \bibinfo {author}
  {\bibfnamefont {S.}~\bibnamefont {Jia}}, \bibinfo {author} {\bibfnamefont
  {R.~J.}\ \bibnamefont {Cava}}, \bibinfo {author} {\bibfnamefont
  {J.}~\bibnamefont {Hone}}, \bibinfo {author} {\bibfnamefont {P.}~\bibnamefont
  {Kim}}, \bibinfo {author} {\bibfnamefont {C.~R.}\ \bibnamefont {Dean}}, \
  and\ \bibinfo {author} {\bibfnamefont {A.~N.}\ \bibnamefont {Pasupathy}}}
  (\bibinfo {year} {{2016}}),\ \href {\doibase {10.1038/NPHYS3579}} {\bibfield
  {journal} {\bibinfo  {journal} {{Nature Physics}}\ }\textbf {\bibinfo
  {volume} {{12}}}~(\bibinfo {number} {{3}}),\ \bibinfo {pages}
  {{208+}}}\BibitemShut {NoStop}%
\bibitem [{\citenamefont {Vojta}\ \emph {et~al.}(2009)\citenamefont {Vojta},
  \citenamefont {Kotabage},\ and\ \citenamefont {Hoyos}}]{Vojta}%
  \BibitemOpen
  \bibfield  {author} {\bibinfo {author} {\bibnamefont {Vojta}, \bibfnamefont
  {T.}}, \bibinfo {author} {\bibfnamefont {C.}~\bibnamefont {Kotabage}}, \ and\
  \bibinfo {author} {\bibfnamefont {J.~A.}\ \bibnamefont {Hoyos}}} (\bibinfo
  {year} {2009}),\ \href {\doibase 10.1103/PhysRevB.79.024401} {\bibfield
  {journal} {\bibinfo  {journal} {Phys. Rev. B}\ }\textbf {\bibinfo {volume}
  {79}},\ \bibinfo {pages} {024401}}\BibitemShut {NoStop}%
\bibitem [{\citenamefont {Wagenblast}\ \emph {et~al.}(1997)\citenamefont
  {Wagenblast}, \citenamefont {van Otterlo}, \citenamefont {Sch\"on},\ and\
  \citenamefont {Zim\'anyi}}]{zimanyi}%
  \BibitemOpen
  \bibfield  {author} {\bibinfo {author} {\bibnamefont {Wagenblast},
  \bibfnamefont {K.-H.}}, \bibinfo {author} {\bibfnamefont {A.}~\bibnamefont
  {van Otterlo}}, \bibinfo {author} {\bibfnamefont {G.}~\bibnamefont
  {Sch\"on}}, \ and\ \bibinfo {author} {\bibfnamefont {G.~T.}\ \bibnamefont
  {Zim\'anyi}}} (\bibinfo {year} {1997}),\ \href {\doibase
  10.1103/PhysRevLett.79.2730} {\bibfield  {journal} {\bibinfo  {journal}
  {Phys. Rev. Lett.}\ }\textbf {\bibinfo {volume} {79}},\ \bibinfo {pages}
  {2730}}\BibitemShut {NoStop}%
\bibitem [{\citenamefont {Wang}\ \emph {et~al.}(2017)\citenamefont {Wang},
  \citenamefont {Tamir}, \citenamefont {Shahar},\ and\ \citenamefont
  {Armitage}}]{Wang2017}%
  \BibitemOpen
  \bibfield  {author} {\bibinfo {author} {\bibnamefont {Wang}, \bibfnamefont
  {Y.}}, \bibinfo {author} {\bibfnamefont {I.}~\bibnamefont {Tamir}}, \bibinfo
  {author} {\bibfnamefont {D.}~\bibnamefont {Shahar}}, \ and\ \bibinfo {author}
  {\bibfnamefont {N.~P.}\ \bibnamefont {Armitage}}} (\bibinfo {year} {2017}),\
  \href@noop {} {\ }\bibinfo {note} {ArXiv:1708.01908}\BibitemShut {NoStop}%
\bibitem [{\citenamefont {Wang}\ \emph {et~al.}(2007)\citenamefont {Wang},
  \citenamefont {Yan}, \citenamefont {Shan}, \citenamefont {Wen}, \citenamefont
  {Tanabe}, \citenamefont {Adachi},\ and\ \citenamefont {Koike}}]{wen}%
  \BibitemOpen
  \bibfield  {author} {\bibinfo {author} {\bibnamefont {Wang}, \bibfnamefont
  {Y.}}, \bibinfo {author} {\bibfnamefont {J.}~\bibnamefont {Yan}}, \bibinfo
  {author} {\bibfnamefont {L.}~\bibnamefont {Shan}}, \bibinfo {author}
  {\bibfnamefont {H.-H.}\ \bibnamefont {Wen}}, \bibinfo {author} {\bibfnamefont
  {Y.}~\bibnamefont {Tanabe}}, \bibinfo {author} {\bibfnamefont
  {T.}~\bibnamefont {Adachi}}, \ and\ \bibinfo {author} {\bibfnamefont
  {Y.}~\bibnamefont {Koike}}} (\bibinfo {year} {2007}),\ \href {\doibase
  10.1103/PhysRevB.76.064512} {\bibfield  {journal} {\bibinfo  {journal} {Phys.
  Rev. B}\ }\textbf {\bibinfo {volume} {76}},\ \bibinfo {pages}
  {064512}}\BibitemShut {NoStop}%
\bibitem [{\citenamefont {Werner}\ \emph {et~al.}(2005)\citenamefont {Werner},
  \citenamefont {Refael},\ and\ \citenamefont {Troyer}}]{troyer}%
  \BibitemOpen
  \bibfield  {author} {\bibinfo {author} {\bibnamefont {Werner}, \bibfnamefont
  {P.}}, \bibinfo {author} {\bibfnamefont {G.}~\bibnamefont {Refael}}, \ and\
  \bibinfo {author} {\bibfnamefont {M.}~\bibnamefont {Troyer}}} (\bibinfo
  {year} {2005}),\ \href {\doibase 10.1088/1742-5468/2005/12/P12003} {\bibinfo
  {journal} {Journal of Statistical Mechanics-Theory and Experiment}\ ,\
  \bibinfo {pages} {P12003}}\BibitemShut {NoStop}%
\bibitem [{\citenamefont {White}\ \emph {et~al.}(1986)\citenamefont {White},
  \citenamefont {Dynes},\ and\ \citenamefont {Garno}}]{WhiteDynes}%
  \BibitemOpen
\bibfield  {journal} {  }\bibfield  {author} {\bibinfo {author} {\bibnamefont
  {White}, \bibfnamefont {A.~E.}}, \bibinfo {author} {\bibfnamefont {R.~C.}\
  \bibnamefont {Dynes}}, \ and\ \bibinfo {author} {\bibfnamefont {J.~P.}\
  \bibnamefont {Garno}}} (\bibinfo {year} {1986}),\ \href {\doibase
  10.1103/PhysRevB.33.3549} {\bibfield  {journal} {\bibinfo  {journal} {Phys.
  Rev. B}\ }\textbf {\bibinfo {volume} {33}},\ \bibinfo {pages}
  {3549}}\BibitemShut {NoStop}%
\bibitem [{\citenamefont {Yao}\ \emph {et~al.}(2009)\citenamefont {Yao},
  \citenamefont {Zhang},\ and\ \citenamefont {Kivelson}}]{hongzhangandme}%
  \BibitemOpen
  \bibfield  {author} {\bibinfo {author} {\bibnamefont {Yao}, \bibfnamefont
  {H.}}, \bibinfo {author} {\bibfnamefont {S.-C.}\ \bibnamefont {Zhang}}, \
  and\ \bibinfo {author} {\bibfnamefont {S.~A.}\ \bibnamefont {Kivelson}}}
  (\bibinfo {year} {2009}),\ \href {\doibase 10.1103/PhysRevLett.102.217202}
  {\bibfield  {journal} {\bibinfo  {journal} {Phys. Rev. Lett.}\ }\textbf
  {\bibinfo {volume} {102}},\ \bibinfo {pages} {217202}}\BibitemShut {NoStop}%
\bibitem [{\citenamefont {Yazdani}(1994)}]{YazdaniThesis}%
  \BibitemOpen
  \bibfield  {author} {\bibinfo {author} {\bibnamefont {Yazdani}, \bibfnamefont
  {A.}}} (\bibinfo {year} {1994}),\ \emph {\bibinfo {title} {Phase Transitions
  in Two-Dimensional Superconductors}},\ \href@noop {} {Ph.D. thesis}\
  (\bibinfo  {school} {Stanford University}, \bibinfo {address} {Stanford, CA
  94305})\BibitemShut {NoStop}%
\bibitem [{\citenamefont {Yazdani}\ and\ \citenamefont
  {Kapitulnik}(1995)}]{Yazdani}%
  \BibitemOpen
  \bibfield  {author} {\bibinfo {author} {\bibnamefont {Yazdani}, \bibfnamefont
  {A.}}, \ and\ \bibinfo {author} {\bibfnamefont {A.}~\bibnamefont
  {Kapitulnik}}} (\bibinfo {year} {1995}),\ \href {\doibase
  10.1103/PhysRevLett.74.3037} {\bibfield  {journal} {\bibinfo  {journal}
  {Phys. Rev. Lett.}\ }\textbf {\bibinfo {volume} {74}},\ \bibinfo {pages}
  {3037}}\BibitemShut {NoStop}%
\bibitem [{\citenamefont {Yazdani}\ \emph {et~al.}(1993)\citenamefont
  {Yazdani}, \citenamefont {White}, \citenamefont {Hahn}, \citenamefont
  {Gabay}, \citenamefont {Beasley},\ and\ \citenamefont
  {Kapitulnik}}]{Yazdani1993}%
  \BibitemOpen
  \bibfield  {author} {\bibinfo {author} {\bibnamefont {Yazdani}, \bibfnamefont
  {A.}}, \bibinfo {author} {\bibfnamefont {W.~R.}\ \bibnamefont {White}},
  \bibinfo {author} {\bibfnamefont {M.~R.}\ \bibnamefont {Hahn}}, \bibinfo
  {author} {\bibfnamefont {M.}~\bibnamefont {Gabay}}, \bibinfo {author}
  {\bibfnamefont {M.~R.}\ \bibnamefont {Beasley}}, \ and\ \bibinfo {author}
  {\bibfnamefont {A.}~\bibnamefont {Kapitulnik}}} (\bibinfo {year} {1993}),\
  \href {\doibase 10.1103/PhysRevLett.70.505} {\bibfield  {journal} {\bibinfo
  {journal} {Phys. Rev. Lett.}\ }\textbf {\bibinfo {volume} {70}},\ \bibinfo
  {pages} {505}}\BibitemShut {NoStop}%
\bibitem [{\citenamefont {Ye}\ \emph {et~al.}(2012)\citenamefont {Ye},
  \citenamefont {Zhang}, \citenamefont {Akashi}, \citenamefont {Bahramy},
  \citenamefont {Arita},\ and\ \citenamefont {Iwasa}}]{Iwasa}%
  \BibitemOpen
  \bibfield  {author} {\bibinfo {author} {\bibnamefont {Ye}, \bibfnamefont
  {J.~T.}}, \bibinfo {author} {\bibfnamefont {Y.~J.}\ \bibnamefont {Zhang}},
  \bibinfo {author} {\bibfnamefont {R.}~\bibnamefont {Akashi}}, \bibinfo
  {author} {\bibfnamefont {M.~S.}\ \bibnamefont {Bahramy}}, \bibinfo {author}
  {\bibfnamefont {R.}~\bibnamefont {Arita}}, \ and\ \bibinfo {author}
  {\bibfnamefont {Y.}~\bibnamefont {Iwasa}}} (\bibinfo {year} {2012}),\ \href
  {\doibase 10.1126/science.1228006} {\bibfield  {journal} {\bibinfo  {journal}
  {Science}\ }\textbf {\bibinfo {volume} {338}}~(\bibinfo {number} {6111}),\
  \bibinfo {pages} {1193}}\BibitemShut {NoStop}%
\bibitem [{\citenamefont {Yu}\ \emph {et~al.}(2016)\citenamefont {Yu},
  \citenamefont {Hirschberger}, \citenamefont {Loew}, \citenamefont {Li},
  \citenamefont {Lawson}, \citenamefont {Asaba}, \citenamefont {Kemper},
  \citenamefont {Liang}, \citenamefont {Porras}, \citenamefont {Boebinger},
  \citenamefont {Singleton}, \citenamefont {Keimer}, \citenamefont {Li},\ and\
  \citenamefont {Ong}}]{ong}%
  \BibitemOpen
  \bibfield  {author} {\bibinfo {author} {\bibnamefont {Yu}, \bibfnamefont
  {F.}}, \bibinfo {author} {\bibfnamefont {M.}~\bibnamefont {Hirschberger}},
  \bibinfo {author} {\bibfnamefont {T.}~\bibnamefont {Loew}}, \bibinfo {author}
  {\bibfnamefont {G.}~\bibnamefont {Li}}, \bibinfo {author} {\bibfnamefont
  {B.~J.}\ \bibnamefont {Lawson}}, \bibinfo {author} {\bibfnamefont
  {T.}~\bibnamefont {Asaba}}, \bibinfo {author} {\bibfnamefont {J.~B.}\
  \bibnamefont {Kemper}}, \bibinfo {author} {\bibfnamefont {T.}~\bibnamefont
  {Liang}}, \bibinfo {author} {\bibfnamefont {J.}~\bibnamefont {Porras}},
  \bibinfo {author} {\bibfnamefont {G.~S.}\ \bibnamefont {Boebinger}}, \bibinfo
  {author} {\bibfnamefont {J.}~\bibnamefont {Singleton}}, \bibinfo {author}
  {\bibfnamefont {B.}~\bibnamefont {Keimer}}, \bibinfo {author} {\bibfnamefont
  {L.}~\bibnamefont {Li}}, \ and\ \bibinfo {author} {\bibfnamefont {N.~P.}\
  \bibnamefont {Ong}}} (\bibinfo {year} {2016}),\ \href {\doibase
  10.1073/pnas.1612591113} {\bibfield  {journal} {\bibinfo  {journal}
  {Proceedings of the National Academy of Sciences of the United States of
  America}\ }\textbf {\bibinfo {volume} {113}}~(\bibinfo {number} {45}),\
  \bibinfo {pages} {12667}}\BibitemShut {NoStop}%
\bibitem [{\citenamefont {Zhang}\ \emph {et~al.}(1989)\citenamefont {Zhang},
  \citenamefont {Hansson},\ and\ \citenamefont {Kivelson}}]{ZHK}%
  \BibitemOpen
  \bibfield  {author} {\bibinfo {author} {\bibnamefont {Zhang}, \bibfnamefont
  {S.~C.}}, \bibinfo {author} {\bibfnamefont {T.~H.}\ \bibnamefont {Hansson}},
  \ and\ \bibinfo {author} {\bibfnamefont {S.}~\bibnamefont {Kivelson}}}
  (\bibinfo {year} {1989}),\ \href {\doibase 10.1103/PhysRevLett.62.82}
  {\bibfield  {journal} {\bibinfo  {journal} {Phys. Rev. Lett.}\ }\textbf
  {\bibinfo {volume} {62}},\ \bibinfo {pages} {82}}\BibitemShut {NoStop}%
\bibitem [{\citenamefont {Zhou}\ and\ \citenamefont
  {Spivak}(1998)}]{ZhouSpivak}%
  \BibitemOpen
  \bibfield  {author} {\bibinfo {author} {\bibnamefont {Zhou}, \bibfnamefont
  {F.}}, \ and\ \bibinfo {author} {\bibfnamefont {B.}~\bibnamefont {Spivak}}}
  (\bibinfo {year} {1998}),\ \href {\doibase 10.1103/PhysRevLett.80.5647}
  {\bibfield  {journal} {\bibinfo  {journal} {Phys. Rev. Lett.}\ }\textbf
  {\bibinfo {volume} {80}},\ \bibinfo {pages} {5647}}\BibitemShut {NoStop}%
\bibitem [{\citenamefont {Zhu}\ \emph {et~al.}(2015)\citenamefont {Zhu},
  \citenamefont {Chen},\ and\ \citenamefont {Varma}}]{varma3}%
  \BibitemOpen
  \bibfield  {author} {\bibinfo {author} {\bibnamefont {Zhu}, \bibfnamefont
  {L.}}, \bibinfo {author} {\bibfnamefont {Y.}~\bibnamefont {Chen}}, \ and\
  \bibinfo {author} {\bibfnamefont {C.~M.}\ \bibnamefont {Varma}}} (\bibinfo
  {year} {2015}),\ \href {\doibase 10.1103/PhysRevB.91.205129} {\bibfield
  {journal} {\bibinfo  {journal} {Phys. Rev. B}\ }\textbf {\bibinfo {volume}
  {91}},\ \bibinfo {pages} {205129}}\BibitemShut {NoStop}%
\bibitem [{\citenamefont {Zhu}\ \emph {et~al.}(2016)\citenamefont {Zhu},
  \citenamefont {Hou},\ and\ \citenamefont {Varma}}]{varma1}%
  \BibitemOpen
  \bibfield  {author} {\bibinfo {author} {\bibnamefont {Zhu}, \bibfnamefont
  {L.}}, \bibinfo {author} {\bibfnamefont {C.}~\bibnamefont {Hou}}, \ and\
  \bibinfo {author} {\bibfnamefont {C.~M.}\ \bibnamefont {Varma}}} (\bibinfo
  {year} {2016}),\ \href {\doibase 10.1103/PhysRevB.94.235156} {\bibfield
  {journal} {\bibinfo  {journal} {Phys. Rev. B}\ }\textbf {\bibinfo {volume}
  {94}},\ \bibinfo {pages} {235156}}\BibitemShut {NoStop}%
\bibitem [{\citenamefont {Zwerger}\ \emph {et~al.}(1986)\citenamefont
  {Zwerger}, \citenamefont {Dorsey},\ and\ \citenamefont {Fisher}}]{zwerger}%
  \BibitemOpen
  \bibfield  {author} {\bibinfo {author} {\bibnamefont {Zwerger}, \bibfnamefont
  {W.}}, \bibinfo {author} {\bibfnamefont {A.~T.}\ \bibnamefont {Dorsey}}, \
  and\ \bibinfo {author} {\bibfnamefont {M.~P.~A.}\ \bibnamefont {Fisher}}}
  (\bibinfo {year} {1986}),\ \href {\doibase 10.1103/PhysRevB.34.6518}
  {\bibfield  {journal} {\bibinfo  {journal} {Phys. Rev. B}\ }\textbf {\bibinfo
  {volume} {34}},\ \bibinfo {pages} {6518}}\BibitemShut {NoStop}%
\end{thebibliography}%
 
\end{document}